\documentclass[12pt,letterpaper]{article}
\pdfoutput=1
\usepackage{soul}
\usepackage{jheppub}
\usepackage{amsfonts}
\usepackage{amsmath}
\allowdisplaybreaks[4]         
\usepackage{amssymb}   
\usepackage{euscript}       
\usepackage{color}          
\newcommand{\labell}[1]{\label{#1}} %{\label{#1}\qquad\mt{#1}} %
\newcommand{\de}{\delta}
\newcommand{\req}[1]{(\ref{#1})} %{Eq.\thinspace(\ref{#1})}  

\newcommand{\bea}{\begin{eqnarray}}
\newcommand{\eea}{\end{eqnarray}}
\newcommand{\ba}{\begin{eqnarray}}
\newcommand{\ea}{\end{eqnarray}}

\newcommand{\beq}{\begin{equation}}
\newcommand{\eeq}{\end{equation} }
\newcommand{\beqa}{\begin{eqnarray}}
\newcommand{\eeqa}{\end{eqnarray}}
\newcommand{\beqar}{\begin{eqnarray*}}
\newcommand{\eeqar}{\end{eqnarray*}}

\renewcommand{\req}[1]{eq.~(\ref{#1})}

\newcommand{\ssc}{\scriptscriptstyle}
\newcommand{\eg}{{\it e.g.,}\ }
\newcommand{\ie}{{\it i.e.,}\ }

\newcommand{\see}{S_{\ssc \rm EE}}

\newcommand{\ctt}{C_{\ssc T}}

\DeclareMathOperator{\tr}{tr}  
\DeclareMathOperator{\sgn}{sgn}

\arxivnumber{arXiv:1904.11495}
\title{ \boldmath Generalizing the entanglement entropy of singular regions in conformal field theories}

\author[a]{Pablo Bueno,}
\author[a]{Horacio Casini}
\author[b]{and William Witczak-Krempa}

\affiliation[a]{
Instituto Balseiro, Centro At\'omico Bariloche, 8400-S.C. de Bariloche, R\'io Negro, Argentina}
\affiliation[b]{Departement de physique, Universit\'e de Montr\'eal, Montr\'eal (Qu\'ebec), H3C 3J7, Canada}

\emailAdd{pablo.bueno@cab.cnea.gov.ar}
\emailAdd{casini@cab.cnea.gov.ar}
\emailAdd{w.witczak-krempa@umontreal.ca}

\date{\today}
\abstract{
We study the structure of divergences and universal terms of the entanglement and R\'enyi entropies for singular regions.
First, we show that for $(3+1)$-dimensional free conformal field theories (CFTs), entangling regions emanating from vertices 
give rise to a universal contribution $S_n^{\rm univ}= -\frac{1}{8\pi}f_b(n)  \int_{\gamma} k^2 \log^2(R/\delta)$, 
where $\gamma$ is the curve formed by the intersection of the entangling surface with a unit sphere centered at the vertex, and $k$ the trace of its extrinsic curvature.
While for circular and elliptic cones this term reproduces the general-CFT result, it vanishes for polyhedral corners. For those, we argue that the universal contribution, which is logarithmic, is not controlled by a local integral, but rather it depends on details of the CFT in a complicated way.
We also study the angle dependence for the entanglement entropy of wedge singularities in 3+1 dimensions.
This is done for general CFTs in the smooth limit, and using free and holographic CFTs at generic angles. In the latter case, we
show that the wedge contribution is not proportional to the  entanglement entropy of a corner region in the $(2+1)$-dimensional 
holographic CFT.
Finally, we show that the mutual information of two regions that touch at a point is not necessarily divergent, as long as the contact is through a sufficiently sharp corner. Similarly, we provide examples of singular entangling regions which do not modify the structure of divergences of the entanglement entropy compared with smooth surfaces.
 }

\begin{document} 
\maketitle
\flushbottom
%%%%%%%%%%%%%%%%%%%%%%%%%%%%%%%%%%%%%
%\setcounter{tocdepth}{2}
%the line above sets the depth of the table of contents. {2} means it will display section and subsections only.
%{\small
%\setlength\parskip{-0.5mm} 
%\tableofcontents
%}

\newpage
%%%%%%%%%%%%%%%%%%%%%%%%%%%%%%
\section{Introduction}
\label{sec:Introduction}
%%%%%%%%%%%%%%%%%%%%%%%%%%%%%%
%Entanglement entropy (EE)
Given a bipartition of the Hilbert space of a quantum system $\mathcal{H}=\mathcal{H}_V\otimes \mathcal{H}_{\bar V}$, the R\'enyi and entanglement entropies associated with $V$ in some state $\rho$ are defined as
\begin{equation}
S_n(V)=\frac{1}{1-n}\log {\rm Tr} \rho_V^n\, , \quad \see(V)=\lim_{n\rightarrow 1} S_n(V)=-{\rm  Tr} (\rho_V \log \rho_V)\, ,
\end{equation}
where $\rho_V= {\rm Tr}_{\bar V}\rho$ is the partial-trace density matrix obtained by integrating the degrees of freedom in the complement of $V$. 
In the context of quantum field theory, R\'enyi entropies are intrinsically UV-divergent.\footnote{See \cite{Nishioka:2018khk} and \cite{Witten:2018lha} for two interesting recent reviews with somewhat complementary scopes.} In particular, given a smooth
entangling region $V$ on a time slice of a $d$-dimensional conformal field theory (CFT), the R\'enyi entropy takes the generic form ---see \eg \cite{Grover:2011fa,Liu:2012eea},
\begin{equation}\label{REn}
S_n^{(d)}= b_{d-2} \frac{H^{d-2}}{\delta^{d-2}}+ b_{d-4} \frac{H^{d-4}}{\delta^{d-4}}+\cdots  + 
 \begin{cases}
b_1  \frac{H}{\delta} +(-1)^{\frac{d-1}{2}} s_n^{\rm univ}\, , & (\text{odd}\,\, d)\, ,  \\
  b_2  \frac{H^2}{\delta^2}+(-1)^{\frac{d-2}{2}} s_n^{\rm univ} \log\left(\frac{H}{\delta} \right)+b_0\, , & (\text{even}\, \, d)\, .
  \end{cases}
\end{equation}
In this expression, $H$ is some characteristic length of $V$, $\delta$ is a UV regulator, and the (R\'enyi index-dependent) coefficients $b_i$ are non-universal, \ie they depend on the regularization scheme.

 In even dimensions, the universal term is logarithmic and its coefficient, $s_n^{\rm univ}$, is controlled by a linear combination of local integrals on the entangling surface $\partial V\equiv \Sigma$ weighted by certain theory-dependent charges which reduce to the corresponding trace-anomaly coefficients for $n=1$. The simplest case corresponds to a segment of length $H$ in a two-dimensional CFT, for which \cite{Calabrese:2004eu,Calabrese:2009qy}
$
S^{(2)}_n=\frac{c}{6} \left(1+\frac{1}{n} \right) \log\left(\frac{H}{\delta} \right)+\mathcal{O}(\delta^0)\, ,
$
where $c$ is the Virasoro central charge. The next case is that of four-dimensional theories. For those, there are three theory-dependent functions of the R\'enyi index, customarily denoted $f_a(n),f_b(n)$ and $f_c(n)$, which control the linear combination of local integrals characterizing the logarithmic universal contribution   \cite{Solodukhin:2008dh,Fursaev:2012mp} ---see \req{game} below. For $n=1$, they reduce to the usual trace-anomaly coefficients, $f_a(1)=a$, $f_b(1)=f_c(1)=c$. An analogous story holds for $d\geq 6$ ---see \eg \cite{Safdi:2012sn,Miao:2015iba}. % In all cases, the dependence on the geometric features of $\Sigma$ is independent of the CFT under consideration. 
The situation is different in odd dimensions though. For those, no logarithmic contribution is present for smooth entangling surfaces, and the universal contribution is a constant term which no longer corresponds to a simple local integral over $\Sigma$. The simplest case corresponds to three-dimensional CFTs, for which\footnote{Constant terms such as $F_n$ are less robust than their even-dimensional logarithmic counterparts. This is because we cannot resolve the relevant IR scales of the entangling region with more precision than the UV cutoff. If we shift the relevant characteristic scale as $R\rightarrow R+a \delta$, with $a=\mathcal{O}(1)$, we will pollute the putative universal contribution as $F_n\rightarrow F_n(1-b_1 a)$. This pollution ---which does not occur for logarithmic contributions---  can be remedied using mutual information as a geometric regulator \cite{Casini:2006ws,chm2}.}
$
S^{(3)}_n=b_1 \frac{H}{\delta} - F_n\, .
$
For $\Sigma=\mathbb{S}^1$, $F$ actually equals the free energy of the corresponding theory on $\mathbb{S}^3$ \cite{CHM,Dowker:2010yj}, which reveals its non-local nature.\footnote{Note that when we omit the subindex $n$ from the different contributions, we will be referring to the entanglement entropy case, corresponding to $n=1$.}

When geometric singularities are present in $\Sigma$, the structure of divergences in \req{REn} gets modified. The prototypical case is that of an entangling region bounded by a corner of opening angle $\Omega$ for three-dimensional CFTs. In that case, a new logarithmic universal contribution appears % the
%the R\'enyi entropy reads
\begin{equation}\label{corner2}
S_n^{(3)\, \rm corner} =b_1 \frac{H}{\delta}- a^{(3)}_n(\Omega)\log \left(\frac{H}{\delta} \right)+b_0\, ,
\end{equation}    
where $a^{(3)}_n(\Omega)$ is a cutoff-independent function of the opening angle which has been extensively studied in the literature ---\eg for free fields in \cite{Casini:2006hu,Casini:2008as,Casini:2009sr,Bueno3,Dowker:2015tma,Dowker:2015pwa,Elvang:2015jpa,Berthiere:2018ouo}, for large-$N$ vector models in \cite{Whitsitt2017}, for holographic theories in \cite{Hirata:2006jx,Myers:2012vs,Bueno2,Fonda:2015nma,Miao:2015dua,Alishahiha:2015goa,Pang:2015lka,Mozaffar:2015xue,Bianchi:2016xvf,Seminara:2017hhh,Pastras:2017fsy,Bakhshaei:2017qud,Ghasemi:2017pke,Seminara:2018pmr,Caceres:2018luq,Ghasemi:2019mif}, in interacting lattice models in \cite{2011PhRvB..84p5134K,Kallin:2014oka,Helmes:2015mwa,Laflorencie:2015lwa,Helmes:2016fcp,DeNobili:2016nmj}, and for general CFTs in \cite{Bueno1,Faulkner:2015csl,Bueno:2015ofa,Witczak-Krempa:2016jhc,Chu:2016tps}.

%fradkin,ardonne

The explicit dependence of $a^{(3)}_n(\Omega)$ on the opening angle and the R\'enyi index changes from one CFT to another ---\eg compare the relatively simple holographic result \cite{Hirata:2006jx} with the highly complicated  resulting expressions for free fields \cite{Casini:2006hu,Casini:2008as,Casini:2009sr}. The nature of $a^{(3)}_n(\Omega)$ is in stark contrast with that of the analogous coefficient corresponding to a conical entangling surface in four-dimensions. In that case, a similar logarithmic enhancement of the universal term does occur, and the  R\'enyi entropy reads
\begin{equation}\label{conne}
S_n^{(4)\, \rm cone} =b_2 \frac{H^2}{\delta^2}- a^{(4)}_n(\Omega)\log^2 \left(\frac{H}{\delta} \right)+ b_0 \log\left( \frac{H}{\delta}\right)+\mathcal{O}(\delta^0) \, .
\end{equation}
For the cone, however, the universal function $a^{(4)}_n(\Omega)$ is much more constrained than $a^{(3)}_n(\Omega)$. On the one hand, the dependence on the opening angle and the R\'enyi index factorize and, on the other, the explicit angular dependence is the same for all four-dimensional CFTs, namely\footnote{Observe that a remarkable degree of universality was nonetheless shown to hold for  $a^{(3)}(\Omega)$ in  \cite{Bueno1,Bueno2,Faulkner:2015csl}, 
 in the sense that normalizing this function by the stress-tensor two-point function charge $\ctt$ \cite{Osborn:1993cr}, the curves corresponding to very different theories become very close to each other. The agreement becomes exact in the almost smooth limit, namely, $a^{(3)}(\Omega)=\frac{\pi^2}{24}\ctt \left(\pi -\Omega \right)^2+\dots$ for general CFTs. The result generalizes to higher-dimensional (hyper)cones \cite{Bueno4,Miao:2015dua} ---see also \cite{Mezei:2014zla}. The possible generalization to $n\neq 1$ R\'enyi entropies turns out to be trickier \cite{Bueno3,Dowker:2015tma,Dowker:2015pwa,Bianchi:2015liz,Dong:2016wcf,Balakrishnan:2016ttg,Bianchi:2016xvf,Chu:2016tps}.} \cite{Klebanov:2012yf,Myers:2012vs}
\begin{equation}\label{a4c}
a^{(4)}_n(\Omega)=\frac{1}{4}f_b(n)\frac{\cos^2\Omega}{\sin \Omega}\, .
\end{equation}
This contrast can be understood from the fact that both $a^{(3)}_n(\Omega)$ and $a^{(4)}_n(\Omega)$ can be thought of as emerging from the respective contributions $s_n^{\rm univ}$ in \req{REn} ---see \eg \cite{Bueno4}. While in $d=3$ this is a constant and non-local term, in $d=4$ it is a geometric integral over the entangling surface. As we explain here, the origin of this difference can be made extremely manifest in the case of free fields, for which the computation of $a^{(3)}_n(\Omega)$ requires the full evaluation of a spectral function on $\mathbb{S}^2$ with a cut of angle $\Omega$ while $a^{(4)}_n(\Omega)$ arises from a simple local integral on the curve resulting from the intersection of the cone with a unit $\mathbb{S}^2$ centered at its tip  ---see our summary of results below.

There are other interesting singular regions one can think of in $d=4$. For example, we can consider the case of a polyhedral corner of opening angles $\theta_1,\theta_2,\dots,\theta_j$. In particular, this geometry is natural in lattice simulations. In that case, it has been observed that the universal contribution is not quadratically logarithmic, but just logarithmic,  like in the case of smooth entangling regions, namely \cite{Kovacs,Devakul2014,Sierens:2017eun,Bednik:2018iby,Witczak-Krempa:2018mqx}
\begin{equation}\label{polys2}
S_n^{(4)\, \rm polyhedral} =b_2 \frac{H^2}{\delta^2}-w_1 \frac{H}{\delta}+ v_n(\theta_1,\theta_2,\cdots,\theta_j) \log \left(\frac{L}{\delta}\right)+\mathcal{O}(\delta^0)\, .
\end{equation}
In \cite{Sierens:2017eun,Bednik:2018iby}, the idea that $v_n(\theta_1,\theta_2,\cdots,\theta_j)$ might be controlled by a simple linear combination of the functions $f_a(n)$ and $f_b(n)$ was put forward, although a definite conclusion was not reached. Here, we will use free-field calculations to understand the true origin of $v_n(\theta_1,\theta_2,\cdots,\theta_j)$, and rule out this possibility. %: very similarly to $a^{(3)}_n(\Omega)$, computing  $v_n(\theta_1,\theta_2,\cdots, \theta_j)$ for those would require  the evaluation of \comment{full spectral integral}.  

In \req{polys2} we observe the appearance of an additional nonuniversal contribution weighted by some constant $w_1$. While $ v_n(\theta_1,\theta_2,\cdots,\theta_j)$ comes from the corner itself, $w_1$ arises from the presence of wedges in the entangling region. If we consider a simpler setup corresponding to an infinitely extended wedge region of opening angle $\Omega$, the R\'enyi entropy is in turn given by \cite{Klebanov:2012yf,Myers:2012vs}
\begin{equation}\label{kle2}
S_n^{(4)\, \rm wedge} =b_2 \frac{H^2}{\delta^2}- f_n(\Omega)\frac{H}{\delta}+\mathcal{O}(\delta^0)\, ,
\end{equation}
where $f_n(\Omega)$ is a function of the wedge opening angle. Naturally, the overall normalization cannot be well defined, since simple redefinitions of the cutoff modify it. However, using holographic and free-field calculations, it has been suggested \cite{Klebanov:2012yf} that the angular dependence of $f_n(\Omega)$ matches the one corresponding to a corner region in one dimension less, namely $ \partial_\Omega \left(f_n(\Omega)/a^{(3)}_n(\Omega)\right)\overset{(?)}{=}0$. We present a careful study of the relation between both functions that reveals that the angular dependence of $f_n(\Omega)$ differs from $a^{(3)}_n(\Omega)$ in general.

In all cases mentioned so far, the entangling surfaces can be thought of as families of straight lines emanating from a vertex. However, one can consider more general entangling surfaces, such as the ones arising from curved corners. In the second part of the paper we will study the somewhat surprising interplay between those and the structure of divergences and universal terms in the R\'enyi entropy and also the mutual information.  A detailed summary of our results can be found next.  

\subsection{Summary of results}\label{summ}
\begin{itemize}
\item[$\diamond$]
 In section \ref{vertex}, we consider the problem of universal contributions to the R\'enyi entropy induced by the presence of vertices in the entangling region. In particular, for a $d$-dimensional free scalar field, we show that the usual logarithmic contribution can be related to the R\'enyi entropy of a region with the same angular boundary conditions on $(d-1)$-dimensional de Sitter space. Performing a high-mass expansion of the latter, we show that the logarithmic contribution gets enhanced to a quadratically logarithmic term of the form
\begin{equation}\label{log22}
S_{n}|_{\log^2}= \frac{-f_b(n)}{8\pi} \log^2\delta  \int_{\gamma} k^2\, ,
\end{equation}
where $\gamma$ is the curve resulting from the intersection of the entangling region $V$ with the unit $\mathbb{S}^2$ centered at its tip, and $k$ the trace of the extrinsic curvature of $\gamma$. In the case of cones and elliptic cones, we explicitly verify that this term agrees with the result obtained from Solodukhin's formula. For polyhedral corners, however, $\gamma$ always corresponds to the union of great circles, for which $k=0$, which explains the absence of a $\log^2\delta$ term in \req{polys2}. Instead, $v_n(\theta_1,\theta_2,\cdots,\theta_j)$ can be seen to arise from the constant contribution $b_0$ in \req{REn} which would require a complicated calculation of an spectral function on $\mathbb{S}^3$ with a cut, similarly to what happened for the corner function $a^{(3)}_n(\Omega)$. An explicit calculation of $v(\pi/2,\pi/2,\pi/2)$ for a trihedral corner in the so-called ``Extensive Mutual Information Model'' is also provided.

\item[$\diamond$] In section \ref{wvsc}, we consider the entanglement entropy of wedge regions in $d=4$. First, we compute the wedge function $f(\Omega)$, as defined in \req{kle2}, in the nearly smooth limit for general CFTs. Then, we argue that the entanglement entropy of a massive free field in $d=3$ corresponding to a corner region can be related to the one corresponding to the wedge in $d=4$. Using this, we show that the angular dependence of $a^{(3)}(\Omega)$ and $f(\Omega)$ do agree with each other, whereas the overall normalization of $f(\Omega)$ is nonuniversal due to the presence of different IR and UV regulators along the transverse and longitudinal directions. Then, we revisit the holographic results for both entangling regions and show that, contrary to the claim in \cite{Klebanov:2012yf}, the angular dependence of both functions is in fact different in that case, and therefore for general CFTs. 

\item[$\diamond$] In section \ref{singi}, we consider entangling regions with sharpened and smoothened corners. First, we show that, contrary to common belief, the mutual information of two regions which touch at a point is not necessarily divergent as long as the contact occurs through a sufficiently sharp corner. Then, we argue that the R\'enyi entropy of regions containing geometric singularities does not always modify the structure \req{REn} characteristic of smooth entangling regions. On the other hand, when the corners are sharper than in the usual straight corner case, the usual logarithmic divergence gets replaced by a more divergent term which, nonetheless, never surpasses the area law one. 

\item[$\diamond$] In appendix \ref{coneEMI2}, we perform an explicit calculation of the cone function $a^{(4)}(\Omega)$ in the Extensive Mutual Information model and show that it agrees with the result valid for general CFTs.

\item[$\diamond$] In appendix \ref{maxim}, we use the explicit formula obtained in section \ref{elic} for the R\'enyi entropy of elliptic cones, to show that circular cones locally maximize the R\'enyi entropy within the class of (fixed-area) elliptic cones.

\item[$\diamond$] Finally, in appendix \ref{hypers} we show that the dependence on the opening angle of the (hyper)cone function $a_n^{(d)}(\Omega)$ for arbitrary even-dimensional CFTs is given by the four-dimensional result, $\cos^2\Omega/\sin\Omega$ times a linear combination of  the form: $\gamma_{0,n}^{(d)}+\gamma_{1,n}^{(d)}\cos(2\Omega)+ \gamma_{2,n}^{(d)}\cos(4\Omega)+\dots+\gamma^{(d)}_{(d-4)/2,n} \cos\left((d-4)\Omega\right)$ for certain theory-dependent quantities $\gamma_{i,n}^{(d)}$ related to the trace-anomaly charges for $n=1$.

\end{itemize}

\newpage

\section{Vertex-induced universal terms}\label{vertex}
In this section we study the universal contributions to the R\'enyi entropy arising when the entangling region contains vertices. In the case of a free scalar field, a radial dimensional reduction and a mapping of the problem to $(d-1)$-dimensional de Sitter space allows us to identify the appearance of a quadratically-logarithmic term of the form \req{log22} in $d=4$. We show that this term accounts for the result obtained using Solodukhin's formula for (elliptic) cones. In the case of polyhedral corners, our result allows for a proper understanding of the absence of a $\log^2\delta$ contribution, and shows that the nature of the remaining logarithmic coefficient ---$v_n(\theta_1,\theta_2,\cdots,\theta_j)$ in \req{polys2}--- is intrinsically non-local. 
%\comment{words to be added}

\subsection{Dimensional reduction for free fields}
Let us consider a real 1-component scalar field of mass $m$ in $d$ spacetime dimensions.
The R\'enyi and entanglement entropies of a subregion $V$ in the groundstate can be obtained from \cite{Casini:2009sr}
\begin{align}\label{snnn}
S_{n}(V)&=\frac{1}{1-n}\log (\tr \rho_V^n)=\frac{1}{1-n}\sum_{k=0}^{n-1}\log Z[e^{2\pi i \frac{k}{n}}] \, , \\
\see(V) &\equiv \lim_{n\rightarrow 1}S_n(V) =-\int_0^{\infty}dt \frac{\pi}{\cosh^2(\pi t)}\log Z[-e^{2\pi t}]\, , 
\end{align}
where $Z[e^{2\pi i a}]$ is the partition function on $\mathbb{R}^d$ corresponding to a field which picks up a phase $e^{2\pi i a}$ when the entangling region $V$ is crossed.  
The expression for $S_n$ follows from a diagonalization procedure in replica space \cite{Casini:2009sr}.

The partition function  $Z[e^{2\pi i a}]$ can be computed by exploiting the relation between the free energy and the trace of the Green function of the associated Laplacian operator
\begin{equation}\label{zg}
\partial_{m^2}\log Z[e^{2\pi i a}]=-\frac12\int_{\mathbb{R}^d} d^{d}\vec{r}\,  G_a(\vec{r},\vec{r})\, ,
\end{equation}
where 
\begin{align}
(-\nabla^2_{\vec{r}_1}+m^2) G_{a}(\vec{r}_1,\vec{r}_2)&=\delta (\vec{r}_1-\vec{r}_2)\, ,\\ \label{bcc} \lim_{\epsilon\rightarrow 0^+} G_{a}(\vec{r}_1+\epsilon \vec{\eta},\vec{r}_2)&=e^{2\pi i a} \lim_{\epsilon\rightarrow 0^+}G_{a}(\vec{r}_1-\epsilon \vec{\eta},\vec{r}_2)\, , \quad \vec{r}_1 \in V\, ,
\end{align}
where $\vec{\eta}$ is orthogonal to $V$. On general grounds, the Green function can be written in terms of the eigenfunctions and eigenvalues of the Laplacian as
\begin{equation}\label{Gr1r2}
G_a(\vec{r}_1,\vec{r}_2)=\int d\lambda \frac{\psi_{\lambda}(\vec{r}_1) \psi_{\lambda}^*(\vec{r}_2)}{\lambda^2+m^2}\, , \quad \text{where} \quad \nabla^2_{\vec{r}_1}\psi_{\lambda}(\vec{r}_1)=-\lambda^2\psi_{\lambda}(\vec{r}_1)\, .
\end{equation}
In situations in which the boundary conditions \req{bcc} are implemented along the angular directions, it is possible to perform a separation of variables between the radial and angular components of $\psi_{\lambda}(x)$. This will be the case for cones of different sections or polyhedral corners. We can write
\begin{equation}\label{laps}
\psi_{\lambda}(\vec{r})=\sum_{\ell} F_{\ell,\lambda}(r) \Phi_{\ell}(\Omega_{\mathbb{S}^{d-1}})\, ,  \quad \text{with} \quad \nabla^2_{\mathbb{S}^{d-1}} \Phi_{\ell}(\Omega_{\mathbb{S}^{d-1}})=-\ell(\ell+d-2) \Phi_{\ell}(\Omega_{\mathbb{S}^{d-1}})\, ,
\end{equation}
where the eigenfunctions of $%\nabla_{\Omega}\equiv
 \nabla^2_{\mathbb{S}^{d-1}}$ will satisfy different boundary conditions inherited from \req{bcc} depending on the entangling region. Those boundary conditions will also determine the possible values of $\ell$ appearing in \req{laps}, which will not correspond to integer numbers in general.
In spherical coordinates, the $d$-dimensional Laplace operator in \req{Gr1r2} becomes
\begin{equation}
 \nabla^2_{\vec{r}}=\frac{\partial^2}{\partial r^2}+\frac{(d-1)}{r}\frac{\partial }{\partial r}+\frac{1}{r^2}\nabla^2_{\mathbb{S}^{d-1}}\, . 
\end{equation}
Then, it is straightforward to find the equation for the radial component of $\psi_{\lambda}(\vec{r})$, namely
\begin{equation}
\ddot F_{\ell,\lambda}(r)+\frac{(d-1)}{r}\dot F_{\ell,\lambda}(r)-\frac{[\ell(\ell+d-2)-\lambda^2 r^2]}{r^2}F_{\ell,\lambda}(r)=0\, .
\end{equation}
The general solution to this equation is given by
\begin{equation}
F_{\ell,\lambda}(r)=r^{-\frac{(d-2)}{2}}\left[\alpha_{\ell,\lambda} J_{\ell+\frac{d-2}{2}} [\lambda r]+ \beta_{\ell,\lambda} Y_{\ell+\frac{d-2}{2}}[\lambda r] \right]\, ,
\end{equation}
where $J_s[x]$ and $Y_s[x]$ are Bessel functions of the first and second kind, respectively, and $\alpha_{\ell,\lambda}$ and $\beta_{\ell,\lambda}$ are 
integration constants. $Y_s[x]$ blows up at $x=0$, so we set $\beta_{\ell,\lambda}=0$.\footnote{Observe that $Y_{\alpha}[x]$ can be written as a linear combination of $J_{\alpha}[x]$ and $J_{-\alpha}[x]$: $Y_{\alpha}[x]=J_{\alpha}[x] \cot(\alpha \pi)-J_{-\alpha}[x]/\sin(\alpha \pi)$. Setting $\beta_{\ell,\lambda}=0$  implicitly selects the sign of $\alpha$ in $J_{\alpha}[x]$ to be positive: negative values of $\alpha$ are the ones responsible for the blow up of $Y_{\alpha}[x]$ at $x=0$.} On the other hand, the orthogonality relation\footnote{Here, we use the orthogonality relations
\begin{equation}
\sqrt{\lambda\lambda' }\int_0^{\infty} dr\,  r J_s[r \lambda]  J_s[r \lambda']=\delta(\lambda-\lambda')\, , \quad \text{and}\quad  \int d\Omega\,  \Phi_{\ell}(\Omega) \Phi^*_{\ell'}(\Omega)=\delta_{\ell \ell'}\, .
\end{equation}
}
\begin{equation}
\int_0^{\infty} r^{d-1} dr \int d\Omega \, \psi_{\ell,\lambda}(\vec{r})\psi^*_{\ell',\lambda'}(\vec{r})=\delta_{\ell\ell'}\delta(\lambda-\lambda')\, ,
\end{equation}
fixes the remaining integration constant to $\alpha_{\ell,\lambda} =\sqrt{\lambda}$. Using this information, we are ready to rewrite the Green function in \req{Gr1r2} as
\begin{equation}
G_a(\vec{r}_1,\vec{r}_2)=\sum_{\ell ,\ell' } r_1^{-\frac{(d-2)}{2}}r_2^{-\frac{(d-2)}{2}} \Phi_{\ell}(\Omega_1) \Phi^*_{\ell'}(\Omega_2) \int d\lambda \frac{\lambda}{\lambda^2+m^2}\, J_{\ell+\frac{d-2}{2}} [\lambda r_1]J_{\ell'+\frac{d-2}{2}} [\lambda r_2]\, .
\end{equation}
We can now write the trace of the Green function appearing in \req{zg} as
\begin{align}
-\int_{\mathbb{R}^d} d^{d}\vec{r}\,  G_a(\vec{r},\vec{r})&=- \sum_{\ell} \int_0^{\infty} dr\, r I_{\ell+\frac{d-2}{2}} [m r] K_{\ell+\frac{d-2}{2}} [m r] \\  
\label{kak} &=-\sum_{\ell}\left(\int_0^{\infty}\frac{dr}{2m}\right) + \frac{1}{2m^2}\sum_{\ell}\left|\ell+\frac{d-2}{2} \right|\, ,
\end{align}
where in the first line we performed the integration over the angles, and used the following result to evaluate the $\lambda$ integral:
\begin{equation}
\int_0^{\infty} d\lambda \frac{\lambda}{\lambda^2+m^2}\, J_{\ell+\frac{d-2}{2}} [\lambda r]^2=I_{\ell+\frac{d-2}{2}} [m r] K_{\ell+\frac{d-2}{2}} [m r]\, ,
\end{equation}
where $I_s[x]$ and $K_s[x]$ are the modified Bessel functions of the first and second kinds, respectively. 
To get \req{kak}, we used the fact that $I_{s} [m r] K_{s} [m r]\to 1/(2mr)$ at large $mr$.

As we can see in \req{kak}, we obtain two terms. The first is manifestly divergent, and will be responsible for non-universal terms, such as the area law. 
We ignore this non-universal contribution from now on ---which, besides, has no dependence on the angles--- and focus on the second. When integrated over $m^2$ to obtain $\log Z[e^{2\pi i a}]$, this term will yield a logarithmically divergent contribution to the R\'enyi and entanglement entropies. Completing squares, we can finally write %\comment{check overall signs in the following expressions}
\begin{equation}
\partial_{m^2} \log Z[e^{2\pi i a}]=\frac{1}{4 m^2}\tr \sqrt{-\nabla^2_{\mathbb{S}^{d-1}}+\frac{(d-2)^2}{4}}\, ,
\end{equation}
which can be alternatively written as
\begin{equation}\label{esdx}
\partial_{m^2} \log Z[e^{2\pi i a}]=\frac{1}{4\pi m^2} \int_0^{\infty} \left[ \xi^{1/2}\tr\left[ \frac{1}{\nabla^2_{\mathbb{S}^{d-1}}-\frac{(d-2)^2}{4}-\xi}\right]+ \xi^{-1/2} \right] d\xi\, .
\end{equation}
The term involving integration over $\xi^{-1/2}$ produces another angle-independent non-universal term which we also ignore.
Integrating over $m^2$, it follows that 
\begin{equation}\label{dfs}
 \log Z[e^{2\pi i a}] |_{\log} = - \frac{\log(\delta/L) }{2\pi}\int_0^{\infty}  \xi^{1/2}\tr\left[ \frac{1}{\nabla^2_{\mathbb{S}^{d-1}}-\frac{(d-2)^2}{4}-\xi}\right] d\xi\, ,
\end{equation}
where ``$|_{\log}$'' makes explicit the fact that we are omitting additional contributions coming from the first term in \req{kak} as well as the last term in \req{esdx}.
Here, $L$ is a long-distance quantity parametrizing the linear size of subregion $V$.

In the prototypical case of a corner region of opening angle $\Omega$ in $d=3$, the problem gets reduced to computing the trace of the Green function on a sphere $\mathbb{S}^2$ with a cut of angle $\Omega$. The corresponding boundary conditions resulting from \req{bcc} read in that case
\begin{equation}
\lim_{\epsilon\rightarrow 0^+}\Phi_{\ell}(\pi/2+\epsilon,\phi)=e^{2\pi i a} \lim_{\epsilon\rightarrow 0^+}\Phi_{\ell}(\pi/2-\epsilon,\phi)\, ,\quad \phi\in [-\Omega/2,\Omega/2]\, .
\end{equation} 
The cut is therefore an angular sector on the equatorial $\mathbb{S}^1$. 
\begin{figure}[t]
	\centering 
	\includegraphics[scale=0.30]{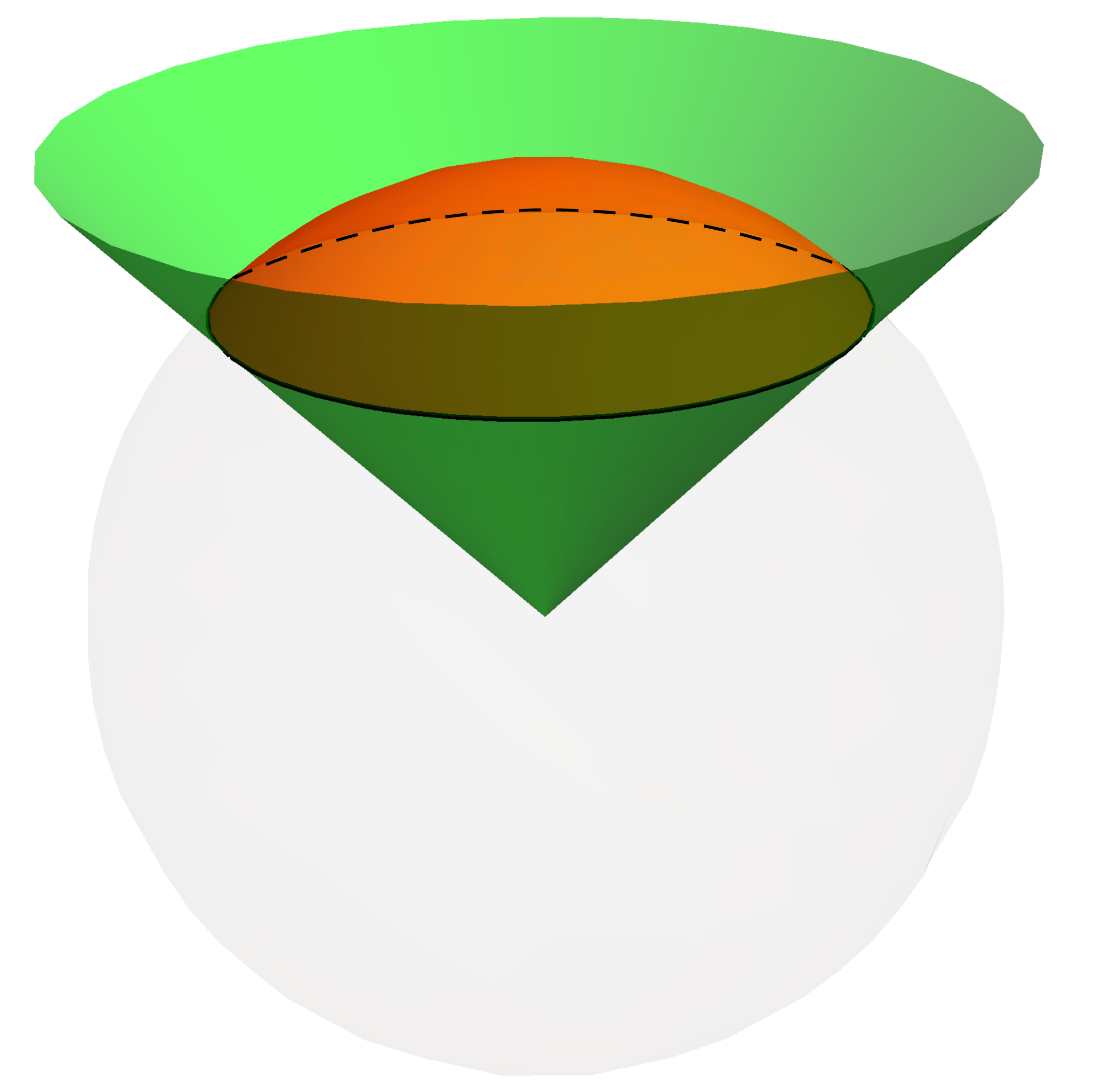}\hspace{0.1cm}
	\includegraphics[scale=0.38]{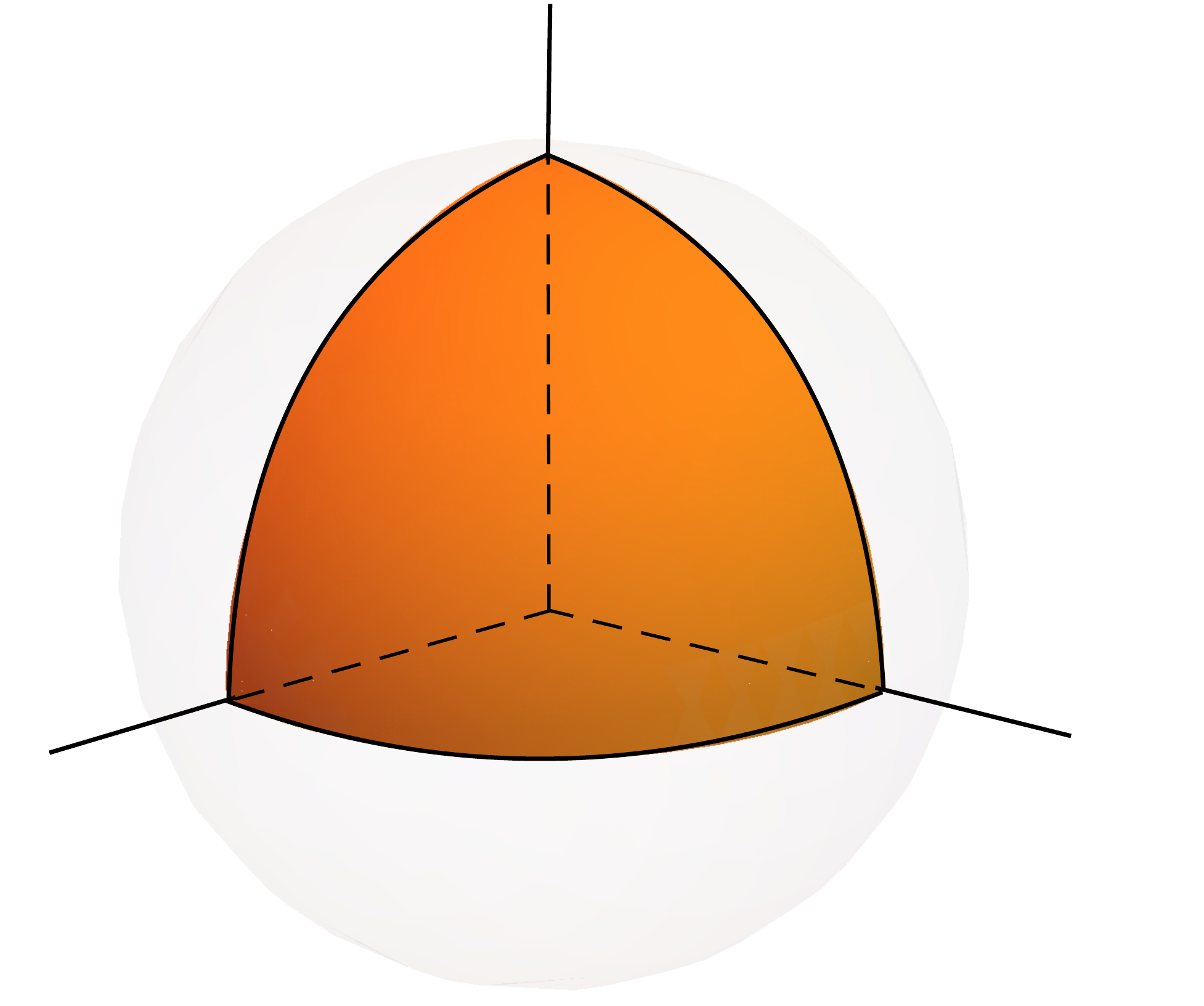}
	\caption{(Left) We plot the $\mathbb{S}^1$ resulting from the intersection of the conical region defined by $ \theta \in [-\Omega,\Omega]$, $\phi \in[0,2\pi)$ with a unit $\mathbb{S}^2$ in the equator of $\mathbb{S}^3$.  The boundary conditions are implemented as we approach the orange region in the equator of $\mathbb{S}^3$ from above and below in the additional angular coordinate (not shown in the figure). %$\psi=\pi/2+\epsilon$ and $\psi=\pi/2-\epsilon$ with $\epsilon\rightarrow 0^+$ (the $\theta$ coordinate is not shown in the figure). 
	(Right) We plot the intersection of a trihedral corner region of right opening angles with the unit $\mathbb{S}^2$. In this case, $\gamma$ is the union of three quarters of great circles, for which the extrinsic curvature vanishes, $k = 0$.}
\label{conefree}
\end{figure}
If we consider entangling regions emanating from a vertex in $d=4$, the corresponding cut on $\mathbb{S}^3$ will correspond to some area on the surface of a $\mathbb{S}^2$. The boundary of the intersection region, which in the case of the $d=3$ corner is just the union of two points, corresponds in this case to some curve $\gamma$ on the surface of the $\mathbb{S}^2$ ---see Fig.~\ref{conefree} for the case of a cone and a trihedral corner.

Interestingly, the trace of the Green function which appears inside the integral can be related to a $(d-1)$-dimensional R\'enyi entropy in de Sitter (dS) space \cite{Casini:2006hu}.  In particular, if one considers the R\'enyi entropy for some region in ${\rm dS}_{(d-1)}$  for a scalar field coupled to the background curvature through a term $g(d) R \phi^2$, where $R$ is the scalar curvature and $g(d)$ some function of the dimension,\footnote{Note that if $g(d)=(d-2)/(4(d-1))$, the scalar is conformally coupled.} the trace of the Green function appearing in \req{zg} can be read straightforwardly, and it follows that
\begin{equation} \label{dS-link}  
\partial_{m^2}  \log Z\left[{\rm dS}_{(d-1)}\right]= \frac12 \tr \left[\frac{1}{\nabla^{2}_{\mathbb{S}^{d-1}}-d(d-1)g(d)-m^2} \right]\, ,
\end{equation}
where we used $R=d(d-1)/ L_{\rm dS}^2$, and we set the dS radius to unity from now on, $L_{\rm dS}=1$. 
Observe that \req{dS-link} is very similar to the quantity appearing in the integrand of \req{dfs}, as long as the boundary conditions imposed in both cases match. Then, if we choose $g(d)=(d-2)^2/(4d(d-1))$, we can use \req{snnn} to write\footnote{If one chooses the conformally coupled value of $g(d)$, the result doesn't change much. The only difference is that the derivative is to be evaluated at some other $m$, $ \int_0^{\infty}  dm'^2 m' \left. \left[ \frac{\partial S_n^{{\rm dS}_{(d-1)}}}{\partial m^2}\right]\right|_{m=m'+f(d)}$. The choice doesn't affect any of the conclusions that follow.}    
\begin{equation}\label{soba}
S_n|_{\log} = -\frac{ \log(\delta/L)}{\pi}  \int_0^{\infty}  dm^2 m\,  \frac{\partial S^{{\rm dS}_{(d-1)}}_n}{\partial m^2 }\, ,
\end{equation}
where in the l.h.s.\ we have the R\'enyi entropy corresponding to the region $V$ in $\mathbb{R}^d$, and in the r.h.s., $S^{{\rm dS}_{(d-1)}}_n$ stands for the R\'enyi entropy in ${\rm dS}_{(d-1)}$ corresponding to a region with the same boundary conditions as those inherited for $V$  from \req{bcc} on the angular coordinates. We can extract some useful information from $ S^{{\rm dS}_{(d-1)}}_n$. The corresponding R\'enyi entropy admits a high-mass expansion of the form  
\begin{equation}\label{expanse}
S^{{\rm dS}_{(d-1)}}_n= c_{n,(d-3)} m^{d-3}+\cdots + c_{n,0} + \frac{c_{n,-1}}{m}+\cdots  
% S^{{\rm dS}_{(d-1)}}_n= c_{n,(d-1)} \Big(\frac{m}{R}\Big)^{d-1} + c_{n,(d-3)} \Big(\frac{m}{R}\Big)^{d-3}
% + \cdots + c_{n,0} + \frac{c_{n,-1}R}{m}+\cdots  
\end{equation}
valid for masses much greater than the de Sitter inverse radius, $m\gg 1$. 
Recall that in our units $L_{\rm dS}=1$.
Observe that in this expansion, we have hidden terms involving various powers of the corresponding UV cutoff $\delta$ inside the coefficients $c_{n,q}$ ---which have dimensions of $(\mbox{length})^{q}$ for each $q$. In principle, one could naively think of various dimensionless combinations of $m$, $\delta$ and local integrals over the entangling surface susceptible of appearing in \req{expanse}. However, divergent contributions involving $\delta$ originate from the (massless) UV theory and they are controlled by local integrals over the entangling surface. The effects of introducing a mass translate into corrections to such contributions which disappear as $m\rightarrow 0$. This forbids the presence of non-vanishing terms involving negative powers of $m$ combined with $\delta$. Terms with negative powers of $m$ not involving $\delta$ can however appear weighted by the appropriate local integrals.\footnote{For discussions on the interplay between mass and UV-divergent terms see \eg \cite{Hertzberg:2010uv,Casini:2014yca}.} This is the case of $\mathcal{O}(m^{-1})$ term in \req{expanse}. 
The coefficient $c_{n,-1}$ has dimensions of $(\mbox{length})^{-1}$ and, as we say, it must be given by an integral over the boundary of the entangling region. In $d=4$, the only term satisfying these properties we can write reads 
\begin{equation}
c_{n,-1}= \alpha_n \int_{\gamma} k^2\, ,
\end{equation}
where $\gamma$ is indeed the boundary of the entangling region in dS$_3$ (which corresponds to the intersection of the original region $V$ with the unit $\mathbb{S}^2$ mentioned above), $k$ is the trace of the extrinsic curvature of $\gamma$, and $\alpha_n$ is a dimensionless constant. It is not difficult to see that an analogous term could be written in general even dimensions as $c_{n,-1}\sim \sum_i \alpha^i_n \int_{\Sigma} k_i^{d-2}$, where $k_i^{d-2}$ would be various independent contractions of extrinsic curvatures of order $(d-2)$ ---or intrinsic curvatures of order $(d-2)/2$. In odd dimensions, however, we are forced to set $c_{n,-1}=0$. This is because any possible contraction of an odd number of extrinsic curvatures produces a term which would change 
sign if instead of considering $V$ we chose its complement as our entangling region. 
However, the purity of the groundstate forbids this ---for related discussions see \eg \cite{Grover:2011fa,Liu:2012eea}. 
In the rest of the section we will thus assume that $d$ is even. Observe also that, in principle, terms involving the curvature of the background manifold could have appeared as well. However, those would be independent of the geometry of $\gamma$. Then, since we know that for a ``cone'' of opening angle $\Omega=\pi$ ---for which $\gamma$ is just the equator of the $\mathbb{S}^2$--- there is no $\log^2\delta$ contribution at all, the corresponding coefficients must be zero.

 As we have said, the local nature of  $c_{n,-1}$ 
prevents it from feeling the curvature of the background geometry. Hence, the coefficient $\alpha_n$ can be for example connected to the result corresponding to a cylinder entangling region in flat space, which fixes $\alpha_n=\frac{1}{8}f_b(n)$, where $f_b(n)$ is one of the coefficients appearing in the universal contribution for smooth entangling surfaces\footnote{For a free scalar and a free Dirac field, the related cylinder dimensional reduction was explicitly carried out for a free scalar in \cite{Huerta:2011qi}, yielding excellent agreement with the general-CFT expectation.} \cite{Solodukhin:2008dh,Fursaev:2012mp} ---see \req{game} below. 

All in all, we are left with a contribution of the form
\begin{equation}
S_n^{({\rm dS}_{3})}|_{m^{-1}}%\sim \int_{m^2}^{\infty} \tr \left[ \frac{1}{\nabla^2_{\mathbb{S}^3}-\xi }\right] d\xi 
= 
\frac{f_b(n)}{8m} \int_{\gamma} k^2\, .
\end{equation}
Inserting this in the integral appearing in \req{soba}, we find
\begin{align}
\int_{{1/L^2}}^{1/\delta^{ 2}}  dm^2 m  \frac{\partial S_n^{({\rm dS}_{3})}|_{m^{-1}}}{\partial m^2} &= -\frac{1}{8} \int_{{1/L^2}}^{1/\delta^{2}} dm^2 m  \frac{f_b(n)}{2m^3}\int_{\gamma} k^2\\  
 &=\frac{1}{16}  f_b(n) \left[ \log(\delta{^2/L^2})+\mathcal{O}(1) \right]\int_{\gamma} k^2\, ,
%&\supset  -\int_0^{1/\delta'} dm'^2 m'  \frac{f_b(n)}{2(m'+1)^3}\int_{\gamma} k^2\\  &\sim f_b(n) \left[ \log \delta'+\mathcal{O}(1) \right]\int_{\gamma} k^2\, .
\end{align}
where we introduced a regulator for high masses. As we can see, this produces an additional logarithmic divergence which, when combined with the one already present in \req{soba}, gives rise to the term
\begin{equation}\label{log2}
S_{n}|_{\log^2}= \frac{-f_b(n)}{8\pi} \log^2\delta  \int_{\gamma} k^2\, .
\end{equation}
%\rd{[[W: shouldn't I get an extra factor of 2 due to the presence of $\epsilon^2$ in 2.27?]]}
Crucially, a contribution like \req{log2} will only be present for entangling regions such that the extrinsic curvature of $\gamma$ is nonvanishing. This will be the case of straight cones with various cross-sections, but not of polyhedral corners, as we explain now.

\subsection{Polyhedral corners}
In the case of entangling regions involving polyhedral corners, the intersection of $\partial V$ with the unit $\mathbb{S}^2$ will be the union of portions of great circles ---see Fig.~\ref{conefree}--- for which $k=0$. This explains why, as previously observed \cite{Kovacs,Devakul2014,Sierens:2017eun,Bednik:2018iby,Witczak-Krempa:2018mqx}, trihedral corners have a universal $\log \delta$ divergence, instead of a $\log^2\delta$ one. The above analysis also reveals that, in the case of polyhedral corners, the corresponding logarithmic contribution will not come from some simple local integral along any curve on $\mathbb{S}^2$. Rather, its calculation would require the full answer for the spectral function on the sphere with a cut appearing in \req{dfs} ---this is completely analogous to what happens for the usual corner universal term in $d=3$ \cite{Casini:2006hu,Casini:2008as,Casini:2009sr}. Hence, the function $v_n(\theta_1,\theta_2,\cdots,\theta_j)$ appearing in the logarithmic contribution to the R\'enyi entropy for polyhedral corners
\begin{equation}
S_{n}^{\rm polyhedral}|_{\rm log}= v_n(\theta_1,\theta_2,\cdots,\theta_j) \log (L/\delta)\, ,
\end{equation}
will generally be a highly non-local term, and there is no reason to expect it to be controlled by a linear combination of trace-anomaly coefficients (or their R\'enyi entropy generalizations, $f_a(n)$ and $f_b(n)$) as suggested in \cite{Sierens:2017eun,Bednik:2018iby}. This observation, while relying on a free-field calculation, should be valid for general CFTs.

On the other hand, one does expect that in the limit of a very open polyhedral corner (``almost smooth''), $v_1(\theta_1,\theta_2,\cdots,\theta_j)$ is nevertheless controlled by the stress-tensor two-point function charge \cite{Osborn:1993cr} $\ctt \propto c$, in line with the results of \cite{Bueno1,Bueno4,Mezei:2014zla,Miao:2015dua,Faulkner:2015csl}. For general CFTs, it was indeed shown that trihedral corners in $d=4$ receive a contribution 
$\propto c\log(L/\delta)$ in the almost smooth limith \cite{Witczak-Krempa:2018mqx}.  In the following subsection, we give an analytical calculation
of the polyhedral contribution away from the flat limit using a special model.

\subsubsection{Cubes in the Extensive Mutual Information model} \label{sec:cube-EMI}
For entangling regions containing vertices that are not near the flat limit, one is left with very few analytical methods. 
One particularly useful tool is the so-called ``Extensive Mutual Information Model'' (EMI) \cite{Casini:2005rm,Casini:2008wt,Swingle2010}.  
The EMI is not defined through a Lagrangian, but instead allows for a simple geometric computation of the entanglement entropy   
consistent with conformal symmetry, and has passed several non-trivial tests in all dimensions \cite{Casini:2005rm,Swingle2010,Bueno1,Bueno3}.     As its name suggests, the characterizing feature of this model is the fact that the mutual information satisfies the extensivity property
\begin{equation}\label{emir}
I(A,B)+I(A,C)=I(A,B\cup C)\, .
\end{equation}
This requirement strongly constrains the form of the entanglement entropy and the mutual information. In particular, the entanglement entropy of a region $A$ in the EMI
 is defined through the following integral:\footnote{A somewhat similar formula for the entanglement entropy of a free Fermi gas was obtained in \cite{Helling:2009qy}.}
\begin{align} \label{emi-def}
  S^{\rm \ssc EMI} =\kappa \int_{\partial A}d^{d-2}{\bf r}_1 \int_{\partial A}d^{d-2}{\bf r}_2 \, \frac{ {\bf n}_1\cdot {\bf n}_2}{|{\bf r}_1-{\bf r}_2|^{2(d-2)}}\, ,
\end{align}
where ${\bf n}$ is the normal vector to the boundary of $A$, $\partial A$. $\kappa$ is a positive parameter.
In this section, we drop the subscript ``EE'', as the EMI ansatz can be trivially adapted to general R\'enyi entropies $S_n$ by replacing $\kappa$ by $\kappa_n$.  \\
% The EE of the EMI has the property that the mutual information $I(A,B)=S(A)+S(B)-S(A\cup B)$ is extensive:
% \begin{align}
%   I(A,B\cup C) = I(A,B) + I(A,C)\,.
% \end{align}

\noindent {\bf Cubic trihedral --- } 
We now analytically calcualte the full entanglement entropy for the cubic trihedral, see Fig.~\ref{conefree}, with $\theta_1=\theta_2=\theta_3=\pi/2$, using the EMI. 
In other words, region $A$ corresponds to an octant of $\mathbb R^3$, say $x,y,z>0$.  
 Since any pair of faces has mutually orthogonal normals,   
we are left with contributions where both ${\bf r}_1$ and ${\bf r}_2$ lie on the same face, which we take to be in the $xy$-plane:
\begin{align}
  S = 3\kappa \int_0^\infty dx_1 dx_2 dy_1 dy_2 \frac{1}{[(x_1-x_2)^2 + (y_1-y_2)^2]^2} \, .
\end{align}
We first perform the $x_1$ integral, then the $x_2$ 
one, for which a finite long-distance cutoff $L$ is needed to avoid the divergence:
\begin{align}
S 
= 3\kappa \int_0^\infty dy_1 dy_2 \,L \frac{2 \tan ^{-1}\left(\frac{L}{y_1-y_2}\right)+\pi 
   \sgn\left(y_1-y_2\right)}{4 \left(y_1-y_2\right)^3}\, .
\end{align}
To perform the integral over $y_1$, we break it up into two parts to avoid the short distance 
divergence occuring when $y_1\to y_2$: $\int_0^\infty dy_1=\int_0^{y_2-\de}dy_1+\int_{y_2+\de}^\infty dy_1$.
The result is
\begin{align}
  I =\int_\de^L dy_2 \frac{1}{4L} \left[\frac{L (L\pi -2 \de )}{\de ^2} \right. &\left.+ 2 \left(\frac{L^2}{\de
   ^2}+1\right) \cot^{-1}(\de/L) + \frac{L}{y_2}\right. \\ \notag & \left. -\frac{L^2\pi
   }{2 y_2^2} - \left(\frac{L^2}{y_2^2} + 1\right) \cot ^{-1}\left(y_2/L\right)\right]\, .
\end{align}
This last integral needs to be performed with both short and long distance cutoffs, which can be done exactly.
% \begin{align}
%   I= \frac{1}{8 \bar\de ^2} \left[ 4 \pi+  2 \bar\de ^2 \log \left(\frac{1}{\bar\de ^2}+1\right)
% -\bar\de \left[\bar\de   (\pi  (\bar\de -3)-4+\log 4)+6 \pi +4\right]
% +2 (\bar\de -1) \left[(\bar\de -1) \bar\de +2\right] \tan ^{-1} \bar\de \right] 
% \end{align}
% where $\bar\de=\de/L$. By Taylor expanding at small $\de$, we find
The final result for $S$, expanded at $\delta/L\ll 1$, reads
\begin{align} \label{eq:S-cubic-tri}
  S =3\kappa I=  3 \kappa\left(\frac{\pi L^2}{2\de^2} - \left(1+\tfrac{3\pi}{4}\right)\frac{L}{\de} + \frac12 \ln(L/\de) + \dotsb\right)\, .
\end{align}
This contains the area law, a negative contribution from the edges $\propto L/\de$, and the positive trihedral contribution governed by
\begin{align}
  v(\pi/2,\pi/2,\pi/2) = \frac32 \kappa\, ,
\end{align}
which is positive since $\kappa>0$.  \\

 \noindent{\bf Finite cube ---}
For a finite cube of linear size $L$ we have 2 types of contributions: ${\bf r}_1$ and ${\bf r}_2$ lying on the same face
or on opposite faces. The latter contribution can be omitted for the purpose of determining
the singular contributions (dependent on $\de$) because $|{\bf r}_1-{\bf r}_2|\geq L$, which implies that
the corresponding integral for $S$ contains no short-distance divergences. It is in fact a pure number, as can be verified by an explicit calculation. 

We thus need to evaluate an integral analogous to the above one but with finite support on a face of the cube:
\begin{align}
  S =6 s_1I = 6s_1 \int_0^L dx_1 dx_2 dy_1 dy_2 \frac{1}{[(x_1-x_2)^2 + (y_1-y_2)^2]^2} \, .
\end{align}
The factor of $6$ counts the number of faces of the cube. After performing the integrals over $x_1,x_2$, we get 
\begin{align}
  I = \int_0^L dy_1 dy_2 \frac{L}{(y_1-y_2)^3}\tan^{-1}\left(\frac{L}{y_1-y_2} \right)\, .
\end{align}
We perform the $y_1$ integral as above by splitting it to avoid the $y_1\to y_2$ divergence. The result reads
\begin{multline}  
  I \!=\!\int_{\bar\de}^{1-\bar\de}\!\!\!  \, \frac{2   \left(\bar\de ^2+1\right)}{2 \bar\de ^2
   \left(\bar y_2-1\right)^2 \bar y_2^2} \bigg\{
\left(\bar y_2-1\right)^2 \bar y_2^2 \cot^{-1}(\bar\de ) %\\
-\bar\de  \left[\bar\de  \left(\left(\bar y_2-2\right)
   \bar y_2+2\right) \bar y_2^2 \cot^{-1}(1-\bar y_2) \right.\\
\left. +\left(\bar y_2-1\right)
   \left(\bar y_2 \left(\bar\de +2 \left(\bar y_2-1\right) \bar y_2\right)+\bar\de  
   \left(\bar y_2-1\right) \left(\bar y_2^2+1\right) \cot^{-1}\left(\bar y_2\right)\right)\right] \bigg\} d\bar y_2\, ,
\end{multline}
where the bar variables are normalized by $L$. In contrast with the semi-infinite trihedral calculation, we note that 
the final integral has divergences at both $y_2=0,L$, which is a consequence of the finiteness of the cube.
Performing the integral and Taylor expanding the answer in powers of $L/\de$, we obtain
\begin{align}
  S =6\kappa I =  6 \kappa \left(\frac{\pi L^2}{2\de^2} - \left(2+\tfrac{3\pi}{2}\right)\frac{L}{\de} + 2 \ln(L/\de) + \dotsb\right)\, .
\end{align}
As expected, we thus find that the logarithmic coefficient of the cube is $8$ times that of a single $\pi/2$ trihedral corner:
\begin{align}
 (\mbox{log coefficient for cube}) = 8\, v(\pi/2,\pi/2,\pi/2) = 12 \kappa\, .
\end{align} 
As a further check, the divergence due to the 12 edges of the cube, $-6\kappa(2+\tfrac{3\pi}{2})L/\delta$, is 
4 times that of the one found for a single cubic trihedral which has 3 edges, see \req{eq:S-cubic-tri}.

We emphasize that for polyhedral corners, even though terms of the form $L/\delta$ due to the presence of edges appear, the contribution from the corner itself is logarithmic, just like for smooth surfaces ---see \req{game} below. We are used to see that the presence of singularities on the entangling surface modifies the structure of divergences in the R\'enyi entropy, so this match on the type of universal divergence  may look surprising at first. In Section \ref{singi} we will provide genuine examples of entangling regions which do not modify the divergences structure. In the present case, however, the coincidence is accidental, as should be clear from the discussion above: while the universal logarithmic term is controlled by a local integral on the entangling surface for smooth regions, we expect the one corresponding to a polyhedral corner  to involve a complicated theory-dependent function. In a setup in which we got to the polyhedral corner from a limiting procedure on a smooth region, the logarithm would not emerge from the already present logarithmic term produced locally near the edges (which would be enhanced to a $\log^2$ term and would vanish for the reasons already explained), but rather from the constant one. In this regard, note that the presence of a new regulator, $\epsilon$, which would smooth out the wedges and trihedral corners would give rise to logarithmic contributions of the form $\log(\epsilon/\delta)$, which in the $\epsilon\rightarrow 0$ limit would contaminate the coefficient of the $\log L$ term, preventing it from being controlled by $f_b(n)$ as one would have naively guessed from the coefficient of the $\log \delta$ term in Solodukhin's formula \req{game}.

\subsection{Elliptic cones}\label{elic}
Let us now turn to the case of (elliptic) conical entangling surfaces, as shown in Fig.~\ref{HwP}. For those, the trace of the extrinsic curvature is non-zero, $k\neq 0$, and a $\log^2\delta$ universal contribution of the form \req{log2} is present. Eq.~(\ref{log2}) suggests that, on general grounds, the dependence on the geometric properties of the corresponding entangling surface is theory-independent, the only information about the theory in question being encoded in the overall $f_b(n)$ coefficient. The result for this universal contribution in the case of cones with circular cross-sections is known  to be obtainable for general CFTs \cite{Klebanov:2012yf} from Solodukhin's formula \cite{Solodukhin:2008dh} ---see \req{game}--- up to a missing $1/2$ factor whose origin we discuss below. Here, we verify that the free-field expression \req{log2} exactly produces the result predicted by Solodukhin's formula both for circular cones, and in the less trivial case of cones with elliptic cross sections, for which we compute the corresponding universal coefficient explicitly in terms of the elliptic cross-sections eccentricity.

% In general, $e^{\prime}$ is a more intuitive and natural quantity to work with.

% For this, the dependence on the R\'enyi index and that on the cone opening angle 

% This can be evaluated for free-fields and cross-checked using Solodukhin's formula \cite{Solodukhin:2008dh}.

\subsubsection{General CFTs}
 
  % This will contribute with a local term to the entanglement entropy controlled by the central charge $c$ which will an extra logarithmic divergence which together with the overall $\log \delta$ builds up to produce a $\log^2 \delta$
	
%	 $\sim \frac{\cos^2\Omega}{\sin\Omega} \log^2\delta$ term in the entanglement and R\'enyi entropies.

%\section{Revisiting cone entanglement}
%\subsection{Elliptic cones} \label{elicon}
%In this section we study the universal contribution to the R\'enyi entropy of general four-dimensional CFTs for entangling regions bounded by (right) elliptic cones. %For these, the 
%We aim to study the entanglement entropy of conical surfaces whose cross-sections are not (hyper)spheres but, more generally, (hyper)ellipsoids. We start with the $d=4$ case. 
When the entangling surface $\Sigma$ is smooth, the corresponding universal contribution can be obtained using Solodukhin's formula\footnote{In this expression, $h_{ab}$ is the induced metric on $\Sigma$ and ${\cal R}$ its associated Ricci scalar. Extrinsic curvatures associated to the two normal vectors $n_\mu^{(i)}$, ($i=1,2$), are denoted by $k^{(i)}_{ab}$, where indices $a,b$ run through the tangent directions to $\Sigma$. Also, $\tr k^2\equiv k^{(i)}_a{}^{\,b}k^{(i)}_b{}^{\,a}$ and $k^2\equiv k^{(i)}_a{}^{\,a} k^{(i)}_b{}^{\,b}$, where indices are raised with the inverse metric $h^{ab}$. Furthermore, $C_{ab}\,^{ab}=0$
is the background Weyl curvature projected on the surface. We will consider a flat background, so we set $C_{ab}\,^{ab}=0$ henceforth. } \cite{Solodukhin:2008dh,Fursaev:2012mp}, 
%For general four-dimensional CFTs, the universal contribution to the R\'enyi entropy for a given entangling surface $\Sigma$ reads
\beq
S_n^{\rm univ}=\int_{\Sigma} d^2y\sqrt{h}\left[ f_a(n)\,{\cal R} + f_b(n)\left( \tr k^2-\frac12 k^2\right)-f_c(n)\, C_{ab}{}^{ab}
\right] \frac{\log \delta }{2\pi}\, .
\labell{game}
\eeq
The only theory-dependent input appears through functions $f_{a,b,c}(n)$, which do not depend on the geometry of the entangling surface or the background spacetime ---in particular, $f_a(1)=a$, $f_b(1)=f_c(1)=c$, where $a$ and $c$ are the usual trace-anomaly charges. As usual, $\delta$ is a UV regulator, which should appear weighted by some finite dimensionful scale $\ell$ characterizing $\Sigma$.
% If $\Sigma$ has some complicated shape, the above formula gives the answer for the coefficient multiplying $\log(\delta)$. 
This scale can depend on the various dimensions characterizing the surface but this does not affect the universal coefficient in front of the logarithmic divergence ---\eg if we have two relevant scales, $R_1$ and $R_2$, the difference between $\log(R_1/\delta)$ and $\log(R_2/\delta)$ is always finite $\sim \log(R_1/R_2)$, so it adds up to subleading $\log \delta$ divergences, but does not affect the $\log^2\delta$ one.

Even though \req{game} is in principle not valid when $\Sigma$ includes geometric singularities, it has been argued that it produces the right answer for (circular) conical entangling regions when properly used ---see below. We will see that exactly the same treatment applies in the case of elliptic cones, thus allowing us to rely on \req{game} also in that case.

\begin{figure}[t]
	\centering 
	\includegraphics[scale=0.34]{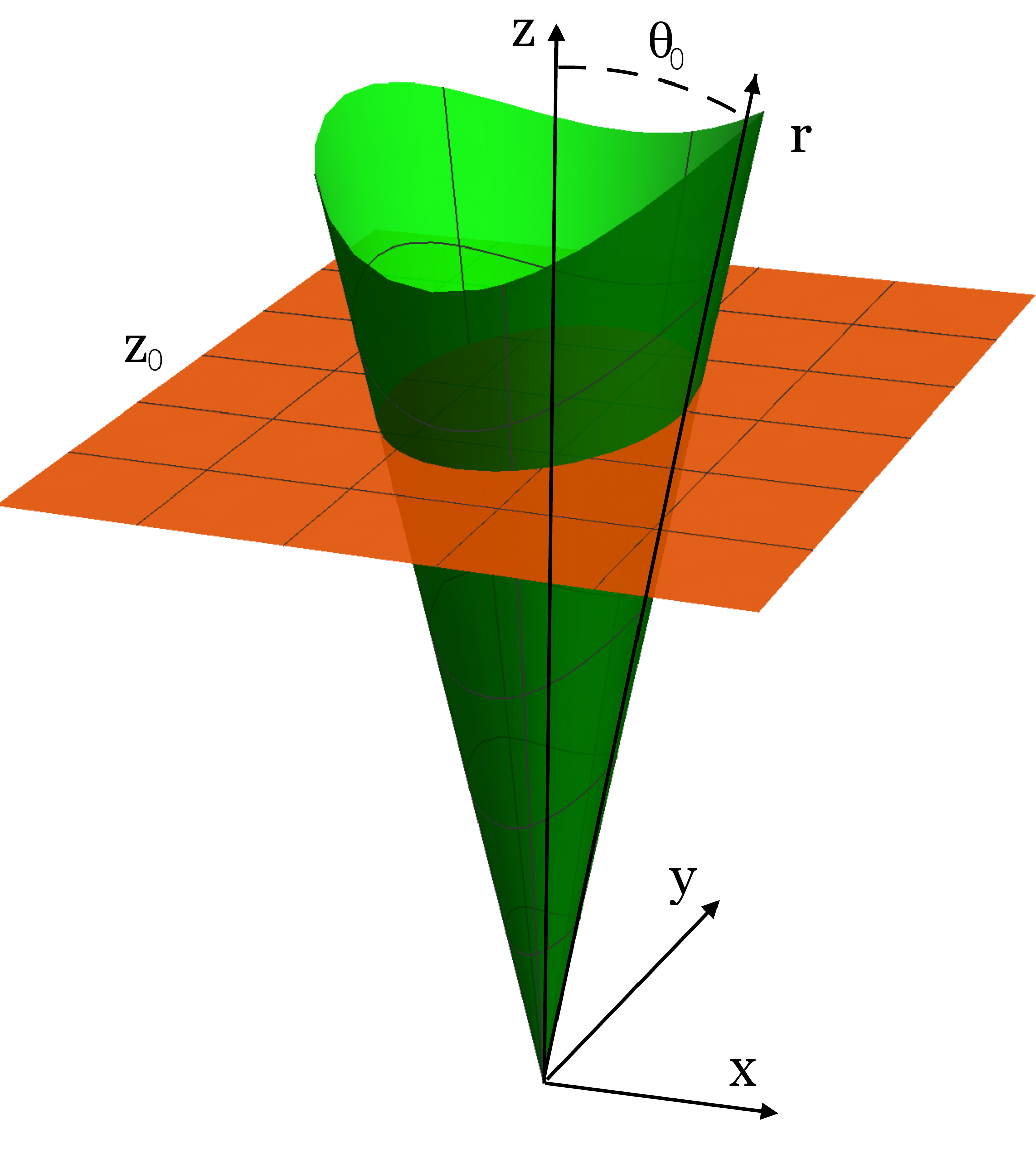}\hspace{0.1cm}
	\includegraphics[scale=0.38]{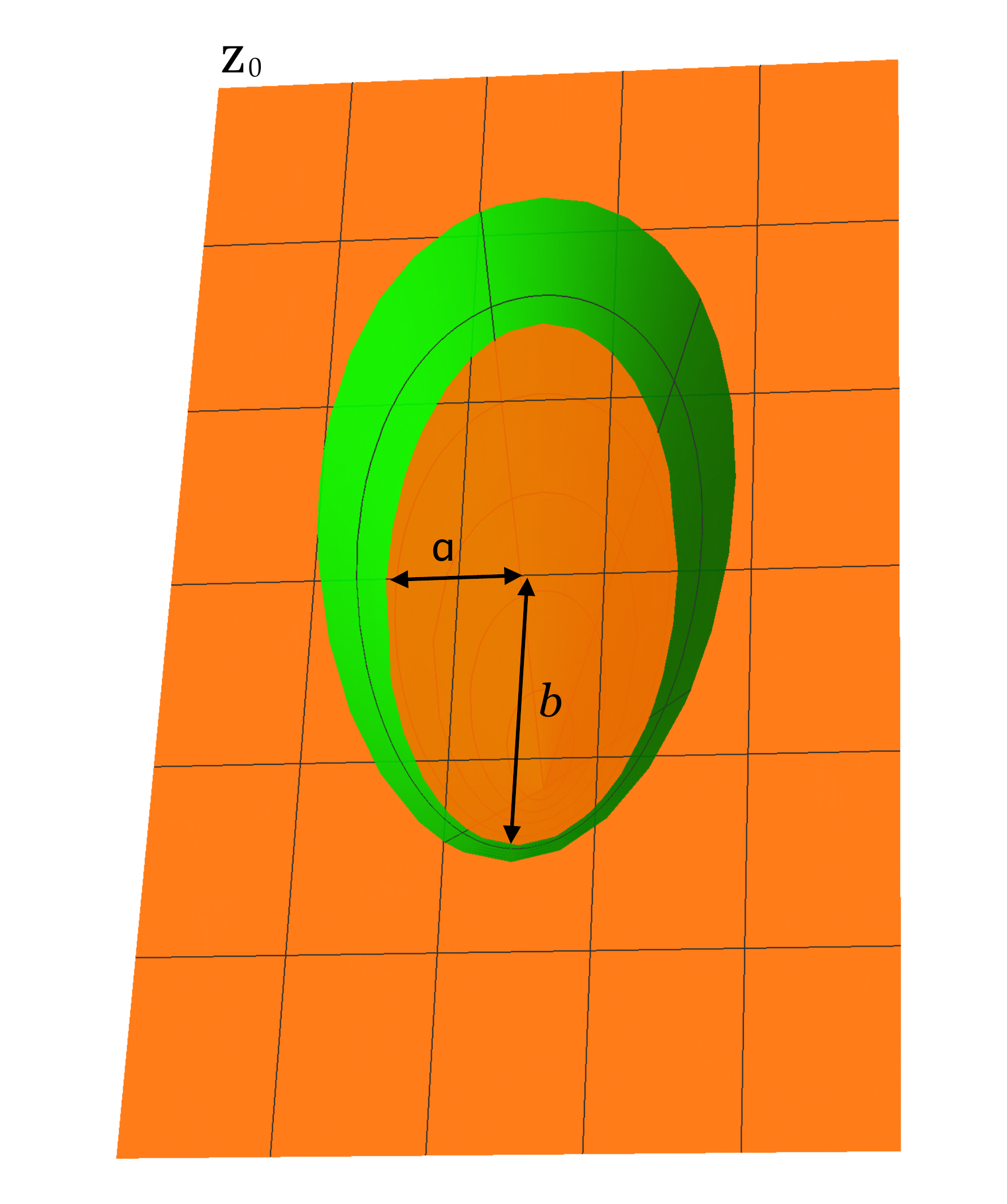}
	\caption{We plot an elliptic cone defined by $\theta=\theta_0$ in the coordinates \req{adf} (left). The intersection of constant height planes $z=z_0$ with the cone form ellipses of semi-minor and semi-major axis $a$ (the direction along  which $\theta_0$ is defined) and $b$, respectively (right).}
\label{HwP}
\end{figure}

In order to study the R\'enyi entropy of elliptic cones, we will use \emph{sphero-conal} coordinates $(r,\theta,\phi)$. These form an orthogonal system of coordinates, and they are defined in terms of the usual Cartesian ones as
\begin{align}\label{adf}
x=r\, \sin \theta\, \cos\phi\, ,\quad
y=r\,\sqrt{1-\kappa^2\cos^2 \theta}\,\sin\phi \, ,\quad
z=r\,\cos \theta\,\sqrt{1-(1-\kappa^2)\sin^2 \phi}\, ,
\end{align} 
where $r\geq 0$, $0\leq \theta \leq \pi$, $0\leq \phi \leq 2\pi$, and the parameter $\kappa$ is defined such that $0\leq \kappa \leq 1$. For  $\kappa=1$, one recovers the usual spherical coordinates. 

In these coordinates, the metric of four-dimensional Euclidean space reads
\begin{equation}\label{eucl}
ds^2=dt^2+dr^2+r^2\left[\frac{\kappa^2\sin^2\theta+(1-\kappa^2)\cos^2\phi}{1-\kappa^2\cos^2\theta}\right]d\theta^2+r^2\left[\frac{\kappa^2\sin^2\theta+(1-\kappa^2)\cos^2\phi}{1-(1-\kappa^2)\sin^2\phi}\right]d\phi^2\, .
\end{equation}
Coordinate surfaces $\theta=\theta_0$ correspond to semi-infinite elliptic cones with their tip at the origin and their axis along the positive $z$-axis if $\cos\theta_0\geq 0$, which we will assume from now on ---in particular, we take $0\leq \theta_0\leq \pi/2$. Cross-sections of the cone with constant-$z$ planes, $z=z_0$, are homothetic ellipses described by the equation $x^2/a^2+y^2/b^2=1$,
%\begin{equation}
%\frac{x^2}{a^2}+\frac{y^2}{b^2}=1\, ,
%\end{equation}
where the semi-minor and semi-major axis lengths, $a$ and $b$, read, respectively
%\footnote{The eccentricity of the ellipse reads then
%\begin{equation}
%e\equiv \sqrt{1-\frac{b^2}{a^2}}=\sqrt{\frac{1-k^2}{1-k^2\cos^2\theta_0}}\, .
%\end{equation}}
\begin{equation}\label{ab}
 a=\frac{z_0 \sqrt{1-\cos^2\theta_0}}{\cos\theta_0} \, , \quad b=\frac{z_0\sqrt{1/\kappa^2-\cos^2\theta_0}}{\cos\theta_0}\, .
 \end{equation}
Hence, $\kappa$ can be understood as a parameter controlling how much these ellipses differ from circles. Given $\kappa$, the ellipses are fully characterized by $\theta_0$.\footnote{The notion of \emph{opening angle} is not uniquely defined anymore when $\kappa \neq 1$. Indeed, an elliptic cone has a different opening angle for each value of the angular coordinate $\phi$. We can nevertheless define notions of \emph{semi-opening angles}, corresponding to the  angles along the semi-minor and semi-major axes $a$ and $b$. For the elliptic cones defined by $\theta=\theta_0$ in the coordinates above, the semi-opening angle in the ($x,z$)-plane (\ie in the direction of $a$) is precisely given by $\theta_0$. The semi-opening angle $\tilde{\theta}_0$ in the  ($y,z$)-plane (\ie in the direction of $b$) is related to $\theta_0$ through
\beq \label{ssed}
\sin^2 \tilde{\theta}_0=%\frac{({e^{\prime}}^2+1)\sin^2\theta_0}{{e^{\prime}}^2\sin^2\theta_0+1}\, .
1-\kappa^2 \cos^2\theta_0\, .
\eeq
Of course, both $\theta_0$ and $\tilde{\theta}_0$ approach each other as $\kappa\rightarrow 1$, and become equal to the single circular-cone opening angle $\Omega$ in that limit.} Later on it will be convenient to express $\kappa$ in terms of the ellipses' second eccentricity, defined as %$e^{\prime}\equiv \sqrt{b^2/a^2-1}$. The relation between both reads
 %On the other hand, one can fix the lengths of the ellipses axes (relative to $z_0$) in advance, and obtain $k$ from
%\begin{equation}
%k=\frac{\sqrt{1+\left(a/z_0\right)^2}}{\sqrt{1+\left(b/z_0\right)^2}}\, . %\quad \cos \theta_0=\frac{1}{\sqrt{1+\left(a/z_0\right)^2}}\, .
%\quad \text{or alternatively,}\quad k=\frac{1}{\sqrt{\cos^2\theta_0+\left(\frac{b}{a}\right)^2\sin^2\theta_0}}\, .
%\end{equation}

%It turns out to be very convenient to write $k$ in terms of the ellipses' second eccentricity, which is defined as
 \beq\label{ecce}
e^{\prime}\equiv\sqrt{b^2/a^2-1}\, . %\quad \text{which is related to the first eccentricity as} \quad e^{\prime}=\frac{e}{\sqrt{1-e^2}}
\eeq
The relation reads
\beq \label{ke}
\kappa^2=\frac{1}{1+{e^{\prime}}^2\sin^2\theta_0}\, .
\eeq
Naturally, the circular case, $\kappa=1$, is recovered when $e^{\prime}=0$.

%In the case of a circular cone of opening angle $\Omega$, the universal contribution to the R\'enyi entropy is given by  
%\beq
%s^{ \rm univ}_n=-a^{(4)}_n(\Omega) \log^2(\ell/\delta)\, , \quad \text{where} \quad a^{(4)}_n(\Omega)=\frac{1}{4}f_b(n)\frac{\cos^2 \Omega}{\sin \Omega}\, .
%\labell{gamsed}
%\eeq
%Our goal is to study how the universal function $a^{(4)}_n(\Omega)$
%\begin{equation}
%a_n^{(4)}(\Omega)=\frac{1}{4}f_b(n)\frac{\cos^2 \Omega}{\sin \Omega}\, ,
%\end{equation}
% corresponding to a circular cone with an opening angle $\Omega$,
 %gets modified when the cones cross section becomes elliptic. 

%It is important to stress that we will be cutting off our elliptic cones at a fixed value of the spherical radial coordinate $r=R$. In the case of a circular cone, this is equivalent to cutting it off at some constant height $z=z_0$. This is not the case however when $e^{\prime}\neq0$, since in that case different values of $\phi$ correspond to different $z_0$ for a given $R$. In particular, one has the relation
%\begin{equation}
%R=\frac{z_{\phi}\sqrt{1+{e^{\prime}}^2\sin^2\theta_0}}{\cos \theta_0 \sqrt{1+{e^{\prime}}^2\sin^2\theta_0\cos^2\phi}}\, .
%\end{equation}
%Hence, the IR boundary of our cut off cones will not lie in a fixed-$z$ plane, but will be somewhat curly  --- see Fig. \ref{}.

Let us now use Solodukhin's formula to compute the universal contribution to the R\'enyi entropy for an elliptic conical entangling surface. The induced metric on the cone parametrized by $t=0$, $\theta=\theta_0$ trivially follows from \req{eucl}, and reads
\begin{equation}
%ds_h^2=dr^2+r^2\left[\frac{\sin^2\theta_0(1+{e^{\prime}}^2 \cos^2\phi)}{1+{e^{\prime}}^2 \sin^2\theta_0 \cos^2\phi}
ds_h^2=dr^2+r^2\left[\frac{\kappa^2\sin^2\theta_0+(1-\kappa^2)\cos^2\phi}{1-(1-\kappa^2)\sin^2\phi}\right]d\phi^2\, ,
\end{equation}
where we already made use of \req{ke}.
%\begin{equation}
%ds_h^2=dr^2+r^2\left[\frac{k^2\sin^2\theta_0+(1-k^2)\cos^2\phi}{1-(1-k^2)\sin^2\phi}\right]d\phi^2\, .
%\end{equation}
The Ricci scalar of this metric vanishes, as expected. The normal vectors to the cone read $n^{(1)}=\partial_t$, $n^{(2)}=\frac{1}{\sqrt{g_{\theta\theta}}}\partial_{\theta}$, and the only non-vanishing component of the associated extrinsic curvatures reads
%\begin{equation}
%K^{(2)}_{\phi\phi}=\frac{\partial_{\theta}g_{\phi\phi}}{2\sqrt{g_{\theta\theta}}}=\frac{k^2 r \cos\theta_0 \sin\theta_0}{\left[1-(1-k^2)\sin^2\phi\right]}\sqrt{\frac{1-k^2\cos^2\theta_0}{k^2\sin^2\theta_0+(1-k^2)\cos^2\phi}}\, .
%\end{equation}
\begin{equation}
%K^{(2)}_{\phi\phi}=\frac{\partial_{\theta}g_{\phi\phi}}{2\sqrt{g_{\theta\theta}}}=\frac{ r \cos\theta_0 \sin\theta_0 \sqrt{1+{e^{\prime}}^2}}{\left(1+{e^{\prime}}^2\sin^2\theta_0\cos^2\phi \right)\sqrt{1+{e^{\prime}}^2\cos^2\phi}}\, .
k^{(2)}_{\phi\phi}=\frac{\partial_{\theta}g_{\phi\phi}}{2\sqrt{g_{\theta\theta}}}=\frac{ r \kappa^2 \sin\theta_0\cos\theta_0 \sqrt{1-\kappa^2\cos^2\theta_0}}{\sqrt{\kappa^2\sin^2\theta_0+(1-\kappa^2)\cos^2\phi}\left(1-(1-\kappa^2)\sin^2\phi \right)}\, .
\end{equation}
This reduces to 
\begin{equation}
k^{(2)}_{\phi\phi}=\frac{r}{2} \sin(2\theta_0)\, ,
\end{equation}
for $\kappa=1$, as expected.
Using this information, it is easy to find
\begin{equation}
%\tr K^2-\frac{1}{2}K^2=\frac{1}{2}\left(h^{\phi\phi}K^{(2)}_{\phi\phi}\right)^2=\frac{\cos^2\theta_0 (1+{e^{\prime}}^2)}{2r^2\sin^2\theta_0(1+{e^{\prime}}^2\cos^2\phi)^3}\, .
\tr  k^2-\frac{1}{2}k^2=\frac{1}{2}\left(h^{\phi\phi}k^{(2)}_{\phi\phi}\right)^2=\frac{\kappa^4 \sin^2\theta_0 \cos^2\theta_0 (1-\kappa^2\cos^2\theta_0)}{2r^2 \left(\kappa^2\sin^2\theta_0+(1-\kappa^2) \cos^2\phi \right)^3}\, .
\end{equation}
%\begin{equation}
%\tr K^2-\frac{1}{2}K^2=\frac{1}{2}\left(h^{\phi\phi}K^{(2)}_{\phi\phi}\right)^2=\frac{k^4\sin^2\theta_0\cos^2\theta_0(1-k^2\cos^2\theta_0)}{2r^2\left[k^2\sin^2\theta_0+(1-k^2)\cos^2\phi\right]^3}\, .
%\end{equation}
%
%\begin{equation}
%\tr K^2-\frac{1}{2}K^2=\frac{1}{2}\left(h^{\phi\phi}K^{(2)}_{\phi\phi}\right)^2=\frac{k^4\sin^2(2\theta_0)(1-k^2\cos^2\theta_0)}{r^2\left[1-k^2\cos(2\theta_0)+(1-k^2)\cos(2\phi)\right]^3}\, .
%\end{equation}
Now, using \req{game} we find
\begin{equation}\label{gfd}
S^{\rm univ}_n=  \frac{f_b(n) \cos^2\theta_0 (1+{e^{\prime}}^2)}{\pi \sin\theta_0}  \int_0^{\pi/2}  \left[\frac{ (1+{e^{\prime}}^2 \sin^2\theta_0\cos^2\phi )^{-1/2} d\phi}{ \, (1+{e^{\prime}}^2\cos^2\phi)^{5/2}} \right] \int_{\epsilon}^R \frac{dr }{r} \log \delta \, ,
\end{equation}
where we have expressed the parameter $\kappa$ in terms of $e^{\prime}$ using \req{ke}. In the above expression we have also introduced an additional cutoff on the cone at $r=\epsilon$. %This radial dependence is unchanged with respect to the circular cones case. 
This yields
\begin{equation}
\int_{\epsilon}^R \frac{dr }{r} \log(\delta)\rightarrow - \log(\epsilon) \log(\delta)+ \dots\, ,
\end{equation}
where the dots denote subleading terms as $\epsilon\to 0$. In principle, $\epsilon$ can be chosen in different ways, depending on how we regulate the cone at the tip. Combined with $\log \delta $, this yields the quadratically logarithmic divergence characteristic of conical surfaces. In the case there is a single cutoff ---the holographic setup would be an example--- the two regulators must be related. Naively, one would set $\epsilon=\delta$, but this fails to produce the right result by a $1/2$ factor ---this mismatch has been mentioned and studied from different perspectives in various papers \cite{Klebanov:2012yf,Myers:2012vs,Safdi:2012sn,Bueno3,Dorn:2016bkd,Dorn:2016eai}.

One way to see what is going on  involves considering the result for a cylindrical region of (cutoff) length $L$ and radius $\rho$. In that case, the analogous result for the R\'enyi entropy universal term is given by
\begin{equation}\label{gfd0}
S^{\rm univ}_n= -\frac{f_b(n)}{2} \frac{L}{\rho} \log\left(\frac{\rho}{\delta} \right)\, .
\end{equation}
If we now consider a thin elliptic cone, the contributions to the R\'enyi entropy from the infinitesimal cylindrical portions that would build up to produce the cone would be given by 
\begin{equation}\label{gfd1}
dS^{\rm univ}_n= -\frac{f_b(n)}{2} \frac{dr(\rho,\theta,\phi)}{\rho} \log\left(\frac{\rho}{\delta} \right)\, ,
\end{equation}
where we used the same notation for the spherical radial coordinate $r$ which runs along the cone surface from the tip.
As long as the cone is right (\ie lines of fixed $\theta$ and $\phi$ are straight), the dependence on $\rho$ of $r$ will be such that $r(\rho,\theta,\phi)=\rho\, g(\theta,\phi)$ for some function of the angular coordinates. Then, one finds
\begin{equation}\label{gfd2}
S^{\rm univ}_n= -\frac{f_b(n) g(\theta,\phi)}{2} \int_{\delta}^{R} \frac{dr}{r} \log\left(\frac{r}{\delta} \right)+ \dots 
\end{equation}
where the dots refer to terms which will produce logarithmic contributions in $\delta$ (but never quadratically logarithmic). Comparing this expression with \req{gfd}, we observe that the IR scale which should appear weighting $\delta$ inside the logarithm actually depends on the value of the radial coordinate $r$, \ie it is not a fixed scale which we can factor out, and its presence has an impact in the universal coefficient. The radial integral in \req{gfd} yields
\begin{equation}
-\int_{\delta}^R \frac{dr}{r}\log \delta= \log^2 \delta+\dots
\end{equation}
This is one of the contributions that appear in \req{gfd2}, but there is another one given by
\begin{equation}
\int_{\delta}^R \frac{dr}{r}\log r= \frac{1}{2} \log^2 R - \frac{1}{2} \log^2\delta \, ,
\end{equation}
namely, there is an additional contribution proportional to $\log^2\delta$ coming from this, which must be taken into account. Combined with the first, it effectively multiplies the answer one would naively obtain from \req{gfd} by a factor $1/2$.

%When integrating over $r$ in \req{gfd}, we have deliberately left the logarithmic divergence unweighted by an IR scale
%Hence, we can use the same arguments as in that case to deal with the radial integral. Naively, integrating over $r$ yields an additional $\log (R/\delta)$, which together with the one already present in $s^{\rm univ}_n$ build up to produce a universal contribution proportional to $\log^2(R/\delta)$. This procedure fails to produce the right answer by a (missing) factor of $1/2$. The origin of this factor is related to the fact that $\int_{\delta}^R \frac{dr}{r} \log(r/\delta)=\frac{1}{2}\log^2(R/\delta)$ whereas $\log(R/\delta) \int_{\delta}^R\frac{dr}{r}=\log^2(R/\delta)$. One can explicitly show that the first is the radial integral one needs to perform when computing the entanglement entropy for conical surfaces in AdS/CFT using the Ryu-Takayanagi prescription, or for the EMI \comment{?}. When using \req{game} instead, one of the logarithms is already outside the integral, and the fact that, strictly speaking, the formula cannot be used when $\Sigma$ contains singularities manifests itself in the failure to produce the $1/2$ factor which appears when the first logarithm appears inside the integrand.
% The first corresponds to the integral one would need to perform when computing this 
%Just like in the case of circular cones, Solodukhin's formula is not good enough to capture this  missing  $1/2$ factor in the case of elliptic cones, but it is good enough to capture everything else.

Taking this into account, we are left with
\begin{equation}\label{gfd33}
S^{\rm univ}_n=-f_b(n) \log^2\delta\,  \left[ \frac{ \cos^2\theta_0 (1+{e^{\prime}}^2)}{2\pi \sin\theta_0} \right] \int_0^{\pi/2}  \left[\frac{ (1+{e^{\prime}}^2 \sin^2\theta_0\cos^2\phi )^{-1/2} d\phi}{ \, (1+{e^{\prime}}^2\cos^2\phi)^{5/2}} \right]\, .
\end{equation}
At this point, one can weight $\delta$ by any of the IR scales $R$ or $z_0$, depending on whether we cutoff the cone at some radial distance, or at some height $z_0$.\footnote{As opposed to the circular cones case, cutting off an elliptic cone at a fixed value of the radial coordinate $r=R$ is inequivalent from cutting it off at some constant height $z=z_0$. In the first case, different values of $\phi$ correspond to different heights $z(\phi,R)=R \cos\theta_0 \sqrt{1-(1-\kappa^2)\sin^2\phi}$ for a given $R$.
Hence, if we choose to cut off our cones at a fixed $r=R$, the corresponding IR boundary will be somewhat curly ---see Fig. \ref{HwP}. Alternatively, we can cut off the cones at some fixed height $z=z_0$, which in terms of the radial coordinate $r$ translates into integrating up to $R(z_0,\phi)=z_0/ \left(\cos\theta_0\sqrt{1-(1-\kappa^2)\sin^2\phi} \right)$ instead. In any case, the coefficient of the R\'enyi entropy universal contribution is not affected by this choice of IR cutoff.}
The choice does not alter the result for the quadratically logarithmic universal term.

We can write then\footnote{Observe that $\log^2(R/\delta)=(\log R -\log \delta)^2=+\log^2 \delta\, +$ subleading.}
\beq
S^{ \rm univ}_n=-a^{(4)}_n(e^{\prime},\theta_0)\, \log^2(R/\delta)\, , \quad \text{where} \quad a^{(4)}_n(e^{\prime},\theta_0)=\frac{1}{4}f_b(n)\gamma(e^{\prime},\theta_0)\, ,
\labell{gamse}
\eeq
and where we defined
%\beq
%E(k,\theta_0)=\frac{2\cos^2\theta_0(1-k^2\cos^2\theta_0)}{\pi k^2 \sin^3\theta_0}\int_0^{\frac{\pi}{2}}\frac{d\phi }{\left[ 1+%\frac{(1-k^2)}{k^2\sin^2\theta_0}\cos^2\phi\right]^{5/2}\left[1+\frac{(1-k^2)}{k^2}\cos^2\phi\right]^{1/2}}\, .
%\labell{gameww}
%\eeq
%This can be nicely rewritten in terms of the ellipse's second eccentricity,
%\beq
%e^{\prime}=\sqrt{\frac{b^2}{a^2}-1}\, , %\quad \text{which is related to the first eccentricity as} \quad e^{\prime}=\frac{e}{\sqrt{1-e^2}}
%\eeq
%as
\beq
\gamma(e^{\prime},\theta_0)=\frac{2\cos^2\theta_0(1+{e^{\prime}}^2)}{\pi \sin\theta_0} \int_0^{\frac{\pi}{2}}\frac{(1+{e^{\prime}}^2\sin^2\theta_0 \cos^2\phi)^{-1/2}  }{(1+{e^{\prime}}^2\cos^2\phi)^{5/2} } d\phi\, .
\labell{gameww}
\eeq
%\beq
%\gamma(e^{\prime},\theta_0)=\frac{2\cos^2\theta_0(1+{e^{\prime}}^2)}{\pi \sin\theta_0} j(e^{\prime},\theta_0)\, , \quad \text{where}\quad j(e^{\prime},\theta_0)=\int_0^{\frac{\pi}{2}}\frac{(1+{e^{\prime}}^2\sin^2\theta_0 \cos^2\phi)^{-1/2}  }{(1+{e^{\prime}}^2\cos^2\phi)^{5/2} } d\phi\, .
%\labell{gameww}
%\eeq
%where we used the identity
%\beq
%{e^{\prime}}^2=\frac{1-k^2}{k^2 \sin^2\theta_0 }\, ,
%\eeq
Note that for $e^{\prime}=0$, this reduces to 
\begin{equation}\label{cccs}
\gamma(0,\theta_0\equiv \Omega)=\frac{\cos^2\Omega}{\sin \Omega}\, ,
\end{equation}
and we recover the well-known result for the circular cones. In appendix \ref{hypers} we show that this function generalizes in a simple way to circular hypercones in arbitrary even dimensions.

As we mentioned earlier, we could have chosen to present $\gamma(e^{\prime},\theta_0)$ in terms of some other opening angle, defined by the intersection of the cone surface with a different plane containing its axis. For instance, we can express the above function in terms of $\tilde{\theta}_0$, as defined in \req{ssed}.\footnote{Since such opening angle is defined in the direction of the semi-major axis, $b$, our result for $\gamma(e^{\prime},\theta_0)$ should take exactly the same functional form when expressed in terms of $\tilde{\theta}_0$ and $\tilde{e}^{\prime}$, where ${(\tilde{e}^{\prime})}^2=-{e^{\prime}}^2/[1+{e^{\prime}}^2]$, namely, the `eccentricity' one would obtain by flipping $a \leftrightarrow b$ in \req{ecce}. It is a straightforward excercise to show that $\gamma(e^{\prime},\theta_0)$ indeed satisfies this symmetry property.}

It is possible to express $\gamma(e^{\prime},\theta_0)$ in terms of known functions. The result can be written as
%\begin{align}\label{gan}
%j(e^{\prime},\theta_0)=&  \left[3 \cos^2\theta_0\, \alpha_1^2 \,\alpha_2^{1/2} \right]^{-1} \times \left\{\cos^2\theta_0 ({e^{\prime}}^2 \alpha_2+3\cos^2\theta_0)\cdot K\left[1-\frac{\alpha_1}{\alpha_2} \right]    \right. \\&   \left. -2\alpha_3 \, i \, \left[\alpha_2 \cdot E\left[\frac{\alpha_1}{\alpha_2} \right]-(\alpha_1\alpha_2)^{1/2}\cdot E\left[\frac{\alpha_2}{\alpha_1} \right] +{e^{\prime}}^2 \left( \frac{\alpha_2}{\alpha_1}\right)^{1/2} \cos^2\theta_0 \cdot K\left[\frac{\alpha_2}{\alpha_1} \right] \right]  \right\} \,  ,
%\end{align}
\begin{align}\label{gan}
\gamma(e^{\prime},\theta_0)=& \frac{2}{3\pi  \sin\theta_0 \cos^2\theta_0\, \alpha_1 \,\alpha_2^{1/2} }\times \left\{\cos^2\theta_0 ({e^{\prime}}^2 \alpha_2+3\cos^2\theta_0)\cdot K\left[1-\frac{\alpha_1}{\alpha_2} \right]    \right. \\ \notag&   \left. -2\alpha_3 \, i \, \left[\alpha_2 \cdot E\left[\frac{\alpha_1}{\alpha_2} \right]-(\alpha_1\alpha_2)^{1/2}\cdot E\left[\frac{\alpha_2}{\alpha_1} \right] +{e^{\prime}}^2 \left( \frac{\alpha_2}{\alpha_1}\right)^{1/2} \cos^2\theta_0 \cdot K\left[\frac{\alpha_2}{\alpha_1} \right] \right]  \right\} \,  ,
\end{align}
where we defined
\beq
\alpha_1\equiv 1+{e^{\prime}}^2\, ,\quad \alpha_2\equiv 1+{e^{\prime}}^2\sin^2\theta_0\, , \quad \alpha_3\equiv 2 +{e^{\prime}}^2 - (3+2 {e^{\prime}}^2)\sin^2\theta_0\, , 
\eeq
and where $K[x]$ and $E[x]$ are the complete elliptic integrals of the first and second kind, respectively. The expression in brackets in the second line is purely imaginary, so $j({e^{\prime}},\theta_0)$ is real for all physical values of ${e^{\prime}}$ and $\theta_0$. 

%\begin{figure}[t]
%	\centering 
%	\includegraphics[scale=0.4]{elico3D1.pdf}\hspace{0.5cm}
%%	\includegraphics[scale=0.38]{elico3D2.pdf}\hspace{0.5cm}
%	\includegraphics[scale=0.36]{elico2Dd.pdf}
%	\caption{We plot $\gamma$ as a function of the ellipses second eccentricity, $e^{\prime}$, and the semi-opening angle $\theta_0$ along the ellipses major axis. For fixed $\theta_0$, $\gamma$ tends to decrease as we increase the eccentricity from the circular-cones case, $e^{\prime}=0$. A minimum is reached around $e^{\prime}\sim 1$ --- see discussion below \req{extre} --- and then $\gamma$ keeps on growing as $e^{\prime}\rightarrow \infty$. This behavior is evident from the 2D plot shown in the right, where the dotted line, corresponding to $e^{\prime}= 1$ lies below the $e^{\prime}=0$ one (in blue) for all values of $\theta_0$.
%}\label{HP}
%\end{figure}

Expansions around $\theta_0=\pi/2$ and $\theta_0=0$ can be easily performed. The result for the first reads
\begin{align}\notag
\gamma(e^{\prime},\theta_0)=&+\frac{8(1+{e^{\prime}}^2)+3{e^{\prime}}^4}{8(1+{e^{\prime}}^2)^{3/2}}\left(\theta_0-\frac{\pi}{2} \right)^{2}+\frac{16+56{e^{\prime}}^2+34{e^{\prime}}^4+9{e^{\prime}}^6}{96(1+{e^{\prime}}^2)^{5/2}}\left(\theta_0-\frac{\pi}{2} \right)^{4}\\ &+\mathcal{O}\left(\theta_0-\frac{\pi}{2} \right)^{6}\, .
\end{align}
The second one yields a leading divergent term of order $1/\theta_0$, just like in the circular-cones case. The coefficient can be obtained  straightforwardly from \req{gan} as a function of $e^{\prime}$, but it is not particularly illuminating. At leading order in $e^{\prime}$, one finds
\begin{equation}
\gamma(e^{\prime},\theta_0)=\left[1-\frac{{e^{\prime}}^2}{4}+\mathcal{O}({e^{\prime}}^4)\right]\frac{1}{\theta_0}+\mathcal{O}(\theta_0)\, .
\end{equation}
For large values of the eccentricty, $\gamma(e^{\prime},\theta_0)$ diverges linearly with $e^{\prime}$. %j(e^{\prime},\theta_0)=&  \left\{\cos^2\theta_0 ({e^{\prime}}^2 (1+{e^{\prime}}^2\sin^2\theta_0)+3\cos^2\theta_0)
%\cdot K\left[\frac{-{e^{\prime}}^2\cos^2\theta_0}{1+{e^{\prime}}^2\sin^2\theta_0} \right]  \right. \\&   -2(2 +{e^{\prime}}^2 - (3+2 {e^{\prime}}^2)\sin^2\theta_0) \, i \, \left[(1+{e^{\prime}}^2\sin^2\theta_0) \cdot E\left[\frac{1+{e^{\prime}}^2}{1+{e^{\prime}}^2\sin^2\theta_0} \right] \right.  \\ & \left. \left.  -(1+{e^{\prime}}^2)^{1/2}(1+{e^{\prime}}^2\sin^2\theta_0)^{1/2}\cdot E\left[\frac{\alpha_2}{\alpha_1} \right] +{e^{\prime}}^2 \left( \frac{\alpha_2}{\alpha_1}\right)^{1/2} \cos^2\theta_0 \cdot K\left[\frac{\alpha_2}{\alpha_1} \right] \right]  \right\} \,  ,
 For small  $e^{\prime}$, one finds, in turn,
\beq\label{ggas}
\gamma(e^{\prime},\theta_0)=\frac{\cos^2\theta_0}{\sin\theta_0}\left[1-\frac{3-\cos (2\theta_0)}{8}{e^{\prime}}^2+\frac{283-92\cos(2\theta_0)+9\cos(4\theta_0 )}{512}{e^{\prime}}^4 +\mathcal{O}\left({e^{\prime}}^6\right)\right]\, .
\eeq
Since the coefficient of the quadratic term is always negative, circular cones locally maximize $\gamma$ for fixed $\theta_0$. A more meaningful comparison of the whole R\'enyi entropy can be performed by fixing the lateral area of the cones, instead of  $\theta_0$, since in that case the area-law contribution is the same. In appendix \ref{maxim} we show that when we fix the latera area of the cones, the quadratic term in \req{ggas} conspires to disappear, the coefficient of the quartic one being always positive. Since $\gamma$ contributes with a minus sign to $S_n$, the cones that maximize the R\'enyi entropy  within the family of elliptic cones are the circular ones.

\subsubsection{Free fields}
%In the case of a cone the boundary conditions one would need to impose read
%\begin{equation}
%\lim_{\epsilon\rightarrow 0^+}\Phi_{\ell}(\pi/2+\epsilon,\theta,\phi)=e^{2\pi i a} \lim_{\epsilon\rightarrow 0^+}\Phi_{\ell}(\pi/2-%\epsilon,\theta,\phi)\, ,\quad \theta \in [-\Omega,\Omega]\, , \quad \phi \in[0,2\pi) .
%\end{equation} 
As we have seen, for free fields we need to characterize the boundary of the intersection of the conical entangling region with $\mathbb{S}^2$. For circular cones, this is a $\mathbb{S}^1$, as shown in Fig.~\ref{conefree}. The embedding of the $\mathbb{S}^1$ on $\mathbb{S}^2$ is given simply by $\theta=\Omega$. Then, the normal vector is given by $n=\partial_{\theta}$, and the induced metric is $ds^2_{\mathbb{S}^1}=\sin^2\Omega d\phi$, so $\sqrt{h}=\sin\Omega$. The only nonvanishing component of the extrinsic curvature is given by 
 \begin{equation}
 k_{\phi\phi}=\frac{\partial_{\theta} g_{\phi\phi}}{2}=\sin\Omega \cos \Omega\, ,
 \end{equation}
 so we find
 \begin{equation}
 \int_{\gamma} k^2= \int_0^{2\pi} d\phi \sqrt{h} \left(h^{\phi\phi}k_{\phi\phi} \right)^2= \frac{2\pi\cos^2\Omega}{\sin\Omega} \, ,
\end{equation}
 and hence
  \begin{equation}
 S_{n}|_{\log^2} =- \frac{1}{4}f_b(n) \frac{\cos^2\Omega}{\sin\Omega} \log^2\delta\, .
  \end{equation}
 which precisely agrees with the angular dependence expected for a conical entangling region for general CFTs in \req{gamse} and \req{cccs}.
 
% \comment{rearrange this with next section, where calculation of elliptic cones for general CFTs using Solodukhin is made;  }
We can readily verify that this also works for elliptic cones. In that case, the metric on the unit round $\mathbb{S}^2$ in sphero-conal coordinates  reads
\begin{equation}
ds^2_{\mathbb{S}^2}=\left[\frac{\kappa^2\sin^2\theta+(1-\kappa^2)\cos^2\phi}{1-\kappa^2\cos^2\theta}\right]d\theta^2+\left[\frac{\kappa^2\sin^2\theta+(1-\kappa^2)\cos^2\phi}{1-(1-\kappa^2)\sin^2\phi}\right]d\phi^2\, .
\end{equation}
The induced metric on the intersection of the elliptic cone of semi-opening angle $\theta_0$ with the $\mathbb{S}^2$ is given by
\begin{equation}\label{nms}
ds^2_h=\left[\frac{\kappa^2\sin^2\theta_0+(1-\kappa^2)\cos^2\phi}{1-(1-\kappa^2)\sin^2\phi}\right]d\phi^2\, .
\end{equation}
The only non-vanishing component of the extrinsic curvature associated to the normal vector $n=\frac{1}{\sqrt{g_{\theta\theta}}}\partial_{\theta}$ reads
\begin{equation}
k_{\phi\phi}=\frac{\partial_{\theta}g_{\phi\phi}}{2\sqrt{g_{\theta\theta}}}=\frac{ \kappa^2 \sin\theta_0\cos\theta_0 \sqrt{1-\kappa^2\cos^2\theta_0}}{\sqrt{\kappa^2\sin^2\theta_0+(1-\kappa^2)\cos^2\phi}\left(1-(1-\kappa^2)\sin^2\phi \right)}\, .
\end{equation}
Then, 
\begin{equation}
k^2=(h^{\phi\phi}k_{\phi\phi})^2=\frac{\kappa^4 \sin^2\theta_0 \cos^2\theta_0 (1-\kappa^2\cos^2\theta_0)}{ \left(\kappa^2\sin^2\theta_0+(1-\kappa^2) \cos^2\phi \right)^3}\, .
\end{equation}
 Plugging this in \req{log2}  along with the induced metric determinant coming from \req{nms}, we are left with an angular integral and a dependence on $\theta_0$ which is again identical to the one found using Solodukhin's formula \req{gfd33}, so the final result is again given by \req{gamse} and \req{gan}, as it should.

\subsection{Higher dimensions}
As we argued above, the coefficient $c_{n,-1}$ in the high-mass expansion \req{expanse} is forced to vanish for odd-dimensional CFTs. Hence, the R\'enyi entropy for conical entangling regions in such a number of dimensions will not contain a  $\log^2 \delta$ term. Rather, the universal contribution will be logarithmic, the corresponding coefficient having a highly non-local origin.  In the case of five-dimensional theories, this structure was explicitly verified for holographic theories dual to  Einstein and Gauss-Bonnet gravities in \cite{Myers:2012vs}.

% particular models logarithmic contribution was computed explicitly --- for the entanglement entropy --- in the case of $d=5$ holographic theories dual to Einstein and Gauss-Bonnet gravities in \cite{Myers:2012vs}.

On the other hand, for even dimensions larger than four, the local  $\log^2 \delta$ term will be present for conical regions. In appendix~\ref{hypers}, we show that the dependence of the universal function $a^{(4)}_n(\Omega)$ on the opening angle generalizes in a simple way to arbitrarily high even dimensions. Namely, we argue that the universal contribution to the R\'enyi entropy of right circular (hyper)cones is given, for general CFTs, by the simple formula
\beq 
S_n^{ \rm univ}=(-1)^{\frac{d-2}{2}}a_n^{(d)}(\Omega) \log^2\left(\frac{R}{\delta}\right)\, , \quad \text{with} \quad a_n^{(d)}(\Omega) =\frac{\cos^2\Omega}{\sin\Omega} \sum _{j=0}^{\frac{d-4}{2}}\left[\gamma_{j,n}^{(d)}\, \cos (2j\Omega) \right]\, ,\label{gamsed}
\eeq
where the only information about the underlying theory appears through the coefficients $\gamma_{j,n}^{(d)}$, which will be related to the R\'enyi entropy generalizations of the trace-anomaly charges.

%\begin{equation}
%\see = 4\pi^2 \kappa \sin^4\Omega \int d\rho_A \int d\rho_B \frac{\rho_A \rho_B (\rho_A^2+\rho_B^2)}{(\rho_A-\rho_B)^3[\rho_A^2+\rho_B^2-2\rho_A\rho_B \cos(2\Omega)]^{3/2}}
%\end{equation}

%Now, since $\alpha_{\rm tot}$ is always an even number between $2$ and $d-2$, we observe that the

%This includes, for example, all $d=4$ terms, corresponding to $n=0$ and $n=2$, both with  $m=0$ \cite{Solodukhin:2008dh},  and all but one of the $d=6$ terms, corresponding to $n=0$, $n=1$, $n=2$, \cite{Safdi:2012sn,Miao:2015iba}.

%We conjecture that \req{gamsed} captures the $\Omega$-dependence of the cone function for general theories in even dimensions.
%, not only in the case of the entanglement entropy, but for general values of the R\'enyi index --- replace $a^{(d\, \text{even})}(\theta)$ and $\alpha_j^{(d)}$ by $a_n^{(d\, \text{even})}(\theta)$ and $\alpha_{n,\,j}^{(d)}$ in \req{add1}. 
%This conjecture is of course correct in $d=4$, and agrees with all the known results for $d=6$ theories and with the result for holographic theories dual to Einstein gravity for $d=8,10,12,14$. \comment{More evidence for general CFTs in $d=6$ and additional holographic theories in other dims; perhaps from EMI?} \comment{can one characterize $\alpha_0^{(d=6)}$ and $\alpha_1^{(d=6)}$ in general? Probably, check Safdi and Miao papers}\\
%\subsection{Four dimensions}
%In $d=4$, \req{gamsed} is well established \cite{ Klebanov:2012yf}. The relevant terms in

%\subsection{Six dimensions}

\section{Wedge entanglement versus corner entanglement}\label{wvsc}
In this section we study the entanglement entropy of a region bounded by a wedge in $(3+1)$-dimensional CFTs.
  We have already encountered wegdges of opening angle $\Omega=\pi/2$ in our analysis of the entanglement entropy
  of trihedral corners using the EMI. Here, we begin by analyzing the wedge contribution in the nearly smooth limit,
  $\Omega\simeq \pi$, for general CFTs. We then consider the calculation for free fields using dimensional reduction.
  We complement the free scalar calculation in \cite{Klebanov:2012yf} with an analysis of the emergence of $f(\Omega)$ from $a(\Omega)$ for general free fields (scalars and fermions). We confirm that the angular dependence of $f(\Omega)$ is indeed a universal quantity given by $a(\Omega)$, which is the $(2+1)$-dimensional corner function of the corresponding
  lower dimensional free theory.
  We also illustrate how the nonuniversal character of the overall factor is connected to the different possible choices of regulators along the transverse and corner directions.
  Finally, we show that, contrary to a previous claim made in \cite{Klebanov:2012yf}, the wedge and corner functions do differ for holographic theories dual to Einstein gravity in the bulk. This suggests that the
  relation between the wedge and the lower dimensional corner function does not hold for interacting theories.

The setup is the following. 
 In the wedge case, the entangling region at some fixed time slice corresponds to the set $\{(r,\phi,z)$ such that $0\leq  r < \infty, \,  0\leq z < \infty \, , 0\leq \phi \leq \Omega\}$ in cylindrical coordinates. The entanglement entropy takes the form in \req{kle2}, namely:
\begin{equation}\label{kle}
\see =b_2 \frac{H^2}{\delta^2}- f(\Omega)\frac{H}{\delta}+\mathcal{O}(\delta^0)\, ,
\end{equation}
where $H$ and $\delta$ are IR and UV cutoffs respectively, $b_2$ is a nonuniversal constant, and $f(\Omega)$ is a function of the wedge opening angle whose overall normalization depends on the UV cutoff. %not well defined. %\footnote{Indeed, redefinitions of the UV cutoff, \eg $\delta \rightarrow a \delta$, modify such normalization, which therefore has no physical meaning. The angular dependence is in principle well-defined and susceptible of being compared to the corner function one.}.
The analogous entangling region bounded by the corner in one dimension less is given in polar coordinates by  $\{(r,\phi)\, \,  \text{such that:} \,\, 0\leq  r < \infty, 0\leq \phi \leq \Omega\}$. There, the entanglement entropy reads
\begin{equation}\label{corner1}
\see =b_1 \frac{H}{\delta}- a(\Omega)\log \left( \frac{H}{\delta}\right)+\mathcal{O}(\delta^0)\, ,
\end{equation}
where $b_1$ is some other nonuniversal constant, and $a(\Omega)$ is a cutoff-independent function of the opening angle.

\subsection{General CFTs in the nearly smooth limit}
To obtain the wedge function in the nearly smooth limit, $\Omega\simeq \pi$, we will employ the 2nd order entanglement
susceptibility, $\chi^{(2)}$.\footnote{This quantity is sometimes called ``entanglement density'', but we shall not use this convention.} This was for example used \cite{Faulkner:2015csl} to obtain the $d=3$ corner function of a general CFT in 
the nearly smooth limit \cite{Bueno1,Bueno2}: 
\begin{align} \label{eq:smooth-corner}
  a(\Omega)=\frac{\pi^2 \ctt}{24} \left(\pi -\Omega \right)^2+\dots 
\end{align}
The idea is to consider how the entanglement entropy changes as a function of a small
deformation $A\to A+\delta A$:
\begin{align} \label{S-expansion}
  \see(A+\delta A)\!=\! \see(A) + \int_{\partial A}\!\!\!  d^{d-2}{\bf r}  \chi^{(1)}({\bf r}) \zeta({\bf r})
  + \frac{1}{2!}\!\int_{\partial A}\!\!\! d^{d-2}{\bf r}\! \int_{\partial A}\!\! d^{d-2}{\bf r}' \chi^{(2)}({\bf r},{\bf r}')
  \zeta({\bf r})\zeta({\bf r}') + \cdots 
\end{align}
The deformation $\delta A$ is defined by sending a point ${\bf r}$ on $\partial A$ to $\zeta({\bf r}) {\bf n}({\bf r})$,
where ${\bf n}({\bf r})$ is the unit normal at ${\bf r}$ and $\zeta({\bf r})$ is taken to be small. We shall consider the case where the state is the vacuum of a CFT,
and $A$ is the half-space at a fixed time slice. Since we are working with a pure state,
the odd susceptibilities such as $\chi^{(1)}$ vanish. The first contribution will come from $\chi^{(2)}$, where \cite{Faulkner:2015csl}
  \begin{align} \label{eq:chi-plane}
  \chi^{(2)}({\bf r} - {\bf r}') = -\frac{2\pi^2 \ctt}{d+1}\, \frac{1}{|{\bf r}-{\bf r}'|^{2(d-1)}}\, ,
\end{align}
is the non-local or universal part of the susceptibility, which is controlled by the 2-point function coefficient of the stress tensor, $\ctt$.

Setting $d=4$, let us parametrize $A$ as the half-space $z<0$, where the spatial coordinates are ${\bf r}=(x,y,z)$. We then choose the following deformation
\begin{align}
  \zeta({\bf r}) = \Theta(x)\, x \tan\alpha\,, 
\end{align}
where $\Theta(x)$ is the Heaviside step function. 
This introduces a wedge of angle $\Omega= \pi-\alpha$, where $\alpha$ is taken to be small. Using \req{S-expansion}, the variation of the entanglement entropy reads 
\begin{align}
  \delta \see = - \frac{\pi^2 \ctt \tan^2\alpha}{5}\int_0^\infty dx dx' \int_{-\infty}^\infty dy dy' \frac{x x'}{((x-x')^2+(y-y')^2)^3}\, .
\end{align}
We first perform the $x'$ integral; the integrand becomes
\begin{align}
  g(x,\Delta) = \frac{x \left(3 \pi  x \sgn(\Delta )+2 \Delta  \left(\frac{x^2}{\Delta
   ^2+x^2}+2\right)+6 x \tan ^{-1}\left(\frac{x}{\Delta }\right)\right)}{16 \Delta ^5}\, ,
\end{align}
where we defined $\Delta=y-y'$. Before performing the $x$-integral, let us perform a large-$x$ expansion to isolate
the IR divergent term:
\begin{align}
  g(x, \Delta)= \frac{3 \pi  x^2 \sgn(\Delta )}{8 \Delta ^5}+\frac{1}{20
   x^3}+ O(1/x^4)\, .
\end{align}
We can then separate the IR divergent piece:
\begin{align}
  \int dx f(x,\Delta) &=   \int_0^\infty dx \left[ g(x,\Delta)- \frac{3 \pi  x^2 \sgn(\Delta )}{8 \Delta ^5}\right]
                        +\int_0^Hdx\, \frac{3 \pi  x^2 \sgn(\Delta )}{8 \Delta ^5} \\
  &= \frac{1}{24\Delta^2} + \frac{\pi  H^3 \sgn(\Delta )}{8 \Delta ^5} \, ,
\end{align}
where $H$ is the IR cutoff.
We now perform the $y'$ integral using a splitting regularization:
$\int_{-\infty}^\infty dy'=\int_{-\infty}^{y-\delta}dy'+\int_{y+\delta}^\infty dy'$. Finally, the $y$ integral can beformed with
UV an IR cutoffs, and we obtain:
\begin{align}
  \delta \see = - \frac{\pi^2 \ctt (\Omega-\pi)^2}{60}\, \frac{H}{\delta} + \cdots 
\end{align}
where the dots represent not only subleading terms in $\delta$, but also $1/\delta^4$ and $1/\delta^3$ divergences. 
These two unphysical terms result from our regularization scheme, and should be discarded. They do not influence the wedge contribution
$\propto H/\delta$. We thus see that in the nearly smooth limit, the wedge function is:
\begin{align}\label{fomm}
  f(\Omega)=  \frac{\pi^2 \ctt }{60}\, (\Omega-\pi)^2\, ,
\end{align}
up to an overall regularization-dependent prefactor. In this limit, the wedge function behaves in a very similar way to the corner
function $a(\Omega)$ in one lower dimension, \req{eq:smooth-corner}.  
This result also holds for the R\'enyi entropies, but with the replacement of $C_T$ by $f_b(n)$ (up to an unimportant prefactor) \cite{Bianchi:2015liz}. 
   
\subsection{Free fields}
 %\comment{rewrite all this using  general dimensional reduction results}
Certain contributions to the entanglement entropy of $d$-dimensional free-field theories can be obtained from others corresponding to $(d-1)$-dimensional contributions \cite{Casini:2005rm,Casini1}. This is the case, in particular, when the entangling region $W$ takes the form of a direct product, $W=C\times  \mathbb{R}$, where $C$ is some $(d-1)$-dimensional set. %and where we parametrize the additional direction by the coordinate $z$.  
In this situation, one would usually cutoff the extra dimension, which we parametrize here by $z$, at some finite distance $L$ to avoid an IR-divergent result. The idea here is to compactify $z$ imposing periodic boundary conditions, $z=z+L$, and then decompose the corresponding $d$-dimensional field into its Fourier modes along that direction. This reduces the problem to a $(d-1)$-dimensional one. Starting with a mass-$m$ free field in $d$-dimensions, the problem is mapped to the one of infinitely many $(d-1)$-dimensional 
independent fields of masses
\begin{equation}
M^2_k=m^2+p_k^2\, ,
\end{equation}
where $p_k\equiv (2\pi k/L)$ is the momentum of the $k$-th mode along $z$. Then, the entanglement entropy for $W$ in the $d$-dimensional theory can be obtained by summing over all $(d-1)$-dimensional entropies corresponding to the entangling region $C$. In the large-$L$ limit, the sum can be converted into an 
integral of the form \cite{Casini:2005rm,Casini:2009sr,Casini1}
\begin{equation}\label{wedg} 
\see(W)= \frac{c_{\rm f} L}{\pi} \int^{1/\epsilon} dp \, \see(C,\sqrt{m^2+p^2})\, ,
\end{equation} 
where the extensivity of $\see(W)$ on $L$ is manifest, and where\footnote{For the fermion, $c_{\rm f}$ is the quotient between the dimensions of the spinorial spaces of $d$ and $(d-1)$ dimensions, respectively. When $d$ is an odd number, $c_{\rm f}=1$, whereas for even $d$, $c_{\rm f}=2$.}
\begin{equation}
c_{\rm f} = \begin{cases}1 & \text{for a scalar}\, , \\2^{\lfloor \frac{d}{2} \rfloor}/ 2^{\lfloor \frac{d-1}{2} \rfloor}  & \text{for a (Dirac) fermion}\, .\end{cases}
\end{equation}
Note also that we have introduced a spatial cutoff $\epsilon$, so that we do not consider infinitely massive modes, but only those with energies smaller than $1/\epsilon$.

\begin{figure}[t]
	\centering 
	\includegraphics[scale=0.4]{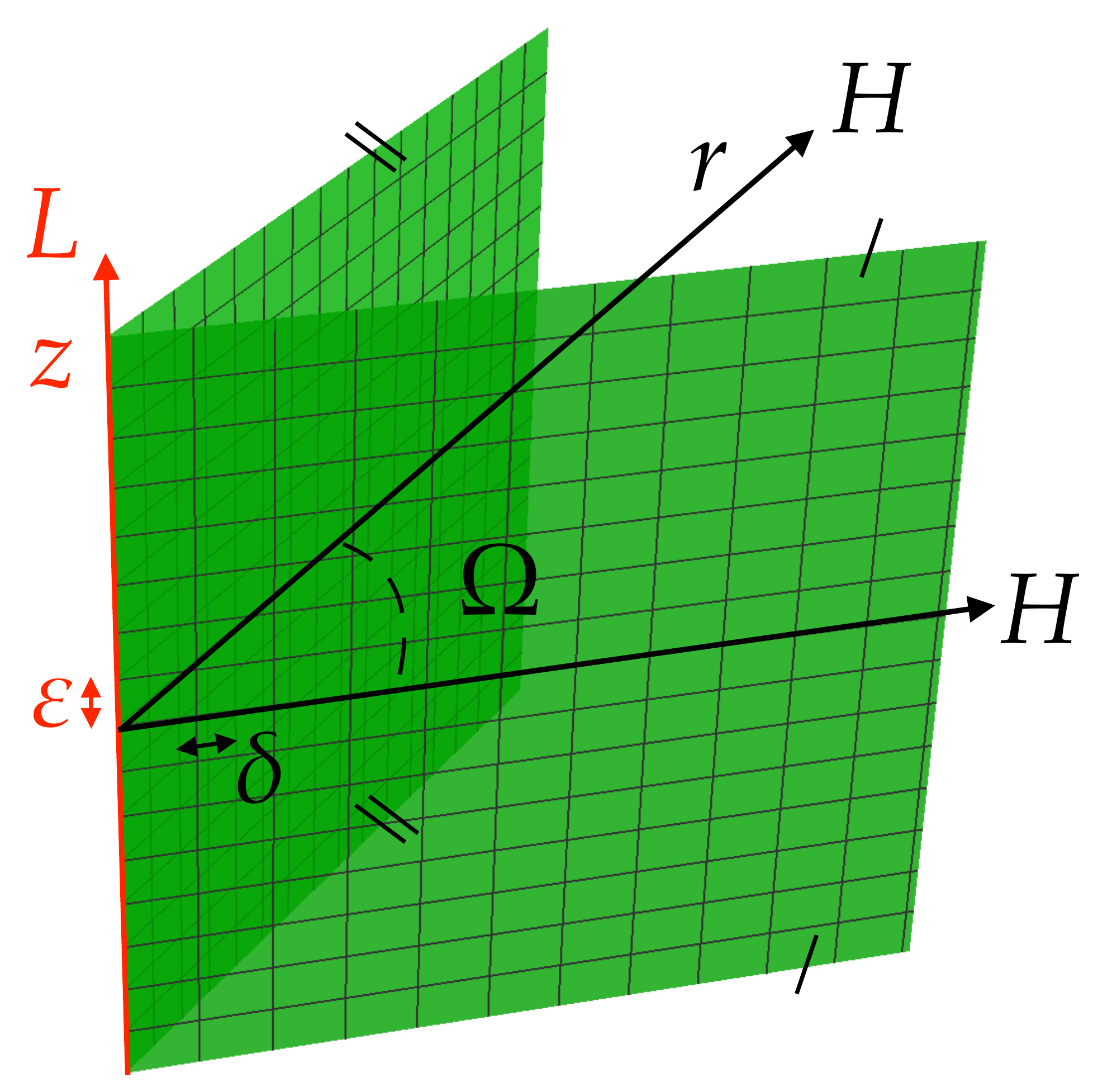}\hspace{0.5cm}
	\caption{We plot a wedge-shaped entangling surface of opening angle $\Omega$ and a corner region in one dimension less resulting from its dimensional reduction. While the wedge contribution to the entanglement entropy -$f(\Omega)H/\delta$ gets polluted by the presence of inequivalent regulators along the radial and transverse directions ---see \req{pollu}--- the dependence on the opening angle is well-defined.  }\label{corwe}
\end{figure}

Of course, in the case at hand, $W$ is the wedge in $(3+1)$-dimensions, and $C$ is the corner in $(2+1)$ ---see Fig. \ref{corwe}. For the latter, the entanglement entropy for a massive field takes the form
\begin{equation}\label{corn}
\see(C,m)=b_1\frac{H}{\delta}+b_0 m H+ a(\Omega) \log(m\delta)+\dots
\end{equation}
in the large-mass limit.
Note that the scale weighting the UV cutoff $\delta$ along the corner directions in the logarithmic contribution controlled by the universal function $a(\Omega)$ is $1/m$ rather than the IR cutoff $H$, which reflects the local character of this term.

Plugging this expression in \req{wedg}, performing the integrals and considering the massless limit, we are left with
\begin{equation}
\see(W)=  \frac{b_1  c_{\rm f}}{\pi} \frac{H L}{\epsilon \delta} + \frac{b_0 c_{\rm f}  }{2\pi} \frac{LH }{\epsilon^2}- \frac{a(\Omega) c_{\rm f}}{\pi} \frac{L }{\epsilon} \left[1-\log\left(\frac{\delta}{\epsilon} \right) \right]\, .
\end{equation}
This is an interesting result. First, we observe that the first two terms, coming from the area-law like term in \req{corn} and $mH$, both contribute to the area law in $\see(W)$. This is manifest if we write the UV cutoff along the $z$ direction in terms of the one along the corner radial directions, $\epsilon=\alpha_{\rm \ssc UV} \delta$, for some $\mathcal{O}(1)$ constant $\alpha$. We can also relate the IR cutoffs $L$ and $H$, $L=\alpha_{\rm \ssc IR} H$. Then, we obtain
\begin{equation}
\see(W)= b_2  \frac{H^2}{\delta^2}- \frac{a(\Omega)  c_{\rm f}}{\pi} \frac{H}{\delta} \left[1+ \log\left(\alpha_{\rm UV} \right) \right] \alpha_{\rm \ssc IR}  \, ,
\end{equation}
where $b_2$ is some nonuniversal constant. This precisely takes the form in \req{kle} expected for general CFTs.  We observe that the function $f(\Omega)$ is indeed related to the corner function $a(\Omega)$ through
\begin{equation}\label{pollu}
f(\Omega)\propto a(\Omega)  \left[1+ \log\left(\alpha_{\rm UV} \right) \right] \alpha_{\rm \ssc IR} \, .
\end{equation} 
There is no physical reason to prefer, say, $\alpha_{\rm \ssc IR}= \pi/ c_{\rm f}$, $\alpha_{\rm \ssc UV}=1$ over any other choice, which illustrates the nonuniversal character of the overall factor in $f(\Omega)$ and how this is polluted by the different choices of regulators.  The angular dependence, however, is physically meaningful, and  inherited from the corner one, as observed in \cite{Klebanov:2012yf} in the particular case of a scalar field. Note also that the above connection extends straightforwardly to general R\'enyi entropies.

Naturally, the dimensional reduction performed here is exclusively valid for free scalars and fermions. It is nonetheless tempting to speculate with the possibility that $a(\Omega)$ and $f(\Omega)$ may be connected in a similar fashion for a larger family of CFTs \cite{Klebanov:2012yf}. As we show in the following subsection, this is not the case in general, as the angular dependence of both functions 
is in fact different for holographic CFTs dual to Einstein gravity in the bulk.

\subsection{Holography}
Consider now holographic theories dual to Einstein gravity. The bulk action is given by
\begin{equation}\label{Einst}
I=\frac{1}{16\pi G} \int d^{(d+1)}x \sqrt{|g|} \left[\frac{d(d-1)}{L^2}+R \right]\, ,
\end{equation}
where $G$ is Newton's constant, and the cosmological constant length-scale $L$ coincides with the AdS$_{(d+1)}$ radius.
Then, the result for the universal function $a(\Omega)$ appearing in the entanglement entropy of a corner region in $d=3$, computed using the Ryu-Takayanagi prescription \cite{Ryu:2006bv,Ryu:2006ef}, is given by
%The result for the entanglent entropy of a corner region in a three-dimensional CFT dual to Einstein gravity is given by
 \cite{Drukker:1999zq,Hirata:2006jx}
%\begin{equation}
%S_{\rm \ssc EE}=\frac{L^2}{2G}\frac{H}{\delta}-a(\Omega) \log\left(\frac{H}{\delta} \right)+c_0\, ,
%\end{equation}
%where
\begin{equation}
a(\Omega)=\frac{L^2}{2G}\int_{g_0}^{\infty} dg\frac{g}{\sqrt{g^2-g_0^2}} \left[1-\sqrt{\frac{1+g^2}{1+g_0^2+g^2}}\right]\, ,
\end{equation}
where $g_0(\Omega)$ is an implicit function of the opening angle, namely
\begin{equation}\label{ope}
\Omega=\int_{g_0}^{\infty}dg \frac{2}{g\sqrt{(1+g^2)\left(\frac{g^2(1+g^2)}{g_0^2(1+g_0^2)}-1\right)}}\, .
\end{equation}
%In the above expression, the leading contribution is the usual area-law term, and $c_0$ is a nonuniversal constant contribution. The universal logarithmic contribution is controlled by the function $a(\Omega)$.
On the other hand, the holographic result for the wedge region $f(\Omega)$ reads \cite{Klebanov:2012yf,Myers:2012vs}
%\begin{equation}
%S_{\rm \ssc EE}=\frac{L^3}{4G}\frac{H^2}{\delta^2}-f(\Omega)\frac{H}{\delta}+c_0\, ,
%\end{equation}
%where
\begin{equation}
f(\Omega)=\frac{L^3}{2G} \left[g_0-\int_{g_0}^{\infty}dg\left(\frac{ g(1+g^2)}{\sqrt{g^2(1+g^2)^2-g_0^2(1+g_0^2)^2}}-1\right)\right]\, ,
%\frac{L^3}{4G}\left[g_0-\int_{g_0}^{\infty}\frac{dg'}{g'^2}\int_{g_0}^{g'}dg \left(\frac{g^2(1+g^2)}{\sqrt{g^2(1+g^2)^2-g_0^2(1+g_0^2)^2}}-g\right)\right]
\end{equation}
where $g_0$ is related to the wedge opening angle through %\req{ope}
\begin{equation}\label{ope}
\Omega=\int_{g_0}^{\infty}dg \frac{2}{g\sqrt{(1+g^2)\left(\frac{g^2(1+g^2)^2}{g_0^2(1+g_0^2)^2}-1\right)}}\, .
\end{equation}
%Now, as opposed to $a(\Omega)$ for the corner, $f(\Omega)$ does not appear in front of a universal contribution. In particular, this means that at least the overall normalization of $f(\Omega)$ is not well defined, as it can be modified by changing the UV cutoff. The angular dependence, however, can in principle be well defined, and one can try to compare it to the one of the corner function. 
As mentioned earlier, in \cite{Klebanov:2012yf} it was observed that ---at least within the numerical resolution considered--- the dependence of both functions on the corresponding opening angles appears to be actually identical, \ie $f(\Omega)=a(\Omega)$ up to a global factor. %Additional evidence supporting the idea that both functions may be universally related was provided  for a free scalar field, for which the result found for the wedge exactly matched the corner one previously obtained in \cite{Casini:2006hu}. 
As we show here, the claim is actually not correct in the holographic case.

\begin{figure}[t]
	\centering 
	\includegraphics[scale=0.57]{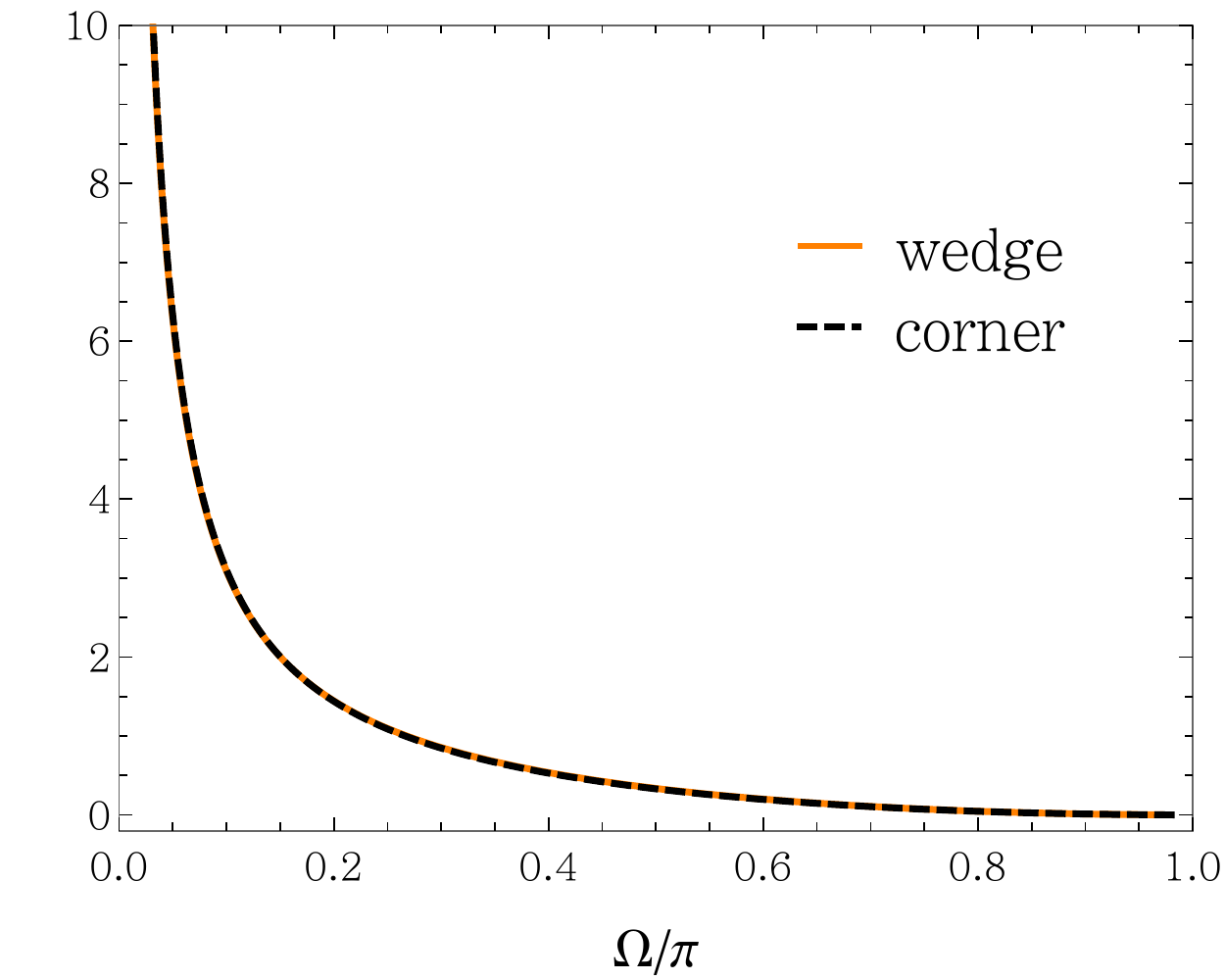}
	\hspace{0.2cm}
	\includegraphics[scale=0.61]{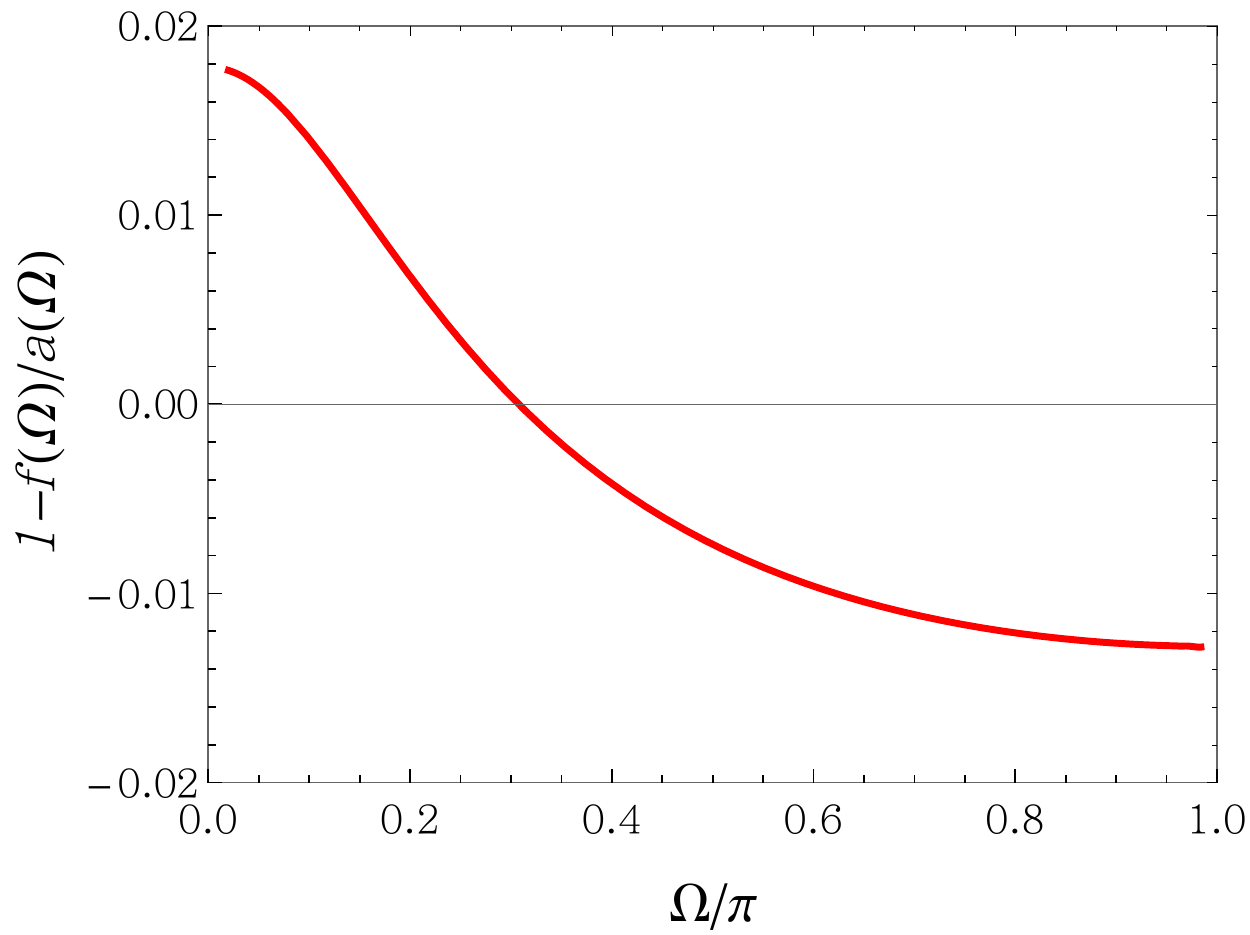}
		\caption{(Left) We plot the wedge (orange) and corner (black dashed) functions $f(\Omega)$ and $a(\Omega)$ normalized so as to make them fall approximately on top of each other, as observed in \cite{Klebanov:2012yf}. (Right) We plot $1-f(\Omega)/a(\Omega)$. As we can see, both functions are in fact slightly different from each other.
}\label{wedgecor}
\end{figure}

In the $\Omega\rightarrow 0$ and $\Omega \rightarrow \pi$ limits, corresponding to very sharp and almost smooth corner/wedges, respectively, it is possible to show that $a(\Omega)$ and $f(\Omega)$ behave as\footnote{Observe that for $\Omega\rightarrow \pi$, the Einstein gravity result for $f(\Omega)$ can be written in terms of $\ctt$ as 
\begin{equation}
f(\Omega)=\frac{3\pi^4 \ctt}{1280}(\Omega-\pi)^2+\dots
\end{equation}
where we used $\ctt=5L^3/(\pi^3 G)$. This disagrees with \req{fomm}, which is not very surprising given that $f(\Omega)$ is only defined up to an overall regulator-dependent coefficient.
 }
\begin{align}
a(\Omega)&\overset{\Omega\rightarrow 0}{=}\frac{\kappa}{\Omega}+\dots\, ,\quad a(\Omega)\overset{\Omega\rightarrow \pi}{=}\sigma\cdot (\Omega-\pi)^2+\dots\, , \\ 
f(\Omega)&\overset{\Omega\rightarrow 0}{=}\frac{\tilde \kappa}{\Omega}+\dots\, ,\quad f(\Omega)\overset{\Omega\rightarrow \pi}{=}\tilde \sigma\cdot (\Omega-\pi)^2+\dots\, , 
\end{align}
where the dots stand for subleading contributions, and where 
\begin{align}
\kappa=\Gamma\left(\tfrac{3}{4}\right)^4\frac{L^2}{2\pi G}\, , \quad  \sigma=\frac{L^2}{8\pi G}\, , \quad \quad \quad
\tilde \kappa=\frac{2^{2/3}\pi^{3/2}\Gamma\left(\tfrac{5}{6}\right) L^3}{ \Gamma\left(\tfrac{1}{6}\right)^2 G}\, , \quad \tilde \sigma=\frac{3\pi L^3}{256 G}\, .
\end{align}
As we mentioned earlier, the overall constant in $f(\Omega)$ is not well defined, so the values of $\tilde \kappa$ and $\tilde \sigma$ are not meaningful by themselves. However, the ratio $\tilde \kappa/\tilde \sigma$ is in principle a meaningful quantity. If the claim in \cite{Klebanov:2012yf} were true, such ratio should agree with the one corresponding to the corner function. We find, however
\begin{equation} 
\frac{\kappa}{\sigma}=4 \Gamma\left(\tfrac{3}{4}\right)^4\simeq 9.0198\, , \quad  \frac{\tilde \kappa}{\tilde \sigma}=\frac{ 2^{2/3} 256 \sqrt{\pi} \Gamma\left(\tfrac{5}{6}\right)}{ 3\Gamma\left(\tfrac{1}{6}\right)^2}\simeq 8.7469 \, ,
\end{equation}
which is close, but obviously different. This discrepancy can also be observed by plotting $[1-f(\Omega)/a(\Omega)]$, as we have done in Fig. \ref{wedgecor}. Remarkably, the overall factor can be chosen so that $f(\Omega)$ and $a(\Omega)$ differ by less than $\sim 2\%$ for each value of $\Omega$.  
%\comment{how difficult would it be to do the free fermion wedge and compare it to the corner?} 

% In view of this result, we expect $f(\Omega)$ and $a(\Omega)$ to be different for general CFTs. \comment{In fact, in most cases it is not even clear in general what would be the $d=3$ counterpart of a given $d=4$ CFT, except for a few cases like free fields.}

%%%%%%%%%%%%%%%%%%%%%%%%%%%%%%%%%%%%%%%%%%%%%%%%%%%%%%%%%%%%
%%%%%%%%%%%%%%%%%%%%%%%%%%%%%%%%%%%%%%%%%%%%%%%%%%%%%%%%%%%%
%%%%%%%%%%%%%%%%%%%%%%%%%%%%%%%%%%%%%%%%%%%%%%%%%%%%%%%%%%%%

\section{Singular geometries versus entanglement divergences}\label{singi}
In this section we analyze the interplay between singular entangling surfaces, and the structure of divergences of entanglement entropies and mutual information. First we show that, contrary to the usual expectations, mutual information $I(A,B)$ does not necessarily become divergent in the limit when $A$ and $B$ have contact points. In particular, whenever the contact is through a sufficiently sharp corner (anything sharper than a straight corner works), $I(A,B)$ remains finite. Then, we consider the entanglement entropy of curved corners. We provide examples of singular regions which do not change the structure of divergences/universal terms with respect to the smooth case, and also how new divergences can appear when the corners become sufficiently sharp.

% As its name suggests, the characterizing feature of this model is the fact that the mutual information satisfies the extensivity property
%\begin{equation}\label{emir}
%I(A,B)+I(A,C)=I(A,B\cup C)\, .
%\end{equation}
%This requirement strongly constrains the form of the entanglement entropy and the mutual information.

In order to extract our conclusions, we make use of the Extensive Mutual Information model again ---see \req{emi-def}.   
In that model,
the mutual information between two entangling regions $A$ and $B$, can be written in terms of a simple local integral as
\begin{equation}\label{emi2}
%\see^{\rm \ssc EMI} = \kappa \int_{\Sigma}d {\bf{r}}' \int_{\Sigma} d {\bf{r}}\,\,  \frac{  {\bf{n}}  ({\bf{r}}^{\prime})\cdot  {\bf{n}} ({\bf{r}})}{|{\bf{r}}-{\bf{r}}^{\prime}|^{2(d-2)}}\, , \quad 
I^{\rm \ssc EMI}(A,B)=-2\kappa  \int_{\partial A} d{\bf r} \int_{\partial B} d{\bf r}' \, \frac{{\bf n} \cdot {\bf n}'}{|{\bf r} - {\bf r}' |^{2(d-2)}}\, .
\end{equation} 
%In these expressions, $\kappa$ is some constant and  ${\bf n}({\bf r}')$ is the  outward-pointing unit normal vector to the corresponding surface at the point ${\bf r}'$.
While a free fermion in two dimensions satisfies the extensivity property \req{emir}, no explicit CFT in $d>2$ is known (at least, for the moment) to do so. 
Nonetheless, the expressions for the entanglement entropy and the mutual information do respect the generic features corresponding to those quantities in general 
dimensions. 
% model captures many generic features involving the  shape dependence of entanglement and R\'enyi entropies. And it does so in a very computationally friendly fashion: computing mutual informations and entanglement entropies in this model is even simpler than for holographic CFTs.
While the exact details of the different terms will depend on the particular theory under consideration, we expect our conclusions regarding the
structure of divergences to hold for general CFTs.

\subsection{Finite mutual information for touching regions}\label{Finitemu}
The mutual information between two regions $A$ and $B$ is typically finite. However, as we move the corresponding entangling surfaces close to each other, it is expected that $I(A,B)$ becomes divergent. For instance, if $A$ is a disk of radius $R$ and $B$ is the exterior of a circular region of radius $R+\varepsilon$ concentric with $A$, the mutual information diverges as $I(A,B)\sim R/\varepsilon$. In fact, if we have some ``interior''  region $A$ and some ``exterior'' region $B$ such that their boundaries are two parallel curves separated by a curved strip of width $\varepsilon$, one can use the mutual information corresponding to this setup to define a regulator for entanglement entropy, with $\varepsilon$ playing the role of UV cutoff \cite{chm2}.  Here we show that if the contact region between $A$ and $B$ is sufficiently sharp, $I(A,B)$ is still finite in the limit in which both surfaces touch at a point.

%For this purpose, we will make use again of the EMI model. The expression for the mutual information is given for that model by
%\begin{equation}\label{miemi}
%I(A,B)=2\kappa \int_{\partial A} d{\bf r}' \int_{\partial B} d{\bf r} \, \frac{{\bf t}({\bf r}') \cdot {\bf t}({\bf r})}{|{\bf r}-{\bf r}'|^{2(d-2)}}\, ,
%\end{equation}

\begin{figure}[t]
	\centering 
	\includegraphics[scale=0.53]{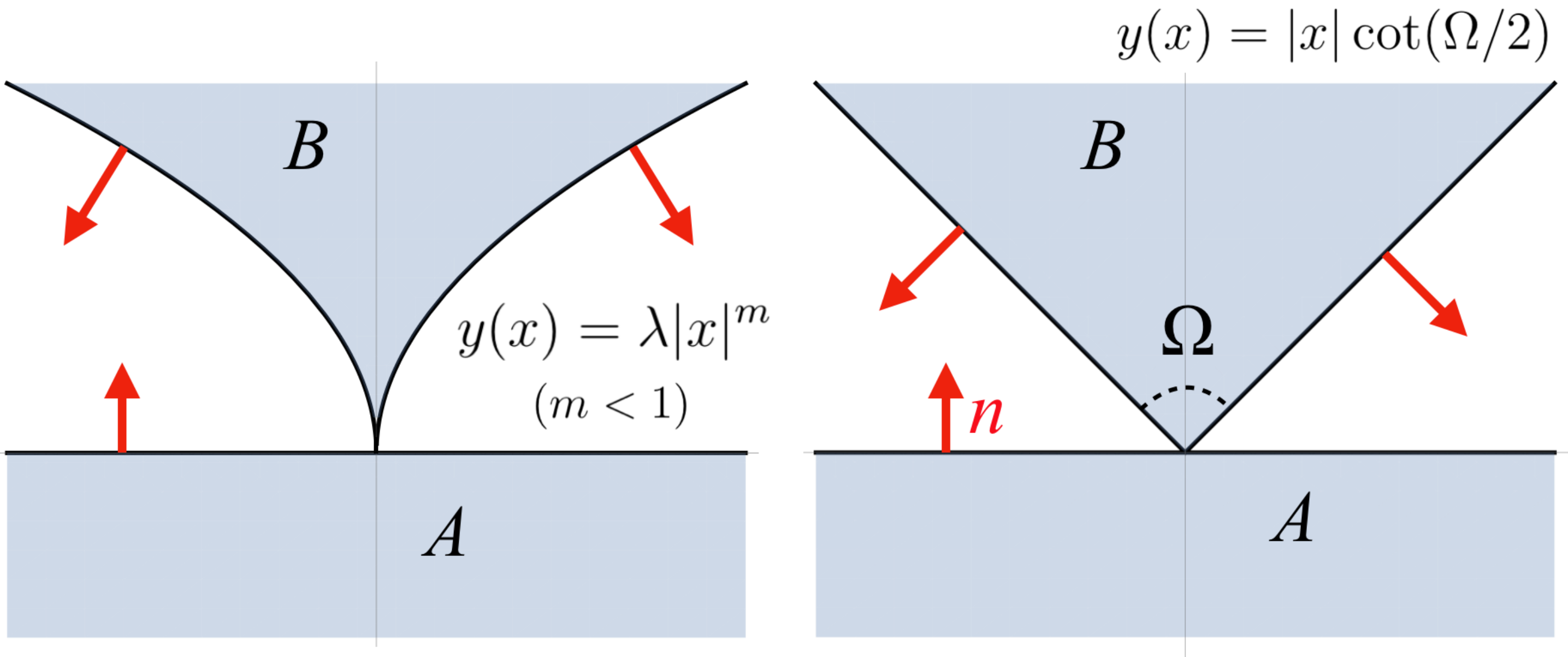}
	\caption{We show two entangling regions on a time slice of a three-dimensional theory corresponding, respectively, to a half plane and a region characterized by a corner defined by the regions $y(x)\geq \lambda |x|^m$, $\lambda\geq 0$ (left) and $y(x)\geq |x| \cot (\Omega/2)$ (right). The red arrows correspond to normal vectors to the entangling surfaces. For the straight corner, the mutual information $I(A,B)$ diverges logarithmically, \req{str}. Whenever $m<1$, however, 
the mutual information remains finite even when $A$ and $B$ touch each other at a point.}  
\label{mutual}
\end{figure}

For concreteness we set $d=3$. We choose $A$ to be the lower half plane ---see Fig. \ref{mutual}. Hence, for the EMI model defined in \req{emi2} we have ${\bf n}({\bf r}')=(0,1)$ and we can write
\begin{equation}\label{iabc}
I(A,B)=2\kappa \int_{\partial B} d{\bf r}\,  n_{y}  \int_{-\infty}^{+\infty} \frac{ dx'  }{ (x'-x)^2+y^2 }= -2\pi \kappa \int_{\partial B}  \frac{  n_{y} }{y} d{\bf r} \, ,
\end{equation}
where we used the notation ${\bf n}({\bf r})\equiv (n_x,n_y)$ and ${\bf r}\equiv (x,y)$. 
Let us now choose region $B$ to be defined by $y(x)\geq \lambda |x|^m$, where $m>0$ and $\lambda$ has units of $(\mbox{length})^{1-m}$. In the limiting case $m=1$, $\lambda\equiv\cot (\Omega/2)$, this corresponds to a region with a straight corner of opening angle $\Omega$. For $m<1$, the corner becomes sharper, and the opposite for $m>1$.
For general $m$ we have $d{\bf r}\, n_y=-dx $. Then, in the case of a straight corner, the result for the mutual information reads
\begin{equation}\label{str}
I(A,B)=4\pi \kappa \tan (\Omega/2) \int_{(\delta/\cot (\Omega/2))}^{L} \frac{dx}{x}=4\pi \kappa \tan (\Omega/2) \log \left(\frac{L}{\delta} \right)+ \mathcal{O}(\delta^0)\, ,
\end{equation}
which is obviously divergent as $\delta \rightarrow 0$.\footnote{The mutual information between two corners in $(2+1)$ dimensions was studied using the AdS/CFT 
correspondence \cite{Mozaffar:2015xue}. However, the calculation was performed for 2 corners of equal opening angles, which is different from our setup.} 
Similarly, for $m\neq 1$ one finds  
\begin{equation}\label{crus}
I(A,B)=\frac{4\pi \kappa}{\lambda} \int_{y^{-1}(\delta)}^{L} \frac{dx}{x^m}= \frac{4\pi \kappa}{(1-m)} \left[ \frac{L^{1-m} }{\lambda }-  \frac{1}{\lambda^{\frac{1}{m}} \delta^{1-\frac{1}{m}}} \right] \, ,
\end{equation}
which is always non-negative.
Note that in this expression we have introduced an IR cutoff $L$ and an UV cutoff as $\partial B$ approaches $\partial A$ a distance $\delta$ in the $y$ direction, \ie at $x=(\delta/\lambda)^{1/m}$. Alternatively, we can make this cutoff go to zero and instead shift the $B$ region vertically a distance $\delta$, so that $\partial B$ is defined as $y(x)=\lambda |x|^m + \delta$. In both cases, the net result is the appearance of a piece $I(A,B) \sim \delta^{1/m-1}$. This term is divergent whenever $m\geq 1$, but vanishes for $m<1$. In that case, the mutual information is UV finite. While the exact coefficients will vary if we consider a different CFT, we expect this phenomenon to hold for general theories.

Note also that the divergences observed approach an area-law as $m\rightarrow \infty$. As $m$ grows, region $B$ tends to be more open, and more points in $\partial B$ become closer to points in $\partial A$ in an increasingly bigger neighborhood of the touching point. In the limit, the situation is similar to the case described above in which $\partial A$ and $\partial B$ are concentric circles, which in the $\varepsilon\rightarrow 0$ limit have an area-law divergent mutual information.
 
\subsection{Entanglement entropy of curved corners}
Typically, the presence of geometric singularities on the entangling surface modifies the structure of divergences ---and universal terms--- of entanglement/R\'enyi entropies. A prototypical example of this phenomenon occurs in $(2+1)$ dimensions. When the entangling surface is smooth, the R\'enyi entropy contains a single (constant) term in addition to the area-law one. However, if a corner of opening angle $\Omega$ is present on the surface, a new logarithmically divergent 
term with a universal prefactor appears ---see \req{corner1} above. As we mentioned above, conical regions in $(3+1)$ dimensions ---and in contradistinction to smooth ones, for which the subleading (universal) term is logarithmic--- produce  universal $\log^2(H/\delta)$ terms, wedges give rise to $H/\delta$  divergences, and so on. However, not all entangling surfaces containing geometric singularities modify the structure of divergences. A first case in which this does not happen was analyzed in Section \ref{vertex} for polyhedral vertices. There, the agreement with the order of divergence of the universal term for a smooth region was however accidental, in the sense that the corresponding universal coefficients had very different origins. 

In this section we present genuine examples of singular entangling regions which do not modify the structure of divergences/universal terms with respect to the smooth case. We also show that in other cases, namely when the entangling region contains sufficiently sharp curved corners, new divergences appear, which approach (but never get more divergent than) an area-law in the limiting case.  Again for simplicity, we restrict ourselves to curved corner regions in $(2+1)$-dimensional CFTs.\footnote{Similar geometric configurations were considered in the context of holographic Wilson loops in \cite{Dorn:2015bfa,Dorn:2018als,Dorn:2018srz,Dorn:2019yms}.} Again, we will use the Extensive Mutual Information model to derive our conclusions.

\begin{figure}[t]
	\centering 
	\includegraphics[scale=0.58]{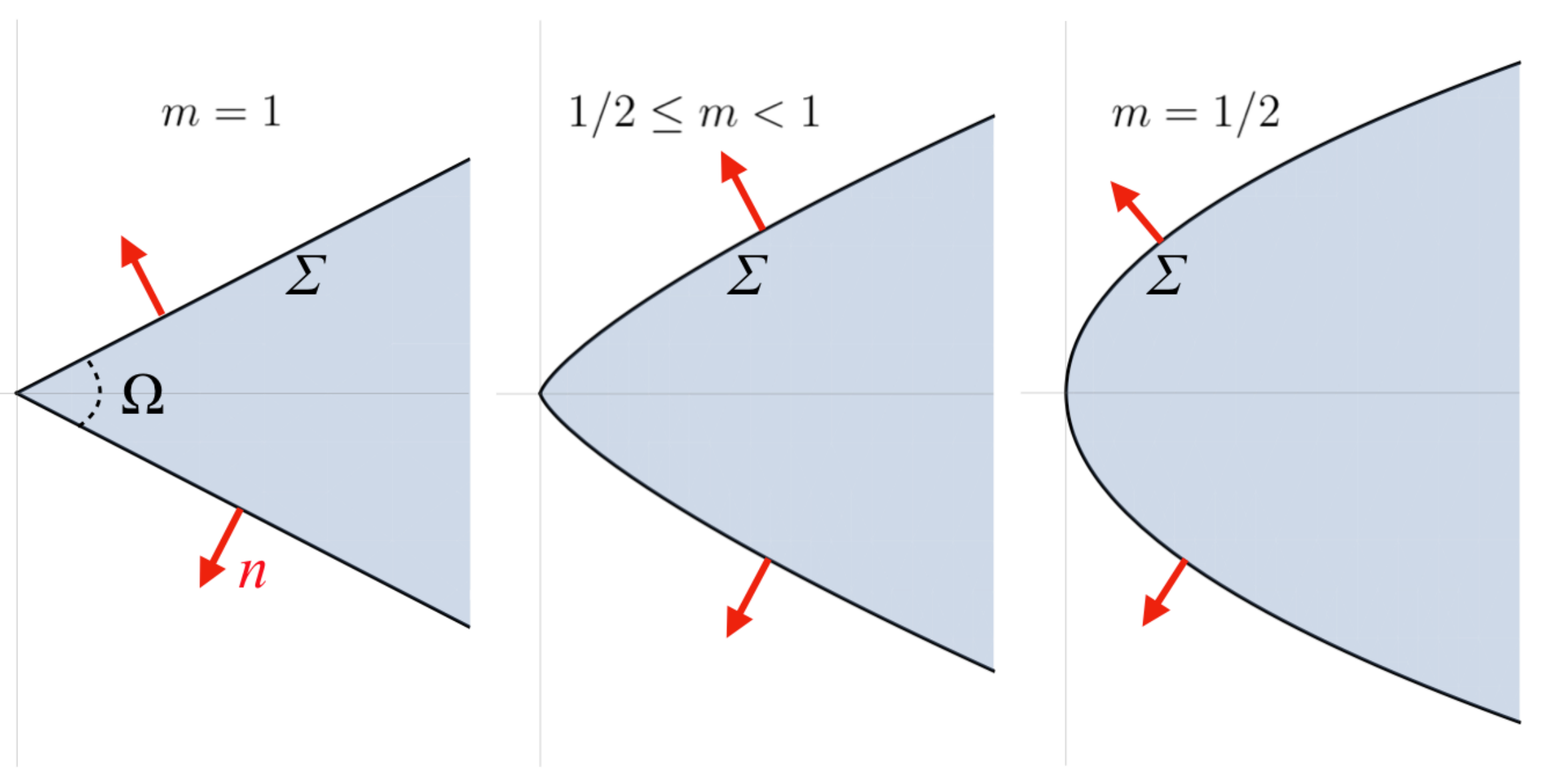}
	\caption{We plot entangling regions on a time slice of a three-dimensional CFT bounded by the curves $y(x)=\pm \lambda x^m $ ($x\geq 0$) with $m=1$ (left), $1/2\leq m<1$ (middle) and $m=1/2$. The usual logarithmic universal term in the entanglement entropy $\sim a(\Omega) \log(H/\delta)$ characteristic of the straight corner $(m=1)$ is no longer present for $m<1$, even though the entangling surface is still singular at the tip for $1/2\leq m<1$. }
\label{popcorn}
\end{figure}

\begin{figure}[t]
	\centering 
	\includegraphics[scale=0.39]{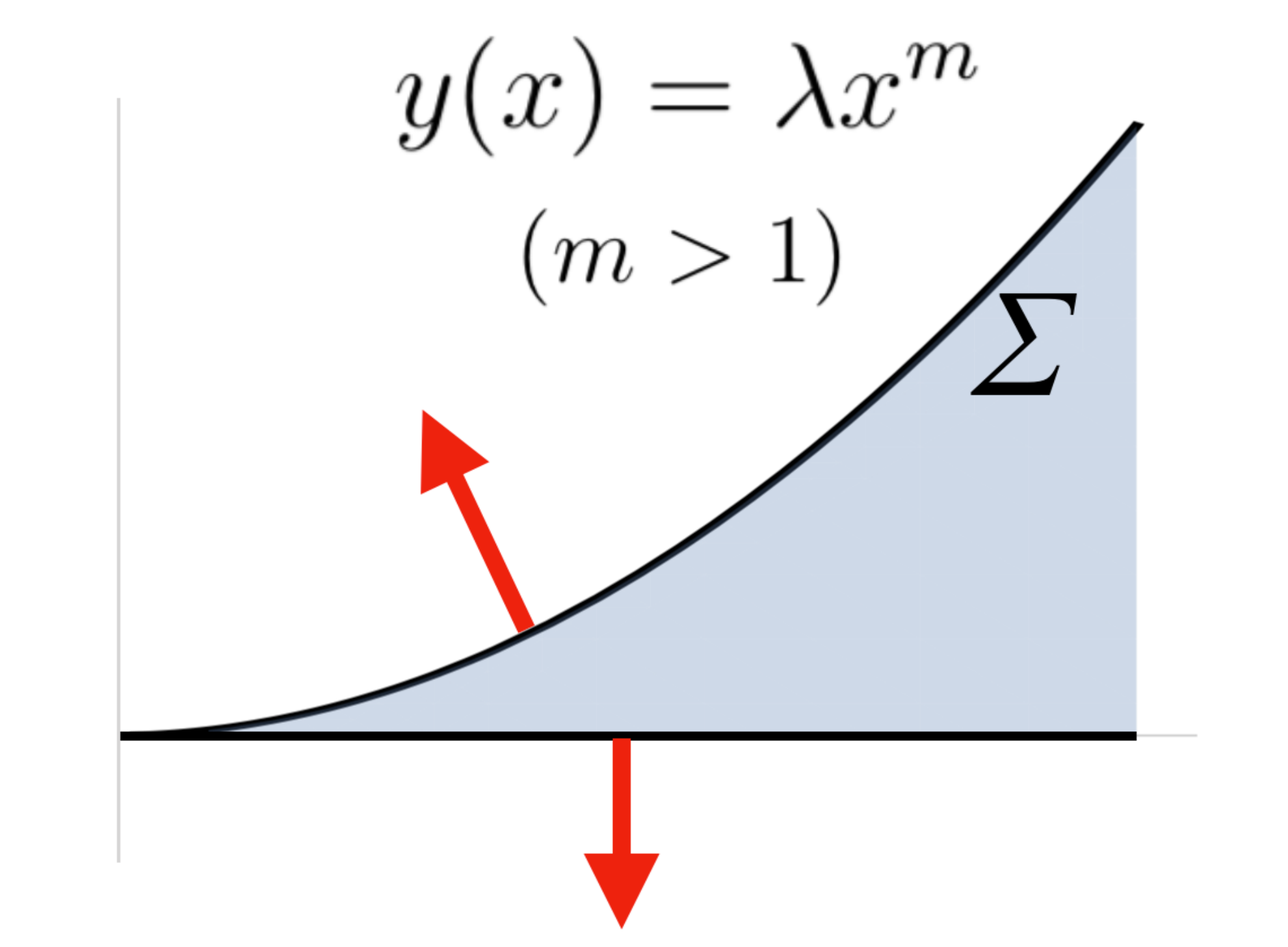}
	\caption{We plot an entangling region on a time slice of a three-dimensional CFT bounded by the curves $y=0$ and $y(x)= \lambda x^m $ ($x\geq 0$) with $m>1$. No logarithmic term is present in the entanglement entropy. Instead, a new non-universal divergence $\sim 1/\delta^{(1-\frac{1}{m})}$ appears. }
\label{popcorn2}
\end{figure}

We will consider entangling regions bounded by the curves
\begin{align}
y(x)=\lambda x^m\, , \quad (x>0)\, ,\\
y'(x')=-\gamma x'^n\, , \quad (x'>0)\, ,
\end{align}
where the exponents $m$ and $n$ are non-negative, and $\lambda,\gamma>0$ are positive constants. Examples are shown in Fig.~\ref{popcorn} for $n=m$ and $\lambda=\gamma$ and in Fig.~\ref{popcorn2} for $\gamma=0$.
 On general grounds, there will be two kinds of contributions to $\see^{\rm \ssc EMI}$ as defined in \req{emi-def}. The first corresponds to the contribution coming from ${\bf{r}}$ and ${\bf{r}}^{\prime}$ lying on the same curve. We assume this to be defined by the equation
$
 y(x)=\lambda x^m\, ,$ $ (x\geq 0)\, .
$ The induced metric on this curve is given by $ds^2_h=\left[1+m^2\lambda^2x^{2(m-1)}\right]dx^2$. We also have $ {\bf{n}} ( {\bf{r}} )=(-m\lambda x^{(m-1)},1)/\sqrt{ 1+m^2\lambda^2 x^{2(m-1)}}$, so
\begin{align}
%&\vec{n}(A)=\frac{(-m\lambda A_x^{m-1},1)}{\sqrt{1+m^2\lambda^2 A_x^{2(m-1)}}}\, , \quad \vec{n}(B)=\frac{(-m\lambda B_x^{m-1},1)}{\sqrt{1+m^2\lambda^2 B_x^{2(m-1)}}} \\ &\rightarrow
 {\bf{n}}  ({\bf{r}})\cdot  {\bf{n}} ({\bf{r}}^{\prime})=\frac{1+m^2\lambda^2 x^{(m-1)}{x^{\prime}}^{(m-1)}}{\sqrt{1+m^2\lambda^2 x^{2(m-1)}}\sqrt{1+m^2\lambda^2 {x^{\prime}}^{2(m-1)}}}\, ,
\end{align}
and
\begin{equation}
| {\bf{r}}- {\bf{r}}^{\prime}|^2=(x-x^{\prime})^2+\lambda^2(x^m-{x^{\prime}}^m)^2\, .
\end{equation}
Using this, we can define one possible contribution to the entanglement entropy $s_1(m,\lambda)$ as
\begin{align}
s_1(m,\lambda)&=\kappa \int_{\delta}^{H}dx^{\prime} \left[\int_{0}^{x^{\prime}-\delta}dx+\int^{\infty}_{x^{\prime}+\delta}dx \right]\frac{[1+m^2\lambda^2 x^{(m-1)}{x^{\prime}}^{(m-1)}]}{(x-{x^{\prime}})^2+\lambda^2(x^m-{x^{\prime}}^m)^2}\, ,
\end{align}
where we introduced UV and IR cutoffs $\delta$ and $H$.
Observe that in the case of a corner formed by straight lines, $s_1(1,\lambda)=s_1(1,0)$. The second possible contribution will arise from ${\bf{r}}$ and ${\bf{r}}^{\prime}$ lying on different curves. 
% Without loss of generality, we can assume the boundaries of the entangling regions to be defined as
% \begin{align}
% y(x)=\lambda x^m\, , \quad (x>0)\, ,\\
% y'(x')=-\gamma x'^n\, , \quad (x'>0)\, ,
% \end{align}
% respectively. 
The contribution from this situation, which must be counted twice (we can flip the labels ${\bf{r}}$ and ${\bf{r}}^{\prime}$) reads then
\begin{align}
s_2(m,\lambda;n,\gamma)&=2\kappa \int_{\delta}^{H}d{x^{\prime}} \left[\int_{0}^{{x^{\prime}}-\delta}dx+\int^{\infty}_{{x^{\prime}}+\delta}dx \right]\frac{[-1+m n \lambda \gamma x^{(m-1)} {x^{\prime}}^{(n-1)}]}{(x-{x^{\prime}})^2+(\lambda x^m+\gamma {x^{\prime}}^n)^2}\, . 
\end{align}
Observe that since the normal vectors must be chosen to point outwards ---this is just a convention, but once chosen, it must be respected--- from the entangling region, $s_2(m,\lambda;m,-\lambda)$ differs by an overall sign from $s_1(m,\lambda)$, for which the normal vectors both point in the same direction. 

Note that both $s_1$ and $s_2$ take the form
\begin{equation}
s= \int_{\delta}^H dx'  \int_0^{\infty}dx   f(x,x') -\int_{\delta}^H dx' \int_{x'-\delta}^{x'+\delta} dx   f(x,x')\, ,
\end{equation}
for some function $f(x,x')$ in each case. With the present regularization, the dependence on the UV cutoff $\delta$ appears only through the integration limits of the integrals. Hence, it follows that
\begin{equation}\label{gsg}
\frac{\partial s}{\partial \delta}=- \int_{2\delta}^{\infty} f(x,\delta) dx - \int_{\delta}^H \left[f(x-\delta,x)+f(x+\delta,x) \right] dx\, ,
\end{equation} 
which is often a much easier expression to use in practice\footnote{A similar approach is proposed in \cite{Myersnew}.} when  extracting the structure of divergences of $s_1(m,\lambda)$ and $s_2(m,\lambda;n,\gamma)$. Of course, $\partial s/\partial \delta$ is blind to $\mathcal{O}(\delta^0)$ contributions, but not to the rest of the terms.

The entanglement entropy for various corner regions will involve  linear combinations of $s_1$ and $s_2$ for different values of $m$, $n$, $\lambda$ and $\gamma$. For example, in the case of a straight corner of opening angle $\Omega$ one finds $\see=2s_1(1,0)+s_2(1,0;1,\tan\Omega)$. These integrals yield
\begin{align}\label{cfg}
2s_1(1,0)&= \frac{4\kappa H}{\delta}-2\kappa \log \left(\frac{H}{\delta}\right)+\mathcal{O}(\delta^0)\, ,\\
s_2(1,0;1,\tan\Omega)&= -2\kappa (\pi-\Omega)\cot\Omega \log\left(\frac{H}{\delta}\right)+\mathcal{O}(\delta^0) \, .
\end{align}
Then, one is left with
\begin{align}
	%\see&=\frac{4 \kappa H}{\delta}-2\kappa\left[1+\frac{\pi+2\arctan(1/\lambda)}{2\lambda}\right] \log \left(\frac{H}{\delta}\right)+\mathcal{O}(\delta^0) \\
	\see=\frac{4\kappa H}{\delta} -a(\Omega )\log \left(\frac{H}{\delta}\right)+\mathcal{O}(\delta^0)\, , \quad \text{where} \quad a(\Omega )=2\kappa\left[1+(\pi-\Omega)\cot\Omega \right]\, .
	\end{align} 
%where we wrote $\lambda\equiv $, where $\Omega$ is the corner opening angle. 
This expression for the corner function $a(\Omega )$ in the EMI model was previously obtained in \cite{Casini:2008wt,Swingle2010,Bueno1,Bueno3}.   

Let us now consider the case of an entangling region defined by $-\lambda x^m\leq y(x)\leq \lambda x^m$, with $m>0$, as shown in Fig. \ref{popcorn}. In that situation, the entangling region contains a geometric singularity at the origin for $m>1/2$. 
For $0<m\leq 1/2$, however, the surface is non-singular. 
%in the sense that the vector normal to the curve is well-defined at the origin, which is not the case for $m>1/2$. 
%\comment{curvatures, nth derivatives, etc}. 
% \rd{[[W: For $m<1/2$, the curve $\Sigma$ when viewed as a function $x=f(y)=|y|^{1/m}$ will in general NOT be infinitely differentiable with respect to $y$ at $y=0$.
% Take for example $m=2/5=0.4$. The $n$th derivative of $f(y)$ diverges at $y=0$ when $n>2$. However, the first and second derivatives of $f(y)$ will be finite everywhere.
% So the curve is not $C^\infty$. The curve is at least $C^2$.]]  
% } 
The result for the entanglement entropy is given in this case by $\see=2s_1(m,\lambda)+s_2(m,\lambda;m,\lambda)$. Using \req{gsg}, one can show that these two contributions behave as
\begin{align}\label{sisi8}
2s_1(m,\lambda)&= \frac{4\kappa H}{\delta}-2\kappa m \log \left(\frac{H}{\delta}\right)+\mathcal{O}(\delta^0)\, ,\\
s_2(m,\lambda;m,\lambda)&=2\kappa m \log\left(\frac{H}{\delta}\right)+\mathcal{O}(\delta^0)\, ,
\end{align}
for $1/2< m<1$. Hence, 
\begin{equation}
\see=\frac{4\kappa H}{\delta}+\mathcal{O}(\delta^0)\, , \quad 1/2< m<1\,.
\end{equation}
in that case. This means that, even though a geometric singularity is present in the entangling region for $1/2 < m<1$, no UV divergence appears in the entanglement entropy besides the usual area-law, and the universal contribution is a constant term, just like in the case of smooth regions. Note that no new UV divergence is expected for $m=1/2$ since the entangling curve $\Sigma$ is non-singular, being a parabola.

The situation changes for $m\geq 1$, corresponding to sharper corners. In order to study those, let us modify the setup slightly and consider the case of corners formed by the intersection of the curves $y(x)=0$ (\ie the $x$ axis) and $y(x)=\lambda x^m$, ($x>0$) ---see Fig.~\ref{popcorn2}. This simplifies computations. 

The result for the entanglement entropy is now given by $\see=s_1(m,\lambda)+s_1(1,0)+s_2(1,0;m,\lambda)$. Again using \req{gsg}, one can show that 
\begin{align}
s_1(m>1,\lambda)&= \frac{2\kappa H}{\delta}-\kappa  \log \left(\frac{H}{\delta}\right)+\mathcal{O}(\delta^0)\, ,\\
s_2(1,0;m>1,\lambda)&= -\frac{2\kappa \pi c_m}{\lambda^{\frac{1}{m}}\delta^{1-\frac{1}{m}}}+2\kappa \log\left(\frac{H}{\delta}\right)+\mathcal{O}(\delta^0)\, ,
\end{align}
 where  the $c_m$ are positive non-universal dimensionless constants. Observe that the coefficient of the logarithmic term in $s_1(m>1,\lambda)$ differs from the one found for $1/2\leq m<1$ in \req{sisi8}. Combining these expressions with \req{cfg} we are left with
\begin{equation}\label{fsf}
\see = \frac{4\kappa H}{\delta}- \frac{2\kappa \pi c_m}{\lambda^{\frac{1}{m}}\delta^{1-\frac{1}{m}}} + \mathcal{O}(\delta^0)\, .
\end{equation}
Hence, the logarithmic divergence conspires to disappear, and we are left instead with a new non-universal divergence $\sim 1/(\delta^{1-\frac{1}{m}})$ which has precisely the same form as the one found in our mutual information computations of the previous subsection. 

As we have mentioned, we expect the results obtained in this section to be (qualitatively) valid for general CFTs. In this sense, note that whenever the entangling region is very sharp ---in the sense that the entangling surfaces $\Sigma_1$, $\Sigma_2$ with $\Sigma=\Sigma_1 \cup \Sigma_2$ are very close to each other along some direction (say, along $y$)--- one could imagine cutting the entangling region in small thin rectangles which would contribute to the total mutual information in an additive way, effectively making the mutual information become extensive ---see also comments at the end of Section \ref{Finitemu}. Heuristically, this would suggest a contribution of the form
\begin{equation}\label{heur}
%\see \sim \int_{\delta}^L \frac{dx}{[y(x)+\delta]} \Leftrightarrow 
\see \sim \int_{y^{-1}(\delta)}^L \frac{dx}{y(x)}\, ,
\end{equation}
which would be the sum of the contributions coming from the small rectangles of width $dx$ and height $y(x)$, and
where the short-distance UV cutoff needs to be imposed along the direction in which the two entangling surfaces are very close to each other ($y$ in this case). This precisely yields the kind of contributions obtained from the EMI model, \eg for $y(x)=\lambda x^m$,
\begin{equation}\label{plw}
 \see \sim \int_{(\delta/\lambda)^{\frac{1}{m}}}^L \frac{dx}{\lambda x^m} \sim \frac{1}{\lambda^{\frac{1}{m}} \delta^{1-\frac{1}{m}}}\, ,
\end{equation}
which is the same kind of divergence we obtained in \req{fsf}. If we consider an even sharper corner like $y(x)=e^{-\frac{1}{x}}$, \req{heur} produces a divergence of the form
\begin{equation}
 \see \sim \frac{1}{\delta \log^2\delta}\, ,
\end{equation}
which is more divergent than the one produced by any of the power-law corners in \req{plw}. The sharper the corner, the closer this contribution gets  to the area-law divergence, without ever reaching it.

%\begin{equation}
%S_n^{\rm smooth}=c_1 \frac{H}{L}-F\, , \quad \text{versus} \quad S_n^{\rm corner}=c_1 \frac{H}{L}-a(\Omega)\log\left(\frac{H}{\delta}\right)+\mathcal{O}(\delta^0)
%\end{equation}

%and $d=3+1$ dimensions correspond to 

%\subsection{Free fields}
%\comment{In this section we explain how the universal $\cos^2\Omega/\sin\Omega$ function arises for free field theories from a line integral of some extrinsic curvatures of the circle resulting from the intersection of the cone with a unit $\mathbb{S}^2$ encircling it}\\
%\comment{Possibly the same happens for elliptic cones, and the function coming from Solodukhin formula can be understood in a similar fashion from a line integral along the curve resulting from the intersection of the cone with a $\mathbb{S}^2$, check}

%%%%%%%%%%%%%%%%%%%%%%%%%%%%%%%%%%%%%%%%%%%%%%%%%%%%%%%%%%%%
%%%%%%%%%%%%%%%%%%%%%%%%%%%%%%%%%%%%%%%%%%%%%%%%%%%%%%%%%%%%
%%%%%%%%%%%%%%%%%%%%%%%%%%%%%%%%%%%%%%%%%%%%%%%%%%%%%%%%%%%%

%\comment{some definitions, to be moved somewhere else
%\begin{equation}
%K_{ab}=t_a^{\mu}t_b^{\nu}K_{\mu\nu}\, , \quad \text{where} \quad K_{\mu\nu}=\nabla_{\mu}n_{\nu}-n_{\mu}n^{\gamma}\nabla_{\gamma}n_{\nu}\, , \quad t_a^{\mu}=\frac{\partial x^{\mu}}{\partial y^a}\,.
%\end{equation}
%which simplifies to
%\begin{equation}
%K_{ab}=t_a^{\mu}t_b^{\nu}\nabla_{\mu}n_{\nu}\, .
%\end{equation}}

\section{Conclusion} \labell{discuss}
In this paper we have presented several new results involving the structure and nature of universal and divergent terms in the von Neumann and R\'enyi entanglement entropies arising from the presence of geometric singularities on the entangling surface. An in-depth summary of our findings can be found in Section \ref{summ}. Let us close the paper with some final words regarding a few possible directions. 

One of our main motivations was trying to gain a better understanding on the nature of the trihedral universal coefficient $v_{n}(\theta_1,\theta_2,\theta_3)$, previously studied using lattice techniques in \cite{Kovacs,Devakul2014,Sierens:2017eun,Bednik:2018iby}, and analytically in the nearly smooth limit \cite{Witczak-Krempa:2018mqx}. Our results indicate that an analytic computation of this coefficient at general angles for free fields would be equivalent to evaluating partition functions on a $\mathbb{S}^3$ with multiplicative boundary conditions on a two-dimensional spherical triangle. The methods required to perform this computation have yet to be developed. An analogous computation for the corner region in one dimension less ---$\mathbb{S}^2$ partition function with boundary conditions on an arc--- gives the result in terms of non-linear ordinary differential equations as a function of the angle. Analogously to this case, it is expected that the trihedral coefficient would be given by the solution to some system of non-linear partial differential equations on the two angular variables describing the trihedral angle. Just like in the case of the $2+1$ corner, we do not see that a particular simplification should occur for any particular value of the angles (except in the almost-smooth limit considered in \cite{Witczak-Krempa:2018mqx}), so the full calculation would need to be addressed, even if one wanted to focus only on the straight-angles case corresponding to $v_{n}(\pi/2,\pi/2,\pi/2)$.

On a different front, let us mention that a behavior similar to the one presented in Section \ref{singi} regarding the interplay between singularities in the entangling region and the structure of divergences of the R\'enyi and entanglement entropies is expected to occur for higher-dimensional CFTs ---\eg for smoothed and sharpened (hyper)conical regions. This could be again tested using the EMI model or, in the four-dimensional case, Solodukhin's formula \req{gfd33}.

Finally, regarding the question addressed in Appendix \ref{maxim}, it should not be difficult to find out whether circular cones globally maximize the entanglement entropy within the family of elliptic cones or, more generally, cones generated by moving a straight line emanating from a vertex in a closed orbit.

\section*{Acknowledgments}
We thank Marco Baggio, Cl\'ement Berthiere, Joan Camps, Rob Myers and Carlos S. Shahbazi for useful discussions.  The work of PB and HC was supported by the Simons foundation through the It From Qubit Simons collaboration. HC also acknowledges support from CONICET, CNEA and Universidad
Nacional de Cuyo, Argentina. WWK was funded by a Discovery Grant from NSERC, a Canada Research Chair, and a ``\'Etablissement de nouveaux chercheurs et de nouvelles chercheuses universitaires'' grant from FRQNT.

\appendix

\section{Cone entanglement in the Extensive mutual information model}\label{coneEMI2}
%While no explicit CFT in $d>2$ is known (at least, for the moment) to satisfy the mutual information extensivity property which defines it, the Extensive Mutual Information (EMI) captures many generic features involving the  shape dependence of entanglement and R\'enyi entropies. And it does so in a very computationally friendly fashion: computing mutual informations and entanglement entropies in this model is even simpler than for holographic CFTs.

%\begin{equation}
%\see=\kappa \int_{\Sigma}d {\bf{r}} \int_{\Sigma} d {\bf{r}}^{\prime}\,\,  \frac{  {\bf{n}}  ({\bf{r}})\cdot  {\bf{n}} ({\bf{r}}^{\prime})}{|{\bf{r}}-{\bf{r}}^{\prime}|^{2(d-1)}}\, ,
%\end{equation}

%\comment{bla bla}
%Many generic features regarding the shape dependence of entanglement and R\'enyi entropies can be understood by considering the Extensive Mutual Information model. 
%\subsection{Cones}

In this appendix we compute the entanglement entropy for a conical entangling surface in $d=4$ in the EMI model.  As we have mentioned, we expect the corresponding universal $\log^2\delta$ term to be controlled by a theory-independent function of the opening angle $4a_n^{(4)}(\Omega)/f_b(n)=\cos^2\Omega/\sin\Omega$. The EMI should not be an exception, and here we explicitly verify that this is indeed the case. 

%For a four-dimensional CFT, the entanglement entropy in the EMI model is given by
%\begin{equation}
%\see = \kappa \int_{\Sigma} dA \int_{\Sigma} dB\,  \frac{\vec{n}(A)\cdot \vec{n}(B)}{|\vec{A}-\vec{B}|^4}\, ,
%\end{equation}
%where \comment{bla bla}. 
As explained in Section~\ref{sec:cube-EMI}, the general expression for the entanglement entropy in the EMI model is given by \req{emi-def}. Let us parametrize the cone surface in cylindrical coordinates by $z=\rho/\tan \Omega$, $t=0$. Naturally, the line element of flat space in these coordinates reads: $ds^2=-dt^2+dz^2+d\rho^2+\rho^2d\phi^2$. The induced metric on the cone surface is given by $ds^2_h=d\rho^2/\sin^2\Omega+\rho^2d\phi$, so $d^2 {\bf{r}} =[ \rho/\sin \Omega] d\rho d\phi$ and analogously for $ {\bf{r}}^{\prime}$. Given the symmetry of the problem, we can just set $\phi^{\prime}=0$ everywhere and multiply the remainder integrals by an overall $2\pi$.
 The unit normal vector to the cone surface is given $\vec{n}=\vec{u}_{\rho}\cos \Omega- \vec{u}_z \sin\Omega$, where $\vec{u}_{\rho}=\cos \phi \vec{u}_x+\sin \phi \vec{u}_y$. Using this, it is straightforward to find
\begin{equation}
 {\bf{n}}  ({\bf{r}})\cdot  {\bf{n}} ({\bf{r}}^{\prime})=\cos^2\Omega \cos \phi + \sin^2\Omega\, .
\end{equation}
Similarly, we find $|{\bf{r}}-{\bf{r}}^{\prime}|^4=\left[\rho^2+{\rho^{\prime}}^2-2\rho{\rho^{\prime}} \cos\phi+(\rho-{\rho^{\prime}})^2/\tan^2\Omega\right]^2$. Then, after some trivial manipulations, using \req{emi2} we are left with the integrals
\begin{equation}
\see =\frac{2\pi \kappa}{ \sin^2\Omega}   \int  \rho  d\rho \int  {\rho^{\prime}} d{\rho^{\prime}} \int_0^{2\pi} d\phi \frac{[\cos^2\Omega \cos \phi + \sin^2\Omega]}{\left[a-b \cos \phi\right]^2}\, ,
\end{equation}
where $a\equiv \rho^2+{\rho^{\prime}}^2+(\rho-{\rho^{\prime}})^2/\tan^2\Omega$, $b\equiv 2\rho{\rho^{\prime}}$, and 
where we regulate the radial integrals as follows
\begin{equation}
\int d\rho\equiv \int_{\delta}^H d\rho \, , \quad \int d{\rho^{\prime}}\equiv \left[ \int_0^{\rho-\delta} d{\rho^{\prime}}+\int_{\rho+\delta}^{\infty} d{\rho^{\prime}}\right]\, .
\end{equation}
The angular integrals can be performed using 
\begin{equation}
\int_0^{2\pi} \frac{d\phi }{(a-b\cos\phi)^2}=\frac{2\pi a}{(a^2-b^2)^{3/2}}\, , \quad \int_0^{2\pi} \frac{d\phi  \cos\phi}{(a-b\cos\phi)^2}=\frac{2\pi b}{(a^2-b^2)^{3/2}}\, .
\end{equation}
We are left with
\begin{equation}
\see=\frac{4\pi^2 \kappa}{\sin^2\Omega} \left[ \cos^2\Omega\,  s_1 +\sin^2 \Omega\,  s_2\right]\, ,
\end{equation}
where
\begin{equation}
s_1=\int    \int  \frac{b\,  \rho {\rho^{\prime}}  }{(a^2-b^2)^{3/2}} d\rho d{\rho^{\prime}}\, ,\quad s_2=\int   \int \frac{ a \, \rho {\rho^{\prime}}}{(a^2-b^2)^{3/2}}d\rho d{\rho^{\prime}}\, ,
\end{equation}
with the help of Mathematica we can perform the radial integrals, and the results read
\begin{equation}
s_1=\frac{1}{64} \sin\Omega \left[\cos(2\Omega)-7 \right] \log^2 \left(\frac{H}{\delta} \right)\, ,
 \quad s_2=\frac{1}{32} \cos^2\Omega \sin\Omega \log^2 \left(\frac{H}{\delta} \right)
\end{equation}
up to nonuniversal contributions. Putting both pieces together, we are left with
\begin{equation}
\see=-\frac{3\pi^2 \kappa }{8}\cdot \frac{\cos^2\Omega}{\sin \Omega} \log^2 \left(\frac{H}{\delta} \right)
\, ,
\end{equation}
which takes the expected form. We can now write the coefficient $\kappa$ in terms of the trace-anomaly charge $c_{\rm \ssc EMI}$, \eg using the result for the  entanglement entropy of a cylinder in the EMI model and comparing it with the general one following from Solodukhin's formula \req{game}. The result reads
\begin{equation}
c_{\rm \ssc EMI}=\frac{3\kappa}{2\pi^2}\, .
\end{equation}
Using this, we find
\begin{equation}
\see=-\frac{c }{4}\cdot \frac{\cos^2\Omega}{\sin \Omega} \log^2 \left(\frac{H}{\delta} \right)
\, ,
\end{equation}
which is the exact result valid for general CFTs ---\ie it already contains the ``famous'' $1/2$ factor that is missing in the  calculation using Solodukhin's formula.

%\subsection{Curved corners}
%\comment{Stuff in this section needs to be completely reorganized}

\section{Which cone maximizes the R\'enyi entropy?}\label{maxim}
In Section \ref{vertex}, we found how the universal cone formula $a_n^{(4)}(\Omega)=f_b(n)\frac{\cos^2\Omega}{4\sin\Omega}$ is modified when the cross-sections of the cones become ellipses rather than circles. We showed that  the full result for the R\'enyi entropy reads in that case
\begin{equation}
S_n=b_2 \frac{H^2}{\delta^2} - \frac{f_b(n)}{4} \gamma(e',\theta_0) \log^2\left(\frac{R}{\delta} \right)
\end{equation}
where $ \gamma(e',\theta_0)$ was given in  \req{gan} as a function of the ellipses second-eccentricity $e'$ and the semi-opening angle $\theta_0$.

A natural question one is led to ask is which cone extremizes the R\'enyi entropy within the family of elliptic cones. In the case of compact smooth surfaces, this question was addressed in \cite{Astaneh:2014uba}, where it was shown that the round sphere $\mathbb{S}^2$ is the one which maximizes it both within the family of genus-$0$ surfaces and for general-genus surfaces.\footnote{For fixed genus, the surfaces maximizing the R\'enyi entropy turn out to correspond to the so-called Lawson surfaces, namely, surfaces which can be minimally embedded in $\mathbb{S}^3$ ---see \cite{Astaneh:2014uba} for details. } A meaningful comparison of this kind can be performed by fixing the lateral area of the cones, so that the area-law terms cancel each other when the difference is considered. 
The lateral area of the elliptic cones is given by 
\begin{equation}\label{conqui}
A=2\int_{0}^{\pi/2}  \frac{R^2 \sin\theta_0\sqrt{1+{e^{\prime}}^2\cos^2\phi}}{\sqrt{1+{e^{\prime}}^{2}\sin^2\theta_0\cos^2\phi}} d\phi\, ,
\end{equation}
which depends on whether we cut it off at a fixed $R$ or at some height $z=z_0$. In the second case, we need to replace $R$ by  $R(z_0,\phi)=z_0\sqrt{1+e'^2\sin^2\theta_0}/ (\cos \theta_0 \sqrt{1-e'^2\sin^2\theta_0\cos^2\phi})$. Then, the result can be expressed in terms of the complete elliptic integral of the second kind $E[x]$ as\footnote{This can be written in terms of the ellipses semi-axes as 
\begin{equation}  A_{(z_0)}=2 a \sqrt{z_0^2+b^2}  E\left[\frac{1-b^2/a^2}{1+b^2/z_0^2}\right]\, , 
\end{equation}  
which is a relatively well-known result. Note that different sources define $E[z]$ in a slightly different way: $E[z]\equiv \int_0^{\pi/2}dt \sqrt{1-z \sin^2 t}$  vs  $E[z]\equiv \int_0^{\pi/2}dt \sqrt{1-z^2 \sin^2 t}$. Here we use the first definition, which is the one implemented in Mathematica.}
\begin{equation}
A_{(z_0)}= \frac{2 z_0^2 \sin\theta_0}{\cos^2\theta_0} \sqrt{1+{e^{\prime}}^2\sin^2\theta_0}\, E\left[\frac{-{e^{\prime}}^2\cos^2\theta_0}{1+{e^{\prime}}^2\sin^2\theta_0} \right]\, .
\end{equation}
Conditions $A_{(R)}=$ constant and $A_{(z_0)}=$ constant cannot be easily converted into explicit relations between $\theta_0$ and ${e^{\prime}}$. They can nonetheless be implemented for small ${e^{\prime}}$. In the first case, one finds
\begin{equation}
\gamma(e^{\prime})|_{A_{(R)}}=\frac{1-\bar A_{(R)}^2}{\bar A_{(R)}}\left[1+\frac{3\left(3+\bar A_{(R)}^2\right) }{32} {e^{\prime}}^{4}+\mathcal{O}({e^{\prime}}^6)\right]\, ,
\end{equation}
where we defined $\bar A_{(R)}\equiv A_{(R)}/(\pi R^2)$, which satisfies $\bar A_{(R)}\leq 1$ for $e^{\prime}<1$.  
Interestingly, the quadratic term in ${e^{\prime}}$ appearing in \req{ggas} disappears when we keep $A_{(R)}$ fixed. As a consequence, circular cones are minima of $\gamma$ when compared to other cones of the same lateral area and cutoff at some radial distance $R$ but different elliptic cross-sections. Since $\gamma$ contributes with a negative sign to the R\'enyi entropy, circular cones are maxima of the R\'enyi entropy within this family.

If we impose $A_{(z_0)}=$ constant instead, we obtain
\begin{equation}
\gamma(e^{\prime})|_{ A_{(z_0)}}=\frac{1}{\bar{A}_{(z_0)}} \left[1+\frac{1+ \left(6 {\bar{A}_{(z_0)}}^2- \sqrt{1+4{\bar{A}_{(z_0)}}^2} \right) \left(1+4{\bar{A}_{(z_0)}}^2 \right)}{64 {\bar{A}_{(z_0)}}^4  }{e^{\prime}}^{4}+\mathcal{O}({e^{\prime}}^6) \right]\, ,
\end{equation}
where we now defined $\bar A_{(z_0)}\equiv A_{(z_0)}/(\pi z_0^2)$. Once again, the quadratic term in \req{ggas} conspires to disappear, and the quartic coefficient is always positive. 
Hence, circular cones are also local maximizers of the R\'enyi entropy within the class of elliptic cones cutoff at a fixed height $z_0$. 
It would be interesting to verify whether this holds globally within this family.

More generally, it would be interesting to find out if circular cones are maximizers of the R\'enyi entropy for ``straight'' cones, i.e.\ those generated
by moving a straight line in a closed orbit.

\section{Hyperconical entanglement in even dimensions}\label{hypers}
%\comment{free fields??}
In this appendix we argue that the universal contribution to the R\'enyi entropy of right circular (hyper)cones in general even-dimensional CFTs is given by the simple formula \req{gamsed}.
%\beq 
%S_n^{ \rm univ}=(-1)^{\frac{d-2}{2}}a_n^{(d)}(\Omega) \log^2\left(R/\delta\right)\, , \quad \text{where} \quad a_n^{(d)}(\Omega) =\frac{\cos^2\Omega}{\sin\Omega} \sum _{j=0}^{\frac{d-4}{2}}\left[\gamma_{j,n}^{(d)}\, \cos (2j\Omega) \right]\, .\label{gamsed}
%\eeq
%For the special case of even-dimensional holographic theories ($d=6,8,10,12,14$) dual to Einstein gravity, we have verified that the $\Omega$-dependence of $a^{(d)}(\Omega)$ can be written in the compact form,
%\begin{align}\label{add1}
%a^{(d)}(\Omega) =\frac{\cos^2\Omega}{\sin\Omega} \sum _{j=0}^{\frac{d-4}{2}}\left[\alpha_j^{(d)}\, \cos (2j\Omega) \right]\,,
%\end{align} 
Namely, the universal contribution to the R\'enyi entropy for a (hyper)conical region, which is quadratically logarithmic in the cutoff, is such that the function of the opening angle consists of the four-dimensional result, $\cos^2\Omega/\sin\Omega$, times a linear combination of  the form: $\gamma_{0,n}^{(d)}+\gamma_{1,n}^{(d)}\cos(2\Omega)+ \gamma_{2,n}^{(d)}\cos(4\Omega)+\dots+\gamma^{(d)}_{(d-4)/2,n} \cos\left((d-4)\Omega\right)$.  
%The coefficients $\gamma_{j,n}^{(d)}$ are given by linear combinations of the trace-anomaly coefficients --- see \req{gamm} for the general $d=4$ and $d=6$ cases --- in the case of the entanglement entropy, and certain functions of the R\'enyi index reducing to those when $n=1$.
The only theory-dependent input appears through coefficients $\gamma_{j,n}^{(d)}$ which, in the entanglement entropy case, are linear combinations of the trace-anomaly charges characterizing  the corresponding CFT ---\eg  for four-dimensional theories, $\gamma_{0,n}^{(4)}=f_b(n)/4$, and for six-dimensional theories, $\gamma_{0,1}^{(4)}$ and $\gamma_{1,1}^{(4)}$ can be obtained from the results in \cite{Safdi:2012sn,Miao:2015iba,Bueno4}.

%In particular, it was previously known that  \req{gamsed} holds for general CFTs in $d=4$ and $d=6$, with \cite{}
%\comment{fix}
%\begin{align}\label{gamm}
%\gamma_{0,n}^{(4)}&=\frac{1}{4}f_b(n)\, , \\ 
%\gamma_{0,n}^{(6)}&=\, ,  \quad \gamma_{1,n}^{(6)}= 
%\end{align}
%where ...

The universal contribution to the R\'enyi entropy of smooth entangling regions on even-dimensional CFTs is logarithmically divergent, the universal coefficient given by sums of local integrals on the corresponding entangling surfaces $\Sigma$, weighted by linear combinations of the trace-anomaly coefficients \cite{Solodukhin:2008dh,Safdi:2012sn,Fursaev:2012mp,Miao:2015iba,Lewkowycz:2014jia}. One of the terms is always controlled by the ``a-type'' charge (or its R\'enyi generalization), whose corresponding local integral is proportional to the Euler characteristic of $\Sigma$ ---namely, it corresponds to the $(d-2)$-dimensional Euler density %$\frac{(d-2)}{2}$-th order Lovelock term 
integrated over $\Sigma$. On the (hyper)conincal surfaces, all these intrinsic-curvature terms vanish, except at the tip, where all curvature is concentrated, in a way such that if we close the cone at some finite distance, the contribution from this term equals the corresponding topological invariant. Since we work with semi-infinite cones, we shall not
be concerned by this contribution. In addition, there are a number of contributions which involve integrals of various combinations of the extrinsic curvature of $\Sigma$. These must be invariant under local diffeomorphisms on $\Sigma$, and their possible linear combinations must be chosen such that conformal invariance is respected. Attending to the first requirement, we can divide the different terms into two categories. The first class of terms takes the generic form
%\begin{equation}\label{ktrk}
%k^{(nm-(m-1)(d-2))} \left[ \tr k^{(d-2)-n} \right]^{m}\, , 
%\end{equation} 
\begin{equation}\label{ktrk}
k^n \left[ \tr k^{m} \right]^{s}\, , 
\end{equation} 
where we use the notation $
(\tr k^m)^s \equiv  (k_{a_1}\,^{a_2} k_{a_2}\,^{a_3} \dots k_{a_{m}}\,^{a_1})^s
$,
and
where $n$, $m$ and $s$ are non-negative integers constrained to satisfy $(n+ms)=(d-2)$, $m\geq 2$.  
Hence, for example, in $d=4$ all possible terms reduce to two: $k^2$ and       
$
\tr k^2$. In $d=6$ there are more options, namely
\begin{equation}
k^4\, , \quad k^2 \tr k^2 \, , \quad k \tr k^3\, , \quad \tr k^4\, , \quad (\tr k^2)^2\, .
\end{equation}
In $d=8$, in turn, we have eight possibilities,
\begin{equation}
k^6\, , \quad k^4 \tr k^2 \, , \quad k^3 \tr k^3\, , \quad k^2 \tr k^4\, , \quad k \tr k^5 \, , \quad  \tr k^6 \, , \quad k^2 (\tr k^2)^2\, , \quad (\tr k^2)^3\, .
\end{equation}
And similarly in higher dimensions.

The second class of terms ---which in fact includes the first as a particular case--- involves covariant derivatives of the extrinsic curvature. The most general form of one of those terms consists of some contraction of
\begin{equation}\label{typ2}
\underbrace{k_{\cdot\cdot} \cdots  k_{\cdot\cdot}}_{\alpha_1} \,  \underbrace{(\nabla_{\cdot} k_{\cdot\cdot})  \cdots  (\nabla_{\cdot} k_{\cdot\cdot})}_{\alpha_2} \,  \underbrace{ (\nabla_{\cdot} \nabla_{\cdot}k_{\cdot\cdot} ) \cdots ( \nabla_{\cdot} \nabla_{\cdot}k_{\cdot\cdot})}_{\alpha_3} \cdots \underbrace{ (\nabla_{\cdot} \nabla_{\cdot}\cdots \nabla_{\cdot} k_{\cdot\cdot})  \cdots   (\nabla_{\cdot} \nabla_{\cdot}\cdots \nabla_{\cdot} k_{\cdot\cdot}  )}_{\alpha_r} \, ,
\end{equation}
with $\frac{1}{2}\sum_i^r (i+1) \alpha_i$ inverse metrics. Observe that in \req{typ2} we have $\alpha_{\rm tot}\equiv \sum_j^r \alpha_j$ terms in total, with $\alpha_j$ of them involving $j-1$ covariant derivatives of the extrinsic curvature. The total number of terms is always an even number, $\alpha_{\rm tot}=$ even, and it is bounded below and above by
$
2\leq \alpha_{\rm tot}\leq d-2
$. Similarly, the $\alpha_j$ are constrained to satisfy 
\begin{equation}
\sum_{i=1}^r i\cdot \alpha_i=d-2\, .
\end{equation}
This captures the intuition that, for dimensionality purposes, one covariant derivative counts as one extrinsic curvature. 
For example, in $d=4$, this condition reads $\alpha_1+2 \alpha_2= 2$, which means that, in principle, we could have contractions of two extrinsic curvatures ($\alpha_1=2$) or, alternatively, one contraction involving $\nabla_a k_{bc}$, ($\alpha_2=1$), but since the total number of terms must be even, the latter case is discarded, and no terms involving covariant derivatives are allowed.

The particular combination of terms appearing in the entanglement/R\'enyi entropy expressions and their respective weights as functions of the coefficients appearing in the trace-anomaly expressions have been worked out explicitly in $d=4$ and $d=6$ \cite{Solodukhin:2008dh,Safdi:2012sn,Miao:2015iba}. As we show now, in the case of entangling surfaces consisting in (hyper)cones, a general pattern exists regarding the functional dependence of $a_n^{(d)}(\Omega)$.

The metric of $d$-dimensional Minkowski space in hyperspherical coordinates can be written as
\begin{equation}
ds^2=-dt^2 + dr^2 + r^2\left[d\theta^2+ \sin^2\theta( d\theta_1^2+\sin^2\theta_1 (d\theta_2^2+\sin^2\theta_2^2(d\theta_3^2+\sin^2\theta_3( \dots\right]\, ,
\end{equation}
where the coordinate ranges are: $\theta,\theta_1,\theta_2,\dots,\theta_{(d-4)} \in [0,\pi]$ and $\theta_{(d-3)}\in [0,2\pi)$, which is the usual $\phi$ coordinate in $d=4$. Hypercones are parametrized by equations: $t=0$ and $\theta=\Omega$. It is then straightforward to obtain the corresponding induced metric, and its determinant, 
\begin{align} 
ds^2_h&=dr^2+r^2\sin^2\Omega \left[d\theta_1^2+\sin^2\theta_1(d\theta_2^2+\dots\right] , \, \\
\sqrt{h}&=\left(r \sin\Omega\right)^{(d-3)}\,    \prod_{j=1}^{d-4} \sin^{(d-3-j)}\theta_{j}\, .
\end{align} 
The only non-vanishing components of the extrinsic curvature associated to the normal vector $n=\frac{1}{\sqrt{g_{\theta\theta}}}\partial_{\theta}$ read\footnote{The extrinsic curvature associated to the other normal, $n^{(1)}= \partial_t$, is trivial in this case and we just ignore it here.}
\begin{equation}\label{kco}
k_{\theta_1 \theta_1}=r \cos \Omega \sin \Omega\, ,\quad k_{\theta_i \theta_i}=k_{\theta_1\theta_1} \prod_{j=1}^i \sin^2\theta_{j}\, , \quad i=2,\dots,d-3\, .
\end{equation}
Now, using the fact that the only non-vanishing components of $h^{ab}$ are given by
\begin{equation}\label{invco}
h^{\theta_1\theta_1}=\frac{1}{r^2\sin^2\Omega}\, , \quad h^{\theta_i\theta_i}=h^{\theta_1\theta_1}  \prod_{j=1}^i \frac{1}{\sin^2\theta_j}\, , \quad i=2,\dots,d-3\, ,
\end{equation}
it is easy to show that 
\begin{align}
\int_{\Sigma} d^{d-2}y \sqrt{h} k^n \left[ \tr k^{m} \right]^{s} &=\frac{ \cos^{(d-2)}\Omega}{\sin\Omega} (d-3)^{(n+s)} \int \frac{dr}{r}  \int_0^\pi 2\pi \prod_{j=1}^{d-4} \sin^{(d-3-j)}\theta_{j} d\theta_j \, , \\ \notag 
&= \frac{2\pi^{\frac{d-2}{2}}(d-3)^{(n+s)} }{\Gamma[\frac{d-2}{2}]} \frac{ \cos^{(d-2)}\Omega}{\sin\Omega}\int \frac{dr}{r}\, .
\end{align}
Here, we observe the appearance of a $\int dr/r$ factor which, independent of the dimension, combines in each case with the overall $\log \delta$ to produce a $\log^2\delta$ divergence in exactly the same way as for the $d=4$ case discussed in Section \ref{vertex} ---and again there will be a missing $1/2$ factor. We also observe that the dependence on the cone opening angle is modified with respect to the $d=4$ case by the appearance of a different power for $\cos \Omega$. We can alternatively think of this as producing additional terms proportional to $\cos (2 j\Omega)$ with $j=1,\dots,(d-4)/2$  multiplying the overall $d=4$ result. Indeed, using the identity
\begin{equation} \label{exps}
 \cos^{(d-2)}\Omega=\frac{\cos^2\Omega}{2^{d-4}} \Big[ \begin{pmatrix} d-4\\ \frac{d-4}{2} \end{pmatrix}+2\sum_{j=1}^{\frac{d-4}{2}} \begin{pmatrix} d-4\\ \frac{d-4-2j}{2} \end{pmatrix} \cos(2j\Omega) \Big]\, ,
\end{equation}
we find
\begin{align}
\int_{\Sigma} d^{d-2}y \sqrt{h} k^n \left[ \tr k^{m} \right]^{s} =&\frac{\pi^{\frac{d-2}{2}}(d-3)^{(n+s)} }{2^{d-5}\Gamma[\frac{d-2}{2}]} \frac{ \cos^{2}\Omega}{\sin\Omega}\Big[ \begin{pmatrix} d-4\\ \frac{d-4}{2} \end{pmatrix} \\ &+2\sum_{j=1}^{\frac{d-4}{2}} \begin{pmatrix} d-4\\ \frac{d-4-2j}{2} \end{pmatrix} \cos(2j\Omega) \Big]\int \frac{dr}{r}\, , 
\end{align}
which takes the form suggested in \req{gamsed}. 

We turn now to the second type of contributions, \ie those of the general form \req{typ2}. We cannot be as explicit in this case, but the general pattern can be also understood. As we explained earlier, integrands for this kind of terms will take the generic form
\begin{equation}\label{jss}
\int_{\Sigma} d^{d-2}y\sqrt{h}\,\underbrace{k_{\cdot\cdot} \cdots}_{a_1}  \,  \underbrace{(\nabla_{\cdot} k_{\cdot\cdot})  \cdots }_{a_2} \,  \underbrace{ (\nabla_{\cdot} \nabla_{\cdot}k_{\cdot\cdot} ) \cdots }_{a_3}\, \cdots \, \underbrace{ (\nabla_{\cdot} \nabla_{\cdot}\cdots \nabla_{\cdot} k_{\cdot\cdot})  \cdots  }_{a_r}\, \underbrace{h^{\cdot\cdot} \cdots h^{\cdot\cdot}}_{\frac{1}{2}\sum_i^r(i+1)a_i}\, .
\end{equation}
Using \req{kco} and \req{invco}, we find that in the case of the hypercones this generically reduces to
\begin{align}
\sim \frac{2\pi^{\frac{d-2}{2}}  }{\Gamma[\frac{d-2}{2}]} \frac{ \cos^{2}\Omega}{\sin\Omega} \cos^{(\alpha_{\rm tot}-2)}\Omega \sin^{(d-2-\alpha_{\rm tot})}\Omega\int \frac{dr}{r}\, ,
\end{align}
up to a constant that depends on the dimension and the specific index structure in \req{jss}. %For example, a contribution of the form $
%k^{bc} k \nabla_{a}k_{bc}\nabla^ak 
%$ in $d=8$ produces 

Now $\sin^{(d-2-\alpha_{\rm tot})}\Omega$ can be expanded in even powers of $\cos \Omega$ so that
\begin{equation}
\cos^{(\alpha_{\rm tot}-2)}\Omega\sin^{(d-2-\alpha_{\rm tot})}\Omega=\sum_{k=0}^{\frac{d-2-\alpha_{\rm tot}}{2}} \begin{pmatrix}  \frac{d-2-\alpha_{\rm tot}}{2} \\ k  \end{pmatrix} (-1)^k \cos^{(2k+\alpha_{\rm tot}-2)}\Omega\, .
\end{equation}
Finally, expanding $\cos^{(2k+\alpha_{\rm tot}-2)}\Omega$ as in \req{exps} we are left with a linear combination of terms of the form $\cos(2j \Omega)$. It is straightforward to show that the condition $2\leq \alpha_{\rm tot}\leq d-2$ constrains the possible values of $2j$ to $2j=0,2,\cdots,d-4$, in agreement with the conjectural relation \req{gamsed}, which therefore holds in general. In passing, we note that we have actually verified explicitly that \req{gamsed} holds for the holographic entanglement entropy of CFTs dual to Einstein gravity in $d=4,6,8,10,12$. In that case, the bulk action is given by \req{Einst}. Closely following the calculations in \cite{Myers:2012vs} for the $d=4,6$ cases, we obtain
\begin{align}
a^{(4)}(\Omega)&= \frac{\cos^2\Omega}{\sin \Omega}\left(\frac{\pi L^3}{32 G}\right)\, , \\
a^{(6)}(\Omega)&= \frac{\cos^2\Omega}{\sin \Omega} \left( \frac{9\pi^2 L^5}{16384 G}\right) \left[31-\cos(2\Omega)\right] \, , \\
a^{(8)}(\Omega)&= \frac{\cos^2\Omega}{\sin \Omega} \left(\frac{25\pi^3 L^7}{95551488 G}\right)\left[22353-964\cos(2\Omega)+11\cos(4\Omega)\right] \, ,  \\
a^{(10)}(\Omega)&= \frac{\cos^2\Omega}{\sin \Omega} \left(\frac{49\pi^4 L^9}{14843406974976 G}\right) \left[449662142 - 21823587 \cos(2\Omega) \right. \\ &\left. + 370802 \cos(4\Omega) - 2669 \cos(6\Omega) \right]\, ,  \\
a^{(12)}(\Omega)&= \frac{\cos^2\Omega}{\sin \Omega} \left(\frac{27\pi^5 L^{11}}{83886080000000000 G}\right)\left[930830869835 \right. \\ &\left. - 48180346664 \cos(2\Omega)+ 977555068 \cos(4\Omega)- 
  11174552 \cos(6\Omega) \right. \\ & \left. + 53881 \cos(8\Omega)\right]\, , 
\end{align}
which indeed respect the aforementioned angular dependence.

%Putting all pieces together, we observe that for general even-dimensional CFTs, 

Observe that, on general grounds, in the sharp and almost-smooth limits, the functions $a_n^{(d)}(\Omega)$ behave similarly to the $d=4$ case (or the $d=3$ corner), namely
\begin{align}
a_n^{(d)}(\Omega)&\overset{\Omega\rightarrow 0}{=} \sum_{j=0}^{\frac{d-4}{2}} \left[\frac{1 }{\Omega} -  \left[\frac{5}{6}+2j^2 \right] \Omega +\mathcal{O}(\Omega^2) \right]  \gamma_{j,n}^{(d)}\, , \\ 
a_n^{(d)}(\Omega)&\overset{\Omega\rightarrow \pi/2}{=}  \sum_{j=0}^{\frac{d-4}{2}} \left[ \left(\Omega-\frac{\pi}{2} \right)^2 +  \left[\frac{1}{6}-2j^2 \right]  \left(\Omega-\frac{\pi}{2} \right)^4 + \mathcal{O}\left(\Omega-\frac{\pi}{2}\right)^6 \right] (-1)^j \gamma_{j,n}^{(d)} \, .
\end{align}

\bibliographystyle{JHEP}
\bibliography{Gravities}
%\label{biblio}

\end{document}